\documentclass{aa}
\usepackage{txfonts} 
\usepackage{lscape}
\usepackage{longtable}
\usepackage{graphicx}

\usepackage{natbib}
\bibpunct{(}{)}{;}{a}{}{,} 

\begin{document}

\title{What produces the extended LINER-type emission in the NUGA galaxy \object{NGC 5850}?\thanks{Based on observations obtained with VIMOS at ESO VLT under program ID 083.B-0906(A).}}

\titlerunning{Extended LINER-Type Emission in NGC 5850}
\authorrunning{Bremer et al.}

\subtitle{}

   \author{M. Bremer\inst{1}
          \and
          J. Scharw\"achter\inst{2}
          \and 
          A. Eckart\inst{1,3}
          \and 
          M. Valencia-S.\inst{1} 
          \and
          J. Zuther\inst{1}
          \and
          F. Combes\inst{2}
          \and
          S.~Garcia-Burillo\inst{4}
          \and
          S. Fischer\inst{1}}
\offprints{M. Bremer (mbremer@ph1.uni-koeln.de)}

   \institute{ I.Physikalisches Institut, Universit\"at zu K\"oln,
              Z\"ulpicher Str.77, 50937 K\"oln, Germany\label{inst1}\\
              \email{mbremer, eckart, zuther, fischer, mvalencias@ph1.uni-koeln.de}
         \and
             Observatoire de Paris, LERMA (CNRS: UMR8112), 61 Av. de l'Observatoire, 75014 Paris, France\label{inst2}\\
	     \email{julia.scharwaechter, francoise.combes@obspm.fr}
         \and
             Max-Planck-Institut f\"ur Radioastronomie, 
             Auf dem H\"ugel 69, 53121 Bonn, Germany\label{inst3}\
         \and
         	 Observatorio Astronómico Nacional (OAN)-Observatorio de Madrid, Alfonso XII, 3, 28014, Madrid, Spain\label{inst4}\\
         \email{s.gburillo@oan.es}}

\date{Received enter date / Accepted enter date}

\abstract{The role of low ionization nuclear emission region (LINER) galaxies within the picture of active galactic nuclei (AGN) has been controversial. It is still not clear whether they host an AGN in a low accretion mode, or whether they are not active at all but dominated by alternative ionization mechanisms, namely shocks, winds/outflows, or photoionization by a post-asymptotic giant branch (p-AGB) stellar population. The detection of extended LINER-like emission was often taken as evidence of ionization by stellar components, but this has not been undisputed.}
{Using optical spectroscopy, we examine the possible ionization mechanisms responsible for the extended LINER-like emission in the central $\sim4~\mathrm{kpc}$ of NGC 5850.}
{We performed integral field spectroscopic observations using VIMOS at the VLT, which provides spatially-resolved spectra for the gas emission and the stellar continuum. We subtract the underlying stellar continuum from the galaxy spectra and fit the emission lines. With these methods, we derive and analyze emission line and kinematic maps. Emission line ratio maps are examined by means of diagnostic diagrams.}
{The central few kpc of NGC 5850 are dominated by extended LINER-like emission. The emission-line ratios that are sensitive to the ionization parameter increase with radial distance to the nucleus. The LINER-like region is surrounded by emission that is classed as `composite' in terms of diagnostic diagrams. Two star-forming (SF) regions are present in the $21\arcsec\times19\arcsec$ field of view. One of them is located approximately in the ring, surrounding the kinematically decoupled core. The second one is close to the nucleus and is the origin of a region of decreased emission line ratios oriented radially outwards. We find the interstellar gas to have a complex kinematic morphology and areas of steep velocity gradients.}
{The extended LINER-like emission in NGC 5850 is dominated by ionization from distributed ionization sources, probably by stars on the p-AGB. The extended `composite' emission is likely due to a mixture of a LINER-like ionization pattern and photoionization by low-level star formation. With the extended region of decreased emission line ratios, we possibly observe enhanced SF or a region that is shielded from the central LINER-like emission by the central \ion{H}{ii}-region. The peculiar gas kinematics are probably caused by the lopsided ($\mathrm{m}=1$) distribution of the gas and make the inflow of gas toward the center appear possible.}


\maketitle

\section{Introduction}
\label{Intro}

Low-ionization nuclear emission regions (LINERs) were first defined by \citet{Heckman_Liner}. They are characterized by high-luminosity emission lines of low-ionization species (e.g. [\ion{N}{ii}] $\lambda6585$, [\ion{S}{ii}] $\lambda\lambda6718,6732$) with respect to the Balmer recombination lines and a moderate [\ion{O}{iii}] $\lambda5008$ luminosity compared to those of Seyfert galaxies. The bolometric luminosity of LINERs is weak, even when compared to the Seyfert galaxies. While Seyfert galaxies have clearly an active galactic nucleus (AGN) and are therefore driven by accretion onto a super-massive black hole (SMBH), it has been a long-standing matter of debate whether LINERs represent the low-luminosity (and/or low-accretion) tail of the AGN family or represent a class on their own. Given their substantial frequency of appearance (more than $30~\%$ of all galaxies and $60~\%$ of Sa/Sab spirals with $B \leq 12.5~\mathrm{mag}$; \citet{Ho_dwarf}) LINERs pose an important piece in the picture of galaxy/AGN evolution. In principal the debate is rooted in the finding that the observed line intensities in the circumnuclear gas of LINERs can be sufficiently powered by compact UV cores. Although not excluded, the presence of a non-stellar source like an AGN would not be mandatory. The necessary radiation can at least partially be provided by a compact stellar cluster \citep[e.g.][]{Pogge_2000, Maoz_UV_LINER}. Furthermore, the AGN scenario is particularly challenged by evidence of extended LINER-like emission in an increasing number of galaxies. Therefore, distributions of photoionizing sources have been taken into consideration. Cooling flows \citep{Fabian_Cooling_Flows, Voit_cooling_flow} and photoionization by post-asymptotic giant branch (p-AGB) stars \citep{Terlevich_warmers, Trinchieri_pAGB, Binette_old_stars, Stasinska_pAGB, Sarzi_pAGB} are possible sources of LINER-like emission. Cooling flows are predominantly found in massive galaxies and clusters. Post-AGB stars can only explain LINERs with relatively weak emission lines \citep{Binette_old_stars, Cid_Fernandes_disconnection}. These two mechanisms are not the only candidates for LINER-like emission. The characteristic emission can be shock induced as well. Several authors have calculated and modelled shocked gas in various configurations \citep{Heckman_Liner, Dopita_1996, Kewley_maximum_starburst, Rich_shock_superwind}. Especially in galaxies showing large scale outflows and in galaxy mergers, shock excitation is important \citep{Veilleux_wind_ratio, Lipari_merger_winds, Monreal_Ibero_LIRG, Sharp_cones} and resembles extended LINER-like spectra. In contrast to these intragalactic shock processes, \citet{Farage_spiral_in} observed extragalactic material spiraling inwards on an inclined trajectory in the brightest cluster galaxy \object{NGC 4696}. This filament structure exhibits LINER-like spectra over its entire extension, which is explained by ram-pressure induced shocks as a result of the transonic passage of a gas-rich galaxy. LINER-like spectra also have been reported for ionization cones emerging from the nucleus \citep[e.g.][]{Kehrig_califa}. However, extended emission can be produced by an unresolved point source as well, despite the number of scenarios alternative to AGN \citep[e.g.][]{Yan_nature_of_LINERS}. The resulting profile of the line emission depends on such parameters as the density, filling factor and spatial distribution of the gas.\

The idea of LINERs being low-luminosity AGN (LLAGN) is supported by the detection of compact radio \citep{Nagar_LLAGN_radio} and hard X-ray cores \citep{Gonzalez_LINER_Xray} in many of these objects. Furthermore, the frequency of detection of compact radio cores in LINERs and Seyferts is similar \citep{Ho_review}. To explain the low luminosity of LINERs, models of radiatively inefficient accretion flows and obscuration have been invoked \citep[e.g.][]{Dudik_LINER_AGN, Gonzalez_LINER_AGN}, although \citet{Ho_review} dismissed the obscuration hypothesis for LLAGNs.\\

We present the data of NGC 5850 observed with the optical integral field spectrograph (IFS) VIMOS. The galaxy NGC 5850 is a prototype double-barred early-type spiral galaxy (SBb(sr) I-II) \citep{Buta_5850, Prieto, Friedli_5850}. It is located at a distance of $34.2~\mathrm{Mpc}$ \citep{Wozniak95}. Its two spiral arms form a ring-like structure, almost resembling the morphology of the center, where the inner bar (projected half-length of $6\farcs2$) is surrounded by a (pseudo-) ring. The spiral arms are distorted $\sim 2\arcmin$ west and southwest from the center, probably due to a recent ($< 200~\mathrm{Myr}$) high-velocity encounter with \object{NGC 5846} \citep{Higdon_encounter}, which is $10\arcmin$ to the northwest of NGC 5850. In optical color images, \citet{Lourenso_stellar} found the large scale (primary) bar and the inner (secondary) bar to differ in their projected position angle by $-67\degr$ $(\mathrm{P.A._{secondary}\approx 49.3\degr}$). The nested bar morphology is not a very rare feature. \citet{Bar_fraction} found in their sample of $112$ nearby ($v_{\mathrm{hel}}<6000~\mathrm{km~s^{-1}}$) galaxies that $(28\pm5)\%$ of the barred spiral galaxies and $(13\pm4)\%$ of non-Seyfert galaxies are double-barred. Observations of NGC 5850 with the optical IFS SAURON  by \citet{sigma_hollows_lorenzo} showed significant drops in the stellar velocity dispersion exactly at the tips of the inner bar, which they attribute to contrast effects between the high dispersion of the bulge and the low dispersion of the bar. The existence of a gaseous polar disk, instead of the secondary bar, was suggested by \citet{doublebarred_structure_moissev}, who also used an optical IFS. They also found the core to be kinematically decoupled. From here on, we refer to the kinematically decoupled core as KDC. The low X-ray luminosity $L_X(0.5-3\ \mathrm{keV})=10^{40.36}\ \mathrm{erg\ s^{-1}}$ suggests NGC 5850 to be in a state of low activity, and \citet{5850_Lx} did not consider it likely to be dominated by an AGN. The high luminosity of [\ion{N}{ii}] lines at the center classifies it as a LINER \citep{Heckman_Liner, Liner_reference}.   

Integral field spectroscopy in the visible wavelength regime allows us to study the excitation mechanisms of the gas and to investigate the gas kinematics in detail. The proximity (high angular resolution), the low luminosity (no outshining by the AGN), and the double-barred structure of NGC 5850 make it an ideal object to study the gas excitation and kinematics in a LINER galaxy.\

After describing the observations and how we treat the data to retrieve the needed information in Sect. \ref{data}, we describe the morphologic, kinematic, and excitational properties of NGC 5850 (Sect. \ref{Results}). In Sect. \ref{discussion}, we discuss the results within different assumed scenarios or under certain aspects (i.e., \ion{H}{ii}-regions). We summarize our results in Sect. \ref{summary}.\

Throughout this paper we assume $\mathrm{H_0}=75~\mathrm{km~s^{-1}~Mpc^{-1}}$ with standard $\mathrm{\Lambda}$CDM cosmology ($\mathrm{\Omega_{matter}}=0.27\mathrm{,~ \Omega_{vacuum}}=0.73$).

\section{Observations, data reduction, line measurements}
\label{data}

\subsection{Observations and data reduction}
\label{Obs}

The observations were undertaken in April and May 2009 at the ESO Very Large Telescope (VLT) unit \emph{Melipal} with the Visible Multi-Object Spectrograph (VIMOS; \citet{VIMOS}) in Paranal, Chile. The VIMOS was configured in IFU mode with high spectral resolution for the blue ($4100-6300~\AA$) and the red ($6500-8650~\AA$) grating and a spectral sampling of $0.54~\AA$ and $0.58~\AA$, respectively. This provides a spectral resolving power of $\mathrm{R} \sim 2700$ at $5500~\AA$ and $\sim 3200$ at $7500~\AA$. The opening angle of the lenslets attached to the fibres is $0\farcs66$, resulting in a field of view (FOV) of $27\arcsec\times27\arcsec$ with $40\times40$ pixels. The observations were conducted for each grating separately with an integration time of $755~\mathrm{s}$ per exposure. Three images per grating were taken, consisting of $1600$ spectra each. For each set of three images, the second one was pointed to an apparently source-free region to observe the sky emission. Overall, we have eight on-source exposures in the blue grating and twelve in the red grating with a total integration time of $6040~\mathrm{s}$ and $9060~\mathrm{s}$, respectively. The exposures are seeing limited with the worst seeing of $\mathrm{FWHM} = 0\farcs86$ (blue) and $0\farcs84$ (red).\

We used the VIMOS pipeline in version 2.3.3 for the basic data reduction steps, which comprises bias correction, spatial extraction, wavelength calibration, relative transmission correction and flux calibration.\ 
The exposures suffered from a `fringing-like' pattern \citep{Lagerholm_fringing}, a spectral and spatial high frequency variation depending on the telescope orientation. The possible origin of fringing in the case of VIMOS is discussed in \citet{Jullo_fringing}. A technique to correct for it has been described in \citet{Lagerholm_fringing} but is only recommended for objects similar to early-type galaxies with spatially slowly changing spectra. For our data, we corrected for the fringing-like pattern based on the flat images taken in the same night as the science exposures. We fitted a medium-order polynomial (on order 4 to 6) to each normalized spectrum of the lamp flat exposures and divided the flat spectrum by the fit. The result is the normalized fringe pattern for each fibre, which shows a standard deviation of about $\sigma \sim 2.5 \%$ in the blue grating and $\sigma \sim 8 \%$ in the red one. This approach assumes that the true flat spectrum is dominant on large wavelength scales and does not change significantly on small scales. Furthermore, the fringing-like pattern itself should not change too much with detector position. We found the fit to be less reliable at the high wavelength end in the red grating. Therefore, we discarded the spectrum redwards of $6800~\AA$. This also comprises parts at the edge of the spectrum that are unreliable (e.g., sudden intensity jumps). We divided the normalized flat images by the fit to retrieve the fringe pattern. Simple division of the science image by the fringe pattern removed the problem quite well. Since we discarded the spectral areas that could not be fitted well in the flat image, the fringing had an influence of only $\simeq 5~\%$.\

\begin{figure}
   \centering
   \includegraphics[width=0.5\textwidth]{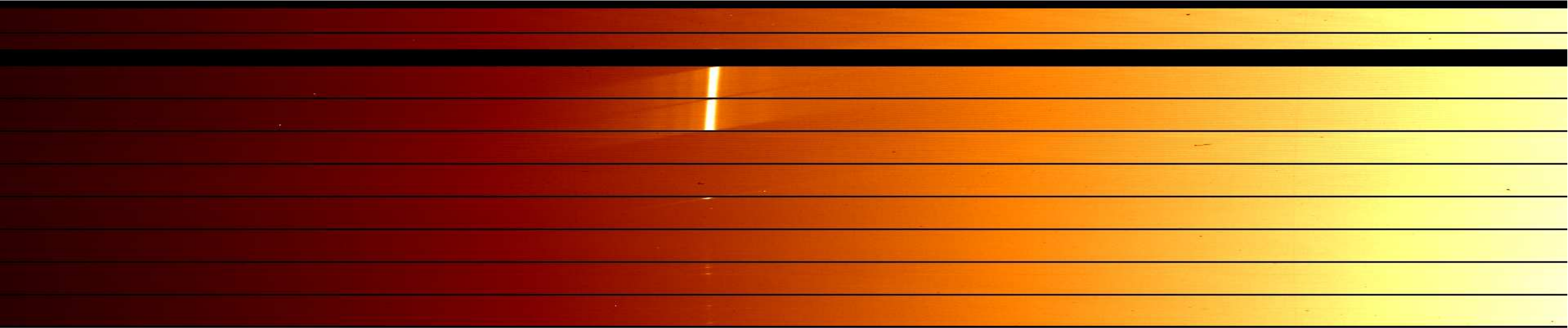}
		\caption{Part of a flat image as extracted by the VIMOS pipeline. Each horizontal line corresponds to a fibre and therefore to a spatial pixel. The wavelength axis extends from left (small $\lambda$) to right (high $\lambda$). The reflection at $\sim5100~\AA$ is clearly visible.} 
   \label{fig:reflection}
   \end{figure}

The flat exposures were contaminated by a strong reflection in quadrant 4 between $5000~\AA$ and $5200~\AA$ (Fig. \ref{fig:reflection}). Since this spectral area contains important emission lines (redshifted [\ion{O}{iii}] $\lambda\lambda 4960,5008$) and influences the fringe correction, we attempted to compensate for this by assuming a perfect detector response in this regime of the spectrum.\

The intensity variation between spatial pixels (spaxels) was corrected by scaling each spectrum to the median emission-line flux of the $5577~\AA$ sky line for the blue grating data and $7316~\AA$ sky line for the red.\

Although dedicated sky observations have been performed, they could not be used because of the unexpected variability of the detected sky lines in intensity and in shape. We estimated the influence of the sky continuum on the galaxy spectrum to be only $5~\%$, and therefore, we decided to correct for contaminations in emission lines only where necessary.\

The individual cubes were combined after aligning them using the galaxy center as reference. We also corrected for differential atmospheric refraction (DAR; \citet{DAR}), which had a maximum influence of about $0.3$ pixels corresponding to $0\farcs2$. Further we calibrated our ground based spectra to the wavelength range in vacuum \citep{Edlen, Morton}. We corrected for Galactic reddening using the extinction law by \citet{Cardelli_extinction} with $E(B-V)=0.056$.\

Up to this point we treated each grating separately. For the latter performed fit of the stellar continuum, it was advantageous to have as many stellar features (e.g., absorption lines) in each spectrum as possible. For this reason we created a common data cube containing both VIMOS gratings as we describe in the following. We found the centroid position of the galaxy center of NGC 5850 to have a slightly different position in the SDSS g-band image compared to the SDSS i-band image. The spatial difference is $0\farcs2$. We used these images as a reference for the blue and red cube, respectively, to spatially align the cubes. The g- and the i-band are not fully covered by our remaining VIMOS spectra but by a dominant fraction. Afterwards, the blue grating spectra were deteriorated to the sampling of the red grating spectra; both were combined and eventually shifted to the rest wavelength.\

\begin{figure}
   \centering
   \includegraphics{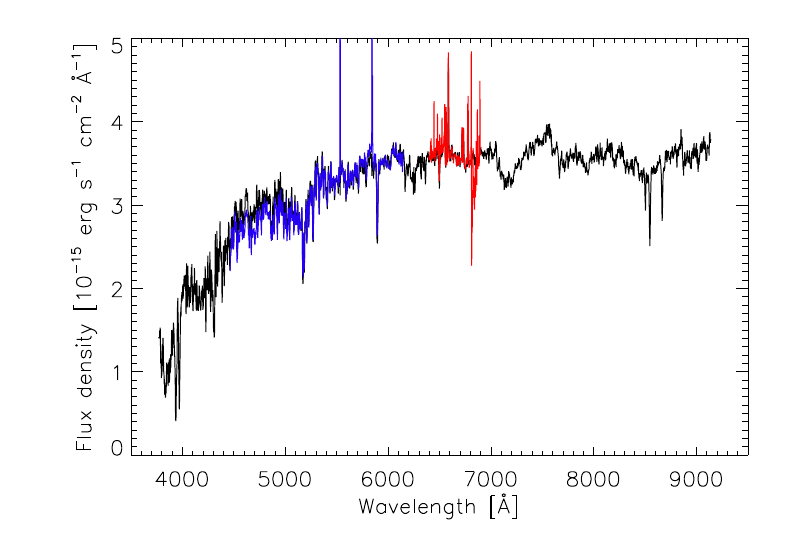}
		\caption{NGC 5850 spectrum from the SDSS with $3\arcsec$ aperture in black. The scaled VIMOS spectrum observed with the blue grating is overplotted in blue, the one observed with the red grating is overplotted in red. All shown VIMOS spectra are extracted from an aperture with a diameter corresponding to the SDSS aperture and considers the seeing conditions. All spectra are shown in the rest-wavelength system. As explained in the text, the spectra are  not corrected for sky contamination.} 
   \label{fig:SDSS_scaling}
   \end{figure}

We used the observed spectrophotometric standard stars for the relative flux calibration of the spectra via the VIMOS pipeline recipe. The absolute flux calibration was performed by scaling the spectra to the available SDSS spectrum. From the VIMOS cube, we extracted a stacked spectrum corresponding to the PSF of the SDSS. The SDSS fibre diameter is $3\arcsec$, and we assumed the worst seeing, given by the 80-percentile seeing information in the header of the SDSS plate with $3\farcs46$. Because there is an observational gap between the gratings ($6145-6400~\AA$) in the VIMOS spectra, we scaled both parts separately. In Fig. \ref{fig:SDSS_scaling}, we show the results of the scaling procedure. Bluewards of $\sim 5000~\AA$, the VIMOS spectrum shows flux densities, which are by $5-10~\%$ lower than the ones of the SDSS spectrum. This is a result of the reflection in the flat spectra mentioned above and its influence on the fringe correction.\

To enhance the S/N, we convolved each cube slice (i.e., image of each wavelength bin) with a Gaussian of $FWHM=2~\mathrm{px}$.\

\subsection{Continuum Subtraction and Emission Line Measurements}
\label{cont_subtract}

\begin{figure*}
   \centering
   \includegraphics[width=\textwidth]{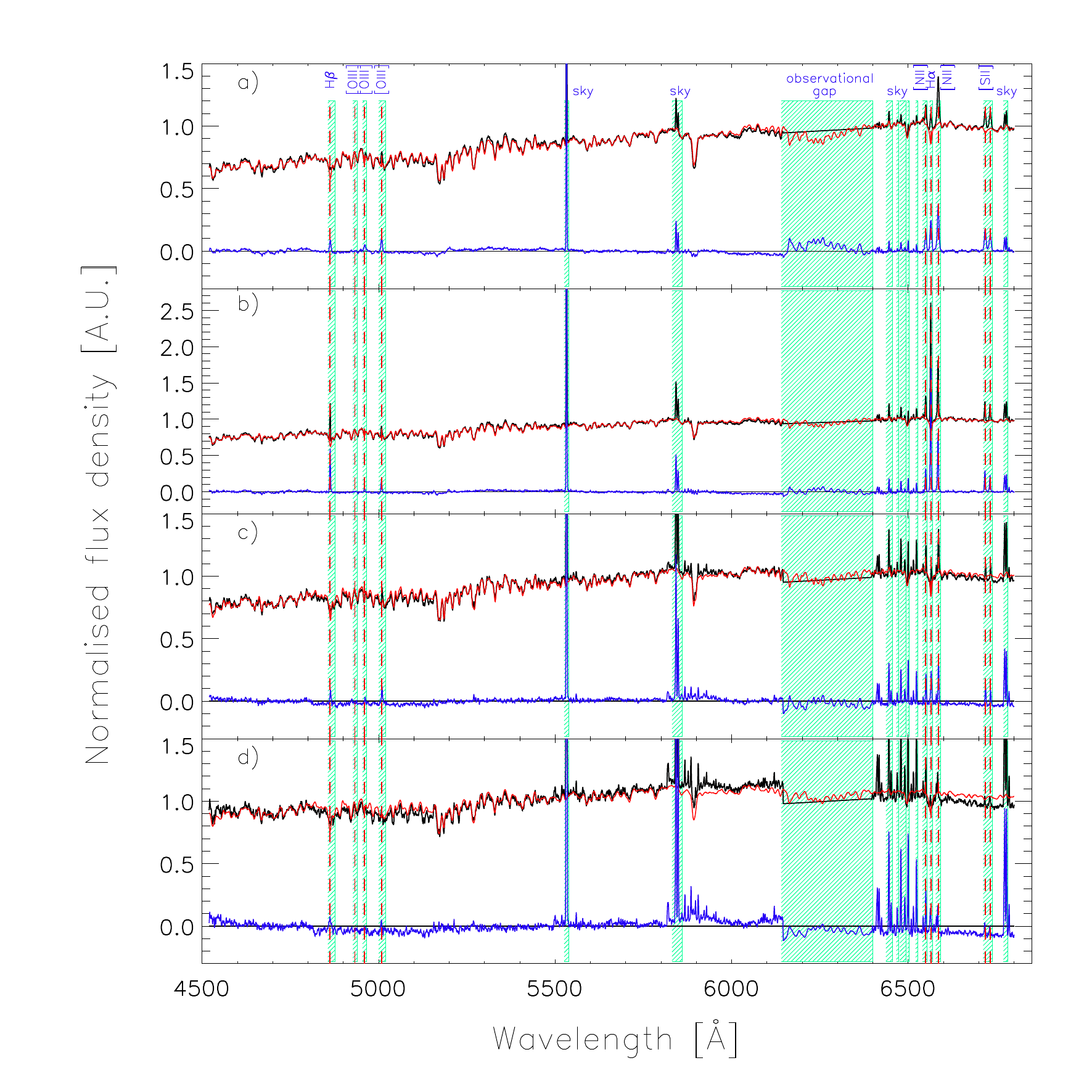}
		\caption{Spectra of different regions (given in arbitrary units; A.U.) within NGC 5850 before stellar subtraction (black) with overplotted stellar continuum fits by STARLIGHT (red) and residual spectrum (blue). The galaxy's emission lines, the most prominent sky emission lines, and the observational gap between the blue and the red grating have been masked and are marked by the green coloured boxes. The panels show the following regions: a) central pixel of NGC 5850, b) a star-formation region, c) a spectrum in a medium distance from the galaxy's center ($\approx 6\arcsec$), and d) a spectrum extracted from a spaxel at the edge of the remaining FOV (ADEV $>4~\%$; see text for details).} 
   \label{fig:stellar_fits}
   \end{figure*}

We used the stellar population synthesis program STARLIGHT \citep{starlight} to subtract the stellar continuum by fitting a linear combination of stellar template spectra to each spaxel in the VIMOS cube. We used the library of theoretical simple stellar population (SSP) models by \cite{BC03}, which is based on a Chabrier initial mass function. We limited the library to a maximum age of $11~\mathrm{Gyr}$. All emission lines, visible sky lines \citep{sky_catalog}, telluric absorption features, artefacts, and the observational gap between the gratings were masked for the STARLIGHT fit. Figure \ref{fig:stellar_fits} shows the spectra, fit results, and residuals for different representative regions. To ensure a minimum quality of the fit, we clipped all spaxels for which STARLIGHT shows an absolute deviation (ADEV) of more than $4~\%$. \

The intensities and the kinematic information of the emission lines were derived by the fitting of a double-component Gaussian to [\ion{N}{ii}] $\lambda6585$ and its sky-line contamination. The peak of the sky-line fit was constrained to a narrow wavelength range instead of a fixed wavelength to consider possible instabilities of the emission line. We derived this wavelength range from spectra in which the sky-line was stronger than the [\ion{N}{ii}] emission. All other emission lines were fitted with a single Gaussian. The [\ion{N}{ii}] $\lambda6550$ emission line was fitted simultaneously with H$\alpha$; both emission lines of the [\ion{S}{ii}] $\lambda\lambda6718+6732$ doublet were fitted together. The fit was additionally constrained by keeping the spectral position difference of the line peaks fixed. We used the IDL routines MPFITEXP and MPFITPEAK for this tasks. The uncertainties were calculated from the fitted parameter errors via error propagation. The input uncertainties are given by the standard deviation of the continuum in a line free region that is close to the respective emission line in the input spectra. As noted by Asari et al. (2007), the SSP models used for the stellar continuum fit have a weak hump around the H$\beta$ line. Consequently, the line base after stellar continuum subtraction is positioned in the negative flux regime. As this is a simple offset, we used the negative continuum as the baseline of the Gaussian fit.\

In general, only emission line detections with a S/N that exceeds $3\sigma$ were used for further analysis (emission line maps, kinematic maps). Additional spaxels were clipped if the uncertainties of the fits were nevertheless high (relative error $ > 33~\%$) or if the contrast in the map benefited from it. One should keep in mind that the use of single Gaussian fits disregards possible line asymmetries. In Sect. \ref{kinematics}, we revisit this topic again. \

The emission line fluxes were corrected for extinction only if necessary - that is, for the calculation of emission line ratios of spectrally distant emission lines, luminosities and values based on them - such as in star-forming rates in high S/N areas. For this, we made use of the Balmer decrement $\mathrm{H\alpha/H\beta}$. We applied the Calzetti extinction law \citep{Calzetti_extinction} and a theoretical Balmer ratio of $2.863$ \citep{Osterbrock_book}, which corresponds to case-B recombination for a region with an assumed temperature of $10^4~\mathrm{K}$ and an electron density of $n_e\sim10^{2}~\mathrm{cm}^{-3}$ as derived from the [\ion{S}{ii}] emission line ratio.\

In summary, the data reduction and the clipping described in Sect. \ref{Obs} and \ref{cont_subtract} leave us with a FOV of $\sim 21\arcsec \times 19\arcsec$ in the largest extensions. The PSF has a size of $FWHM \simeq 2\arcsec$. This value is based on the PSF of a spectrophotometric standard star and has been calculated by considering the worst seeing in our science exposures and smoothing. The spatial sampling is $0\farcs66$. The spectra cover a range from $4520~\AA$ to $6800~\AA$ with an observational gap from $6145-6400~\AA$. The spectral sampling is $0.58~\AA$ and the median spectral resolution is $\mathrm{FWHM}_{\mathrm{line}}\approx 2~\AA$. 

\section{Results}
\label{Results}

\subsection{Morphology}
\label{Maps}

\begin{figure}
  \centering
  \resizebox{\hsize}{!}{\includegraphics{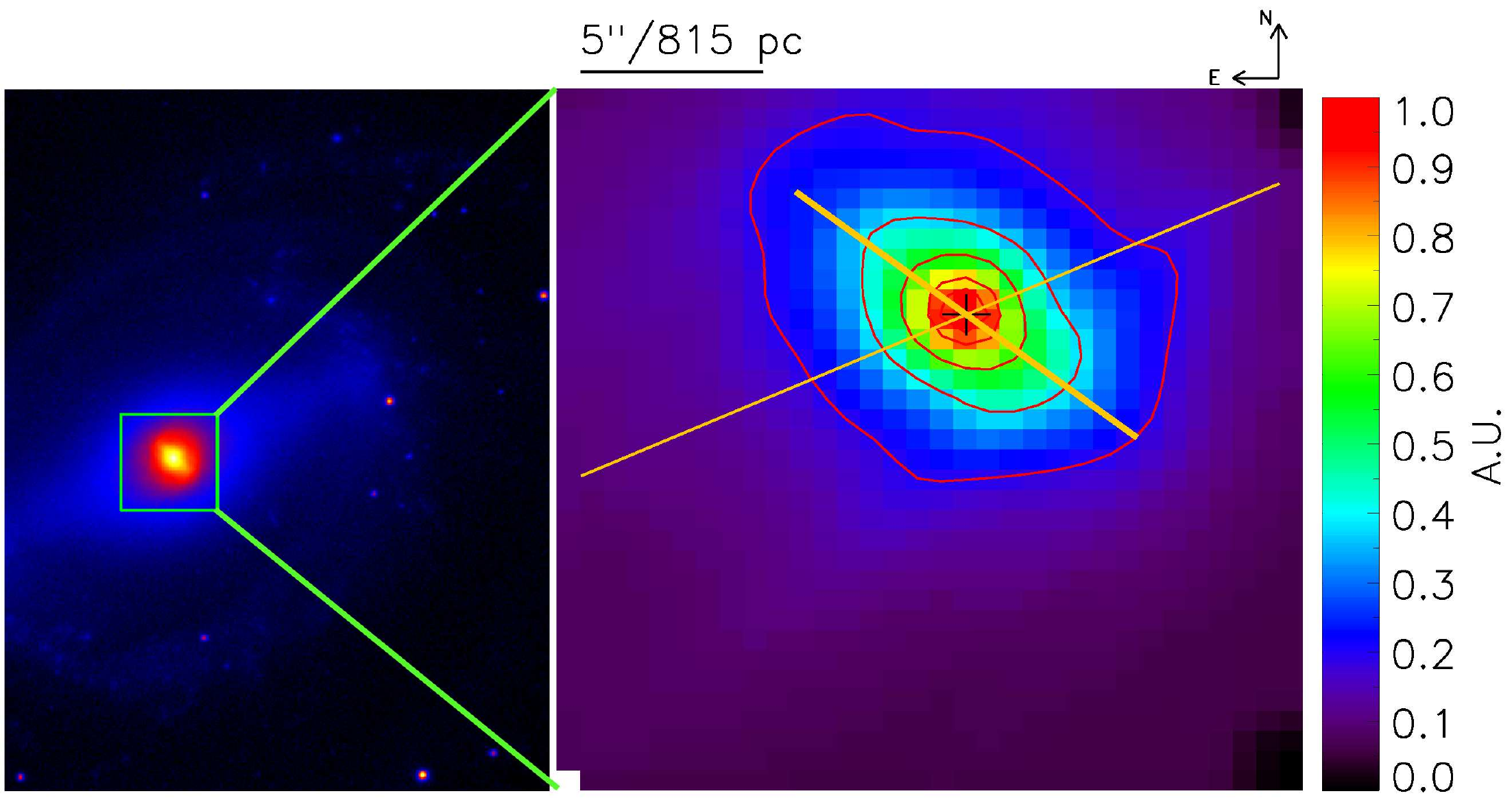}}
  \caption{Left: SDSS r-band image of NGC 5850. Right: Continuum image of NGC 5850 from $6000$ to $6100~\AA$ zoomed in on the central $27\arcsec\times27\arcsec$. The flux density is given in arbitrary units (A.U.). The major axis of the primary (thin line) and the secondary bar (thick line) are overplotted. The contours are given at 20, 40, 60, 80, 90, and 95 percentage of maximum flux density. The continuum peak is marked with a black cross.}
  \label{fig:continuum}
\end{figure}

The $6000$ to $6100~\AA$ continuum image of NGC 5850 (Fig. \ref{fig:continuum}) shows an elongated structure, which strongly peaked in its center, that agrees with the location of the inner (secondary) bar derived by \citet{Lourenso_stellar}. They calculated the position angle (P.A.) of the secondary bar to be $49.3\degr \pm 1\degr$ similar to our own elliptical fits using the IRAF STSDAS ELLIPSE task\footnote{IRAF is distributed by the National Optical Astronomy Observatories, which are operated by the Association of Universities for Research in Astronomy, Inc., under cooperative agreement with the National Science Foundation.}. The locations of the projected primary and secondary bars, as taken from \citet{Lourenso_stellar}, are overplotted in  Fig. \ref{fig:continuum} and all subsequent maps for comparison. The continuum peak is used as the galaxy core throughout the paper.

\begin{figure*}
   \centering
   \includegraphics[width=0.33\textwidth]{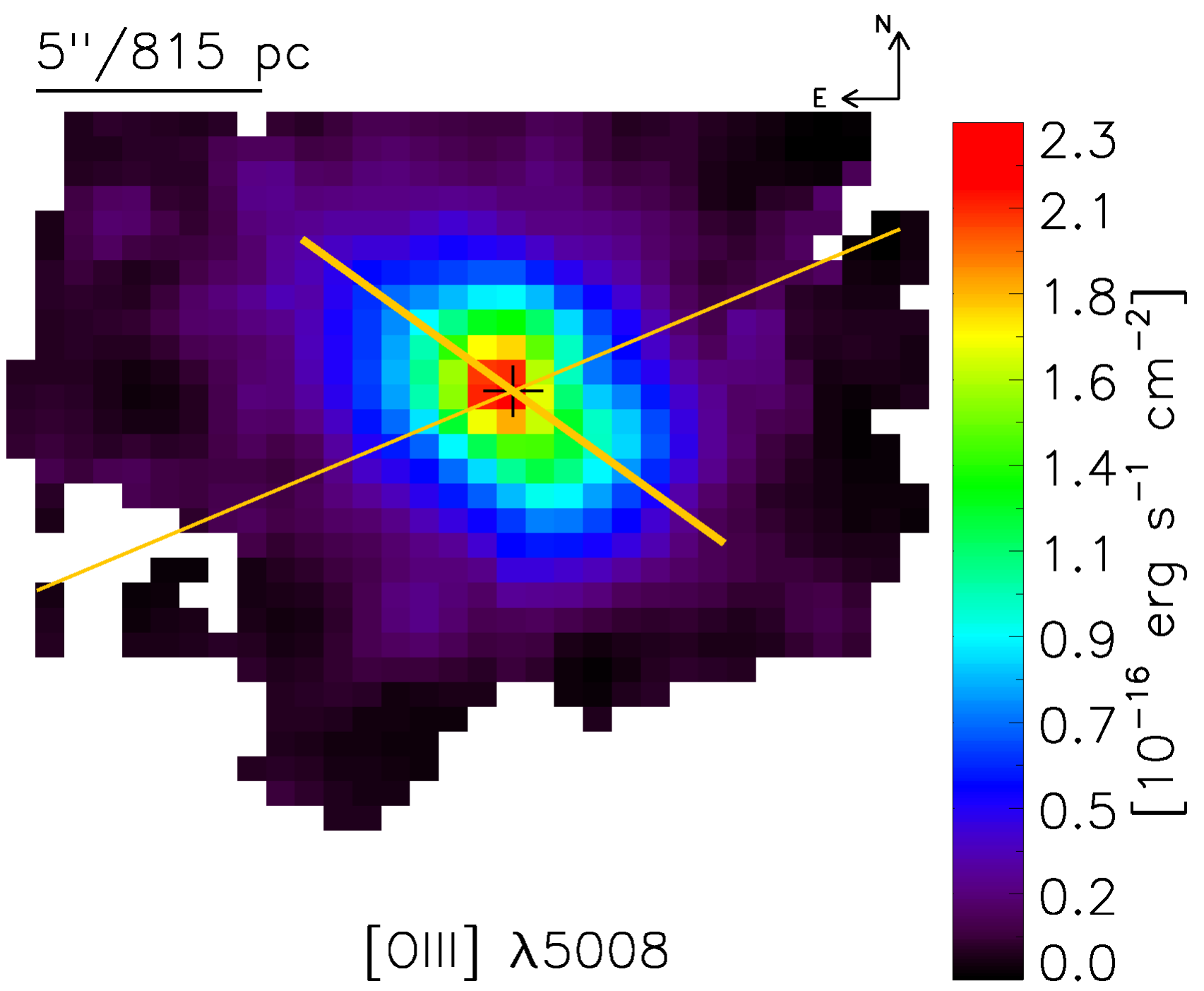}
   \includegraphics[width=0.33\textwidth]{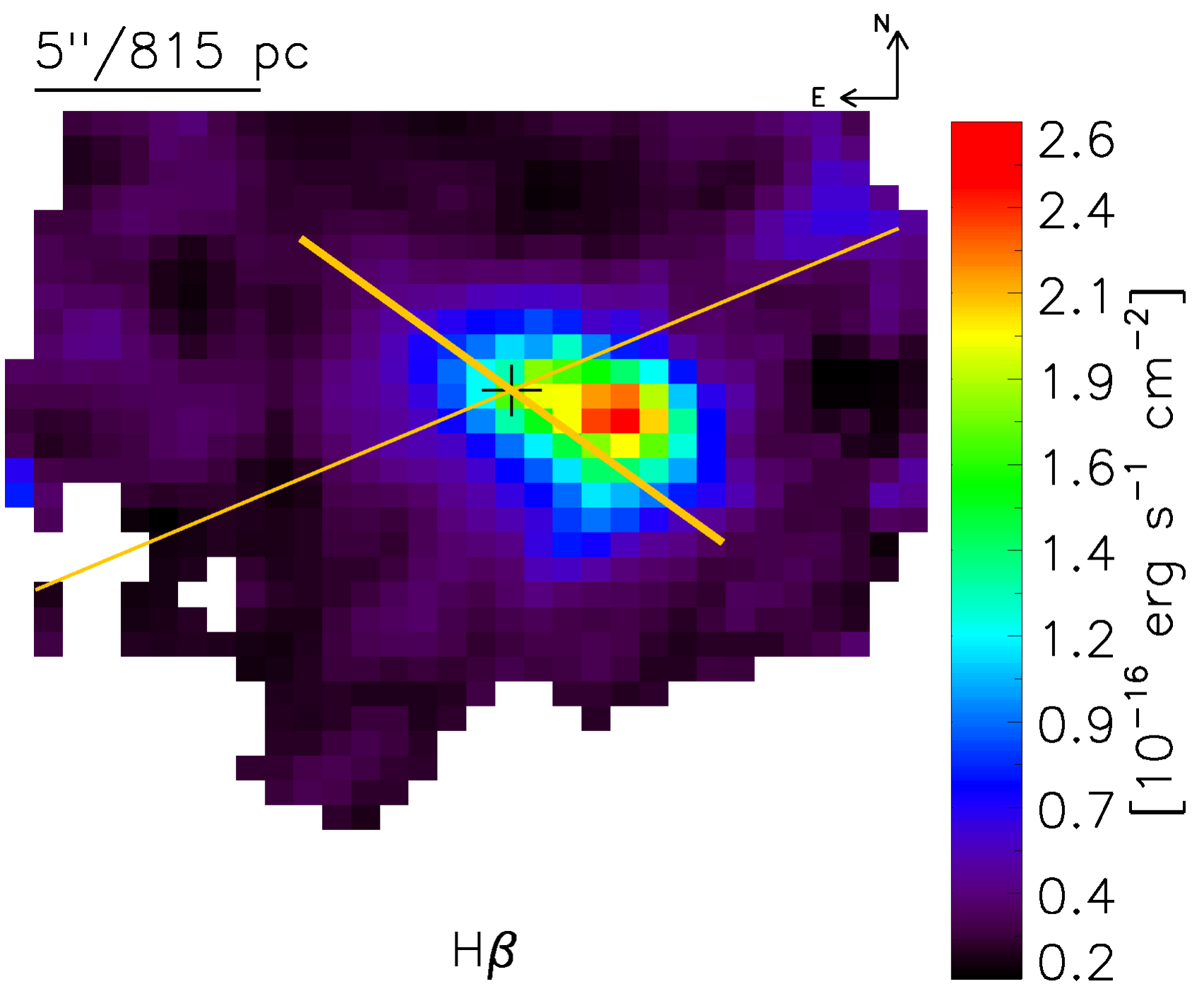}
   \includegraphics[width=0.33\textwidth]{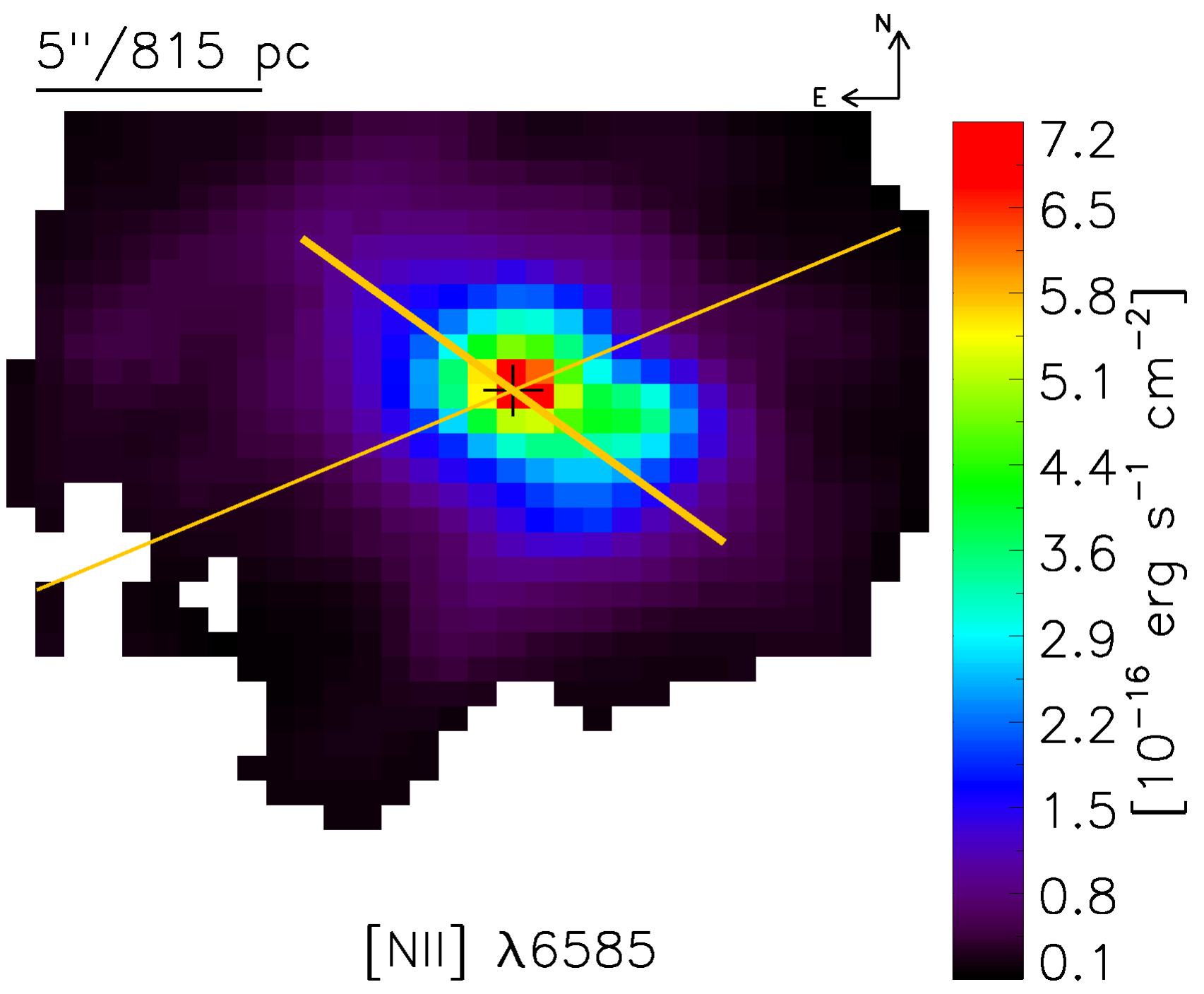}
   \includegraphics[width=0.33\textwidth]{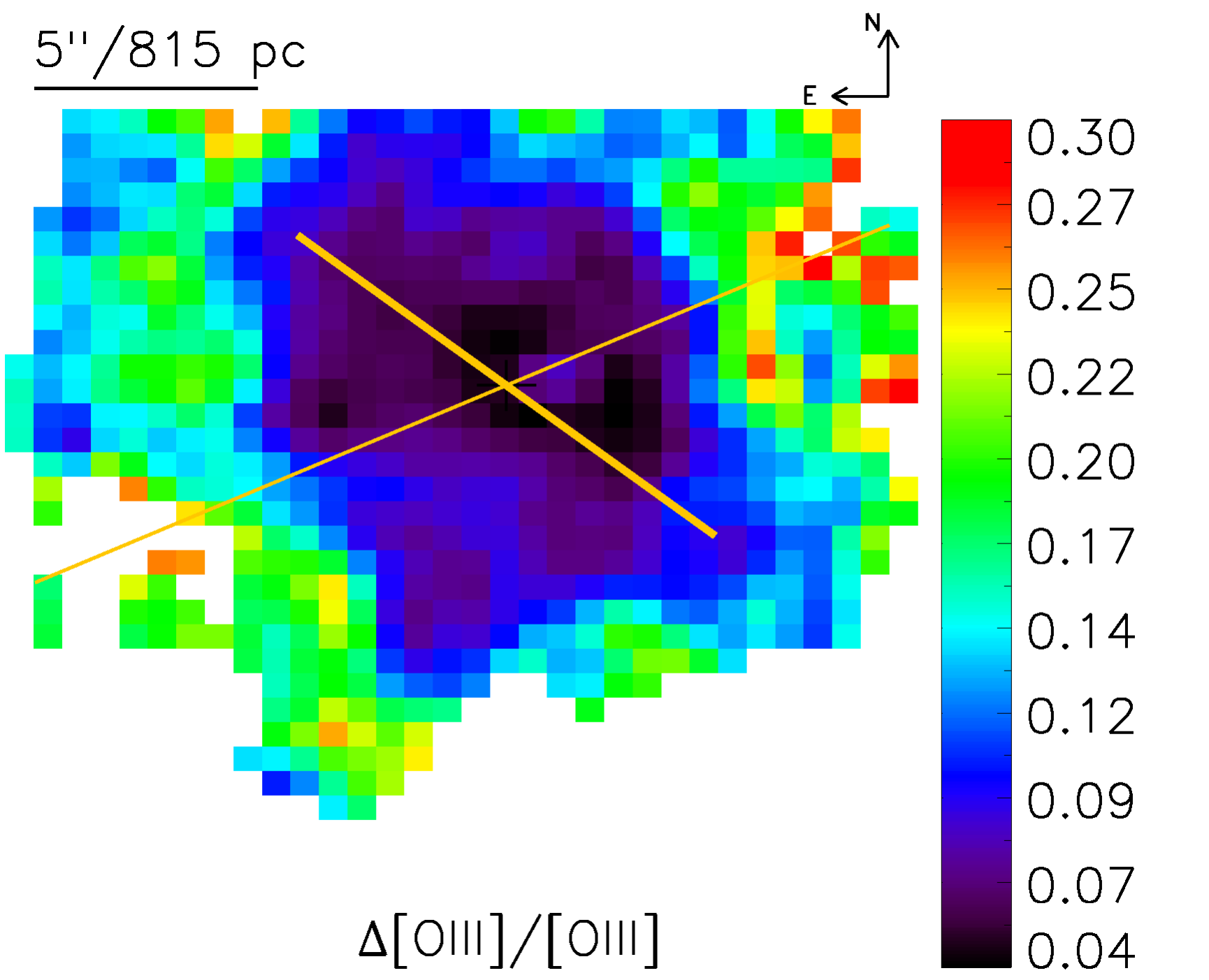}
   \includegraphics[width=0.33\textwidth]{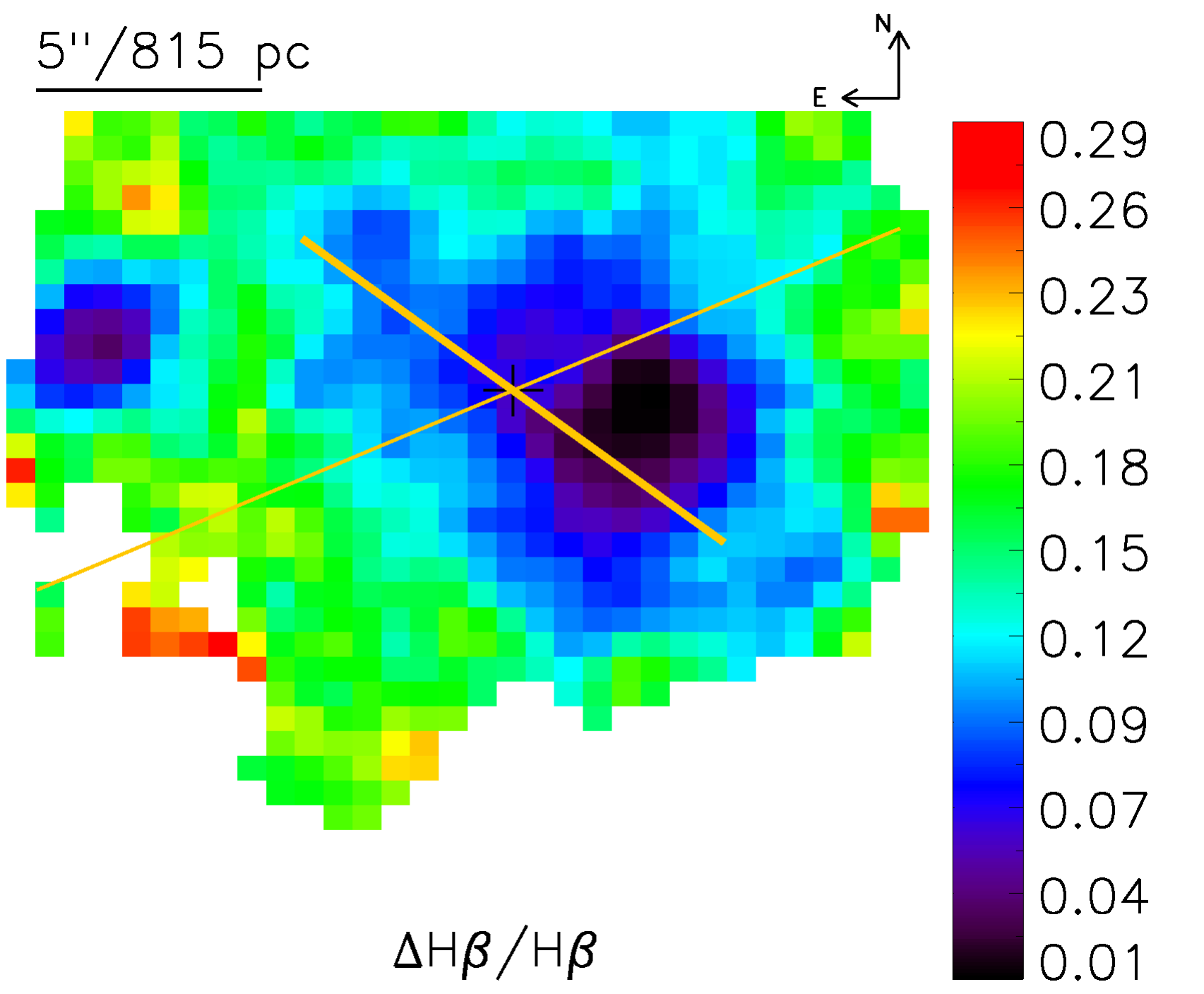}
   \includegraphics[width=0.33\textwidth]{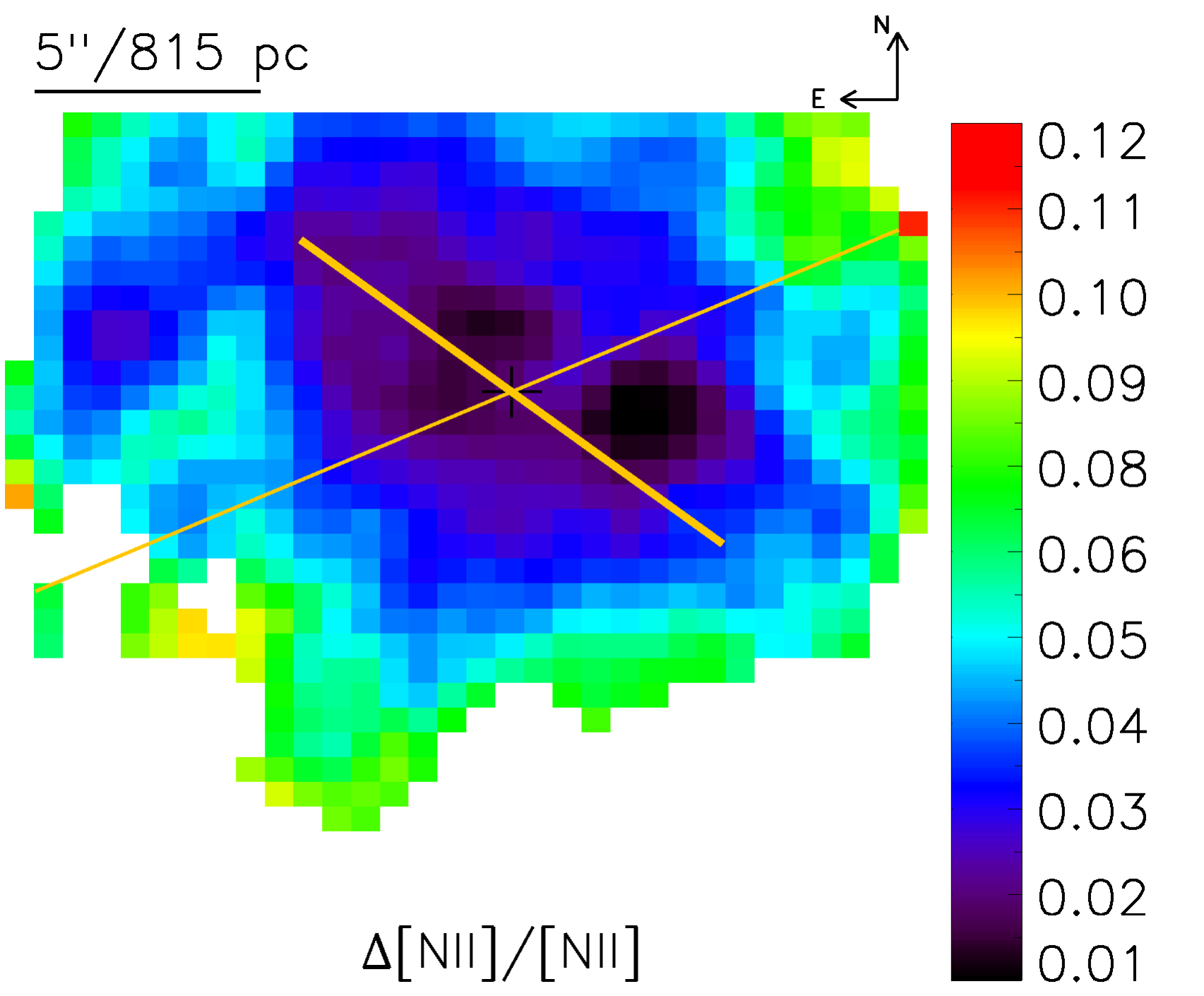}
   \includegraphics[width=0.33\textwidth]{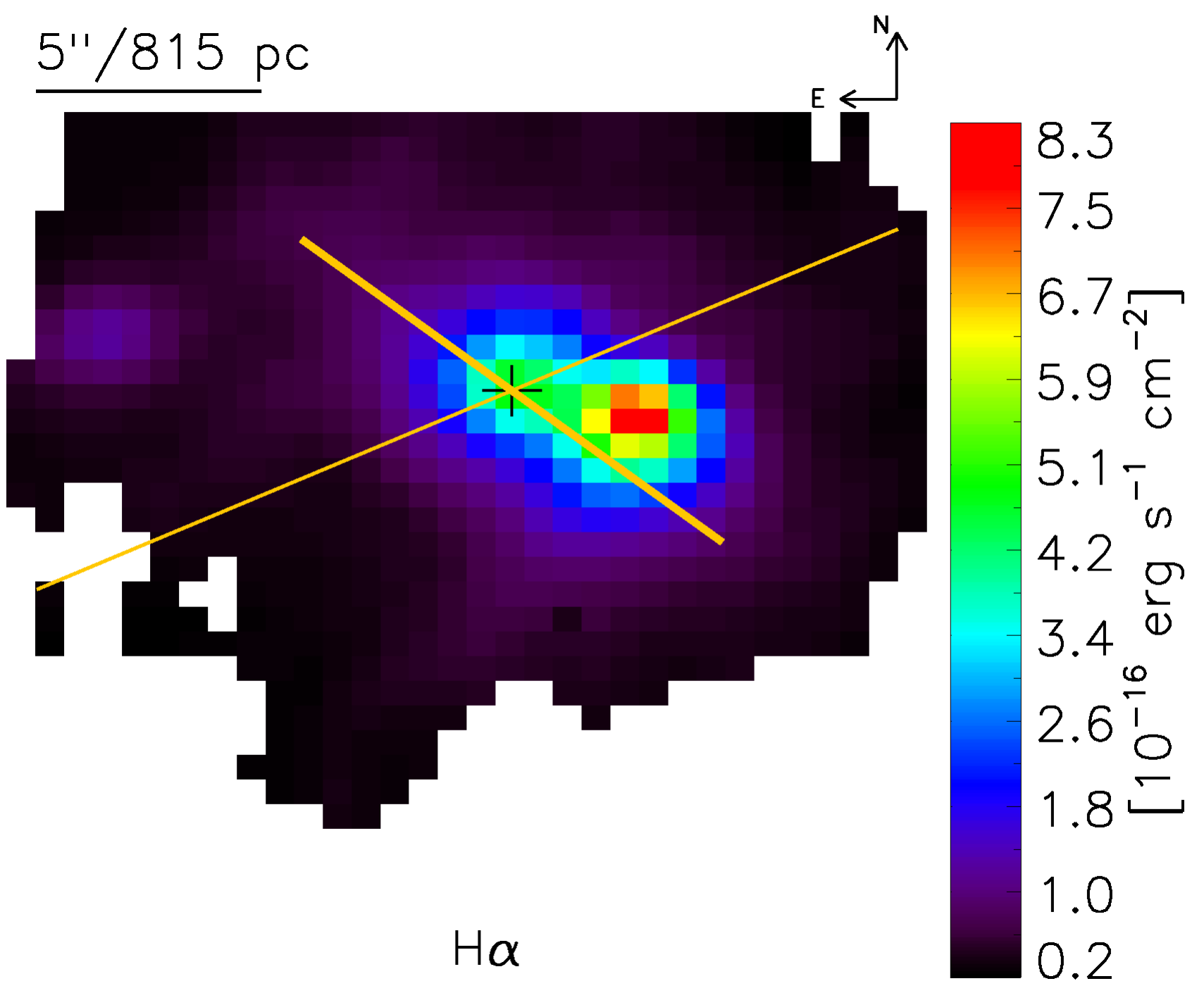}
   \includegraphics[width=0.33\textwidth]{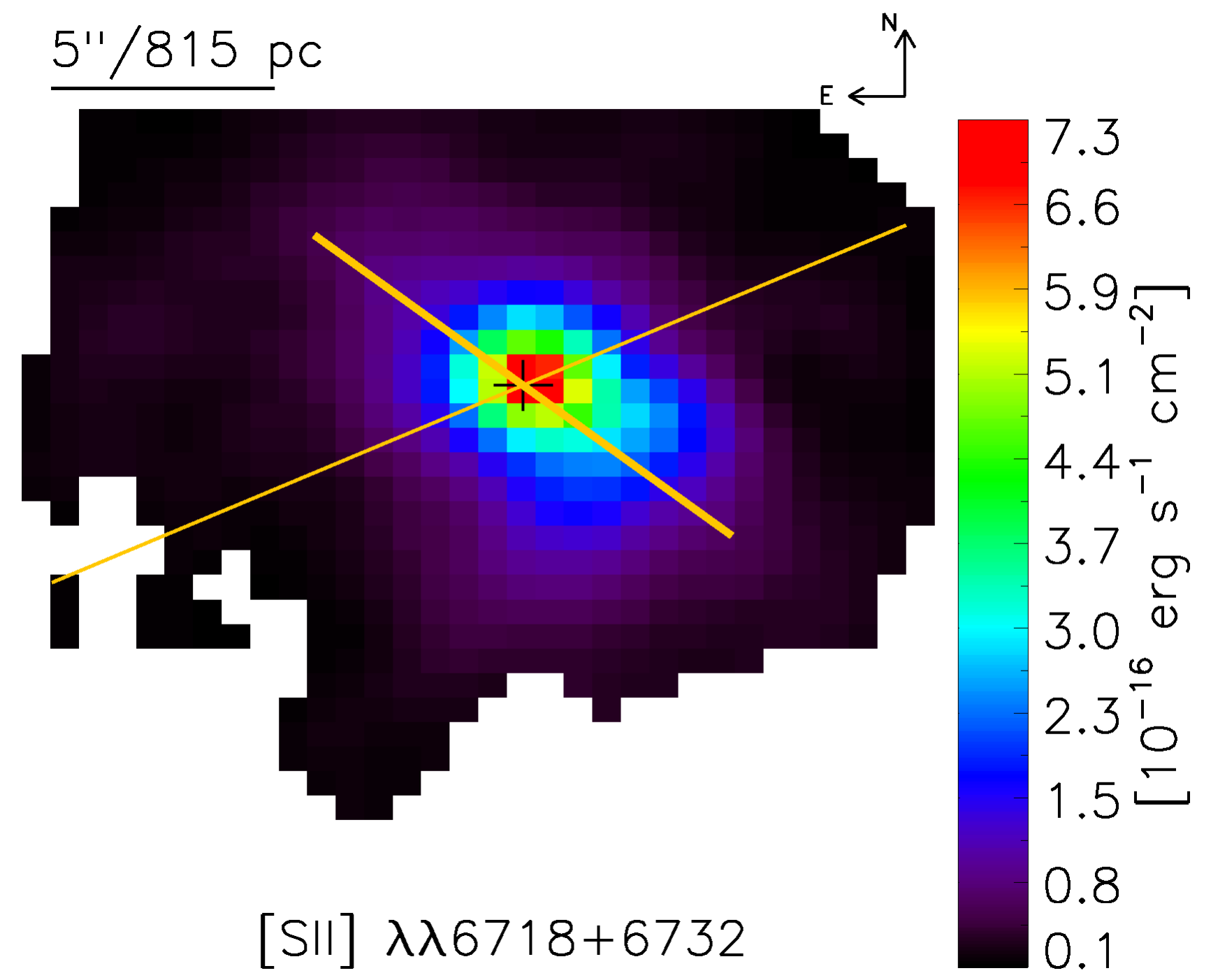}\

   \includegraphics[width=0.33\textwidth]{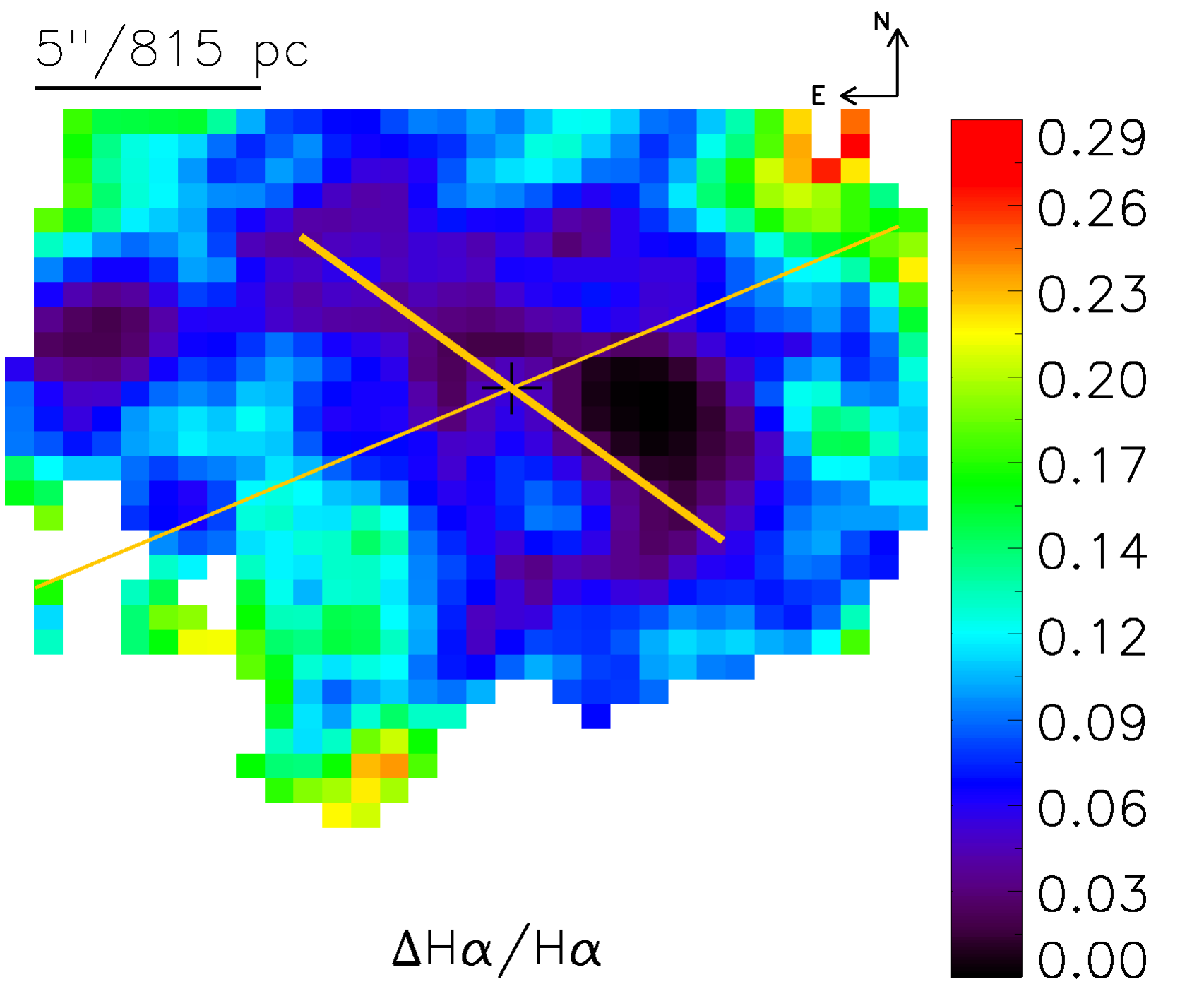}
   \includegraphics[width=0.33\textwidth]{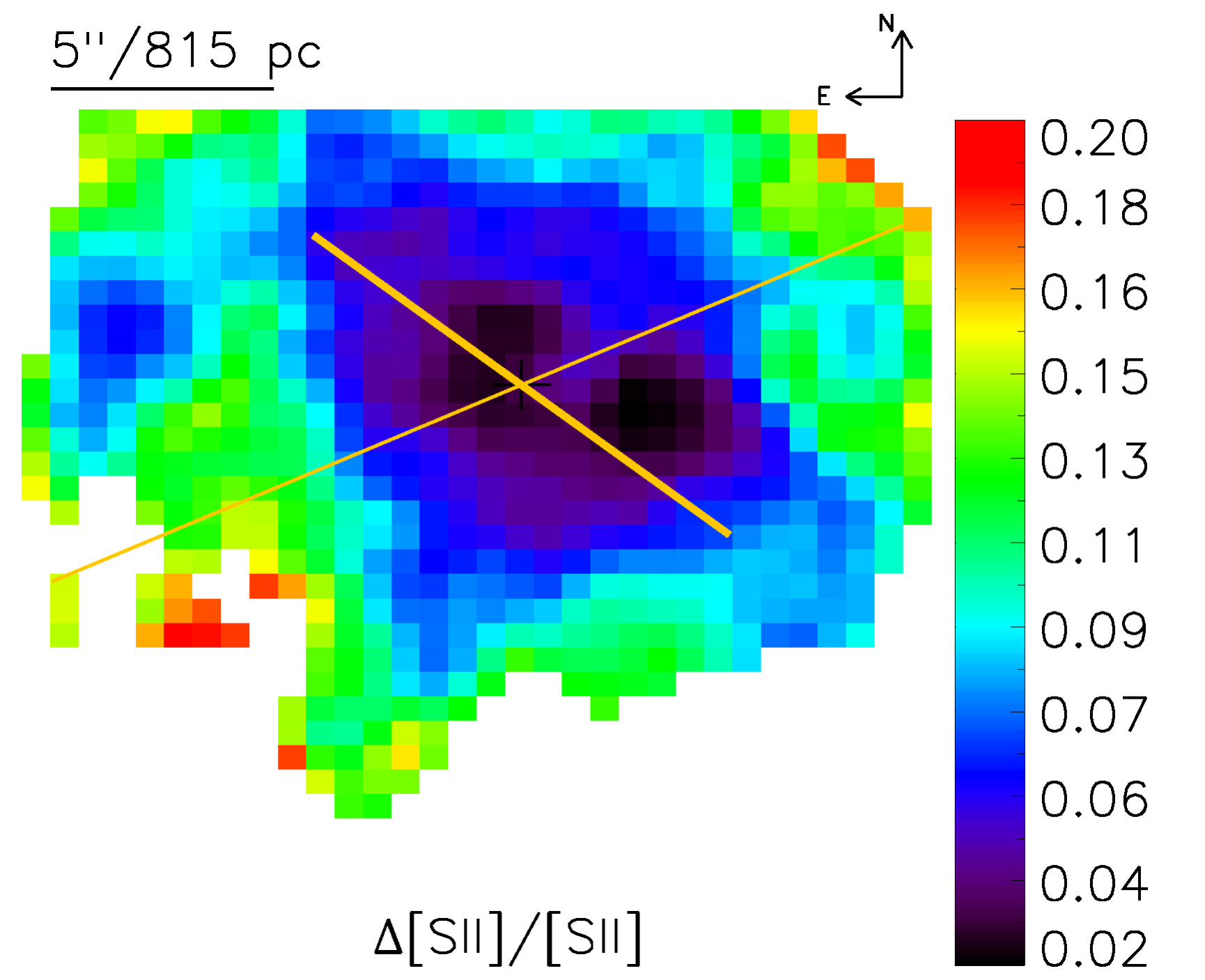}
   \caption{The first and the third row show the emission line maps of the indicated species. The second and the fourth row show the according relative error maps. In all images, the cross denotes the center of the continuum peak. The major axis of the primary (thin line) and the secondary bar (thick line) are overplotted.}
   \label{fig:line_maps_1}
\end{figure*}

Figure \ref{fig:line_maps_1} shows the emission line maps of the strongest emission lines. Though relatively strong in the center, the H$\alpha$ and H$\beta$ emission is more strongly peaked at $\sim 2\farcs1\ \left(\sim 346\ \mathrm{pc}\right)$ southwest of the center, providing evidence of the presence of an \ion{H}{ii}-region complex. A second but faint \ion{H}{ii}-region complex can be found at a distance of $\sim 10\arcsec\ \left(\sim 1627\ \mathrm{pc}\right) $ to the east of the center. It does not lie on the axis of the secondary bar but might be part of the circumnuclear ring.\

In the emission line maps of the forbidden low ionization species ([\ion{N}{ii}] ${\lambda}6585$, [\ion{S}{ii}] ${\lambda\lambda}6718 + 6732$), we see that the location of the emission peak coincides with that of the continuum emission, confirming this to be the location of the galaxy's center. Furthermore, the emission in both forbidden lines is extended along the major axis of the bar. We also note a contribution from the inner \ion{H}{ii}-region. The forbidden high ionization transition [\ion{O}{iii}] ${\lambda}5008$ is mostly extended along the secondary bar.\\

\subsection{Kinematics}
\label{kinematics}

\begin{figure*}
   \centering
   \includegraphics[width=0.33\textwidth]{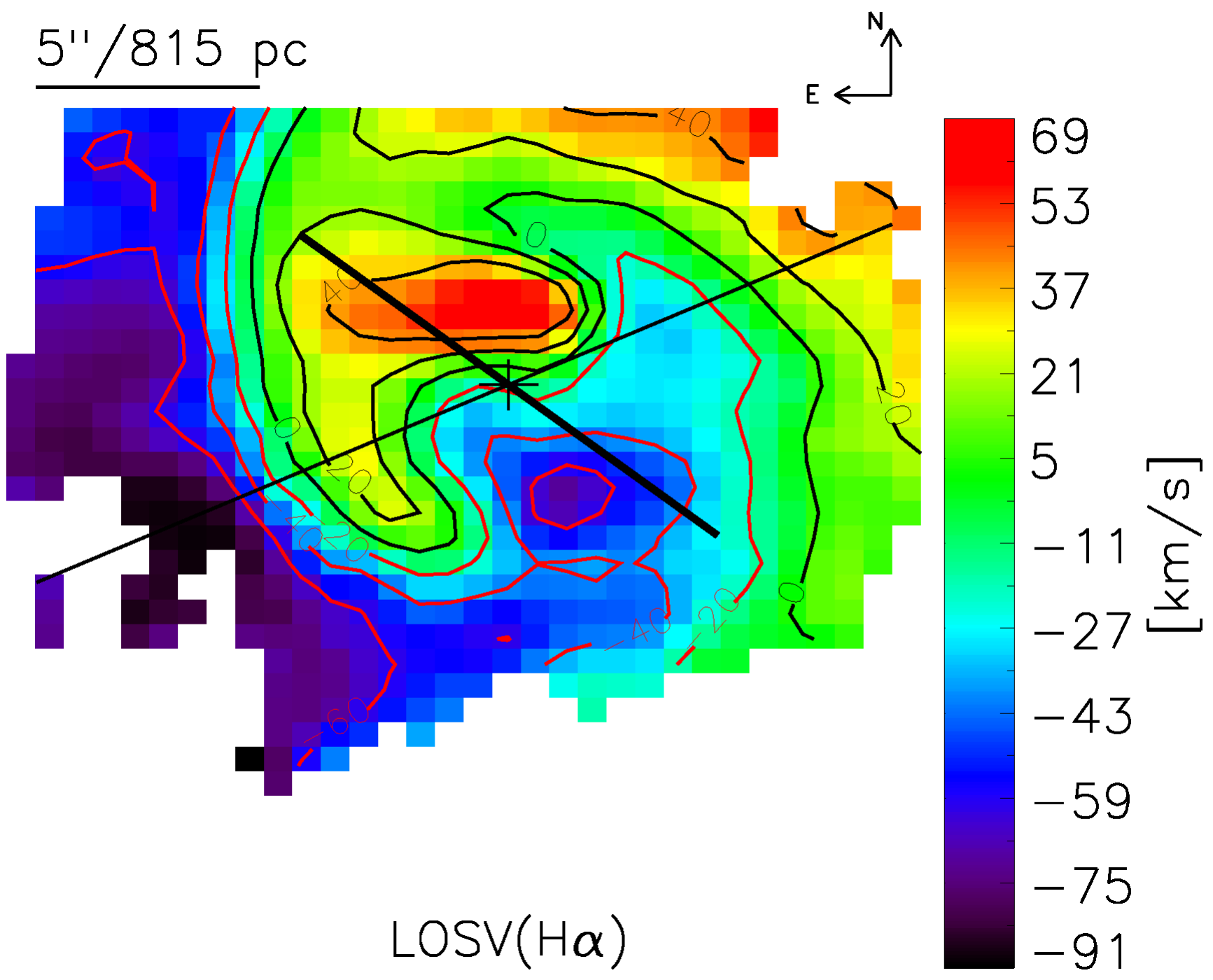}
   \includegraphics[width=0.33\textwidth]{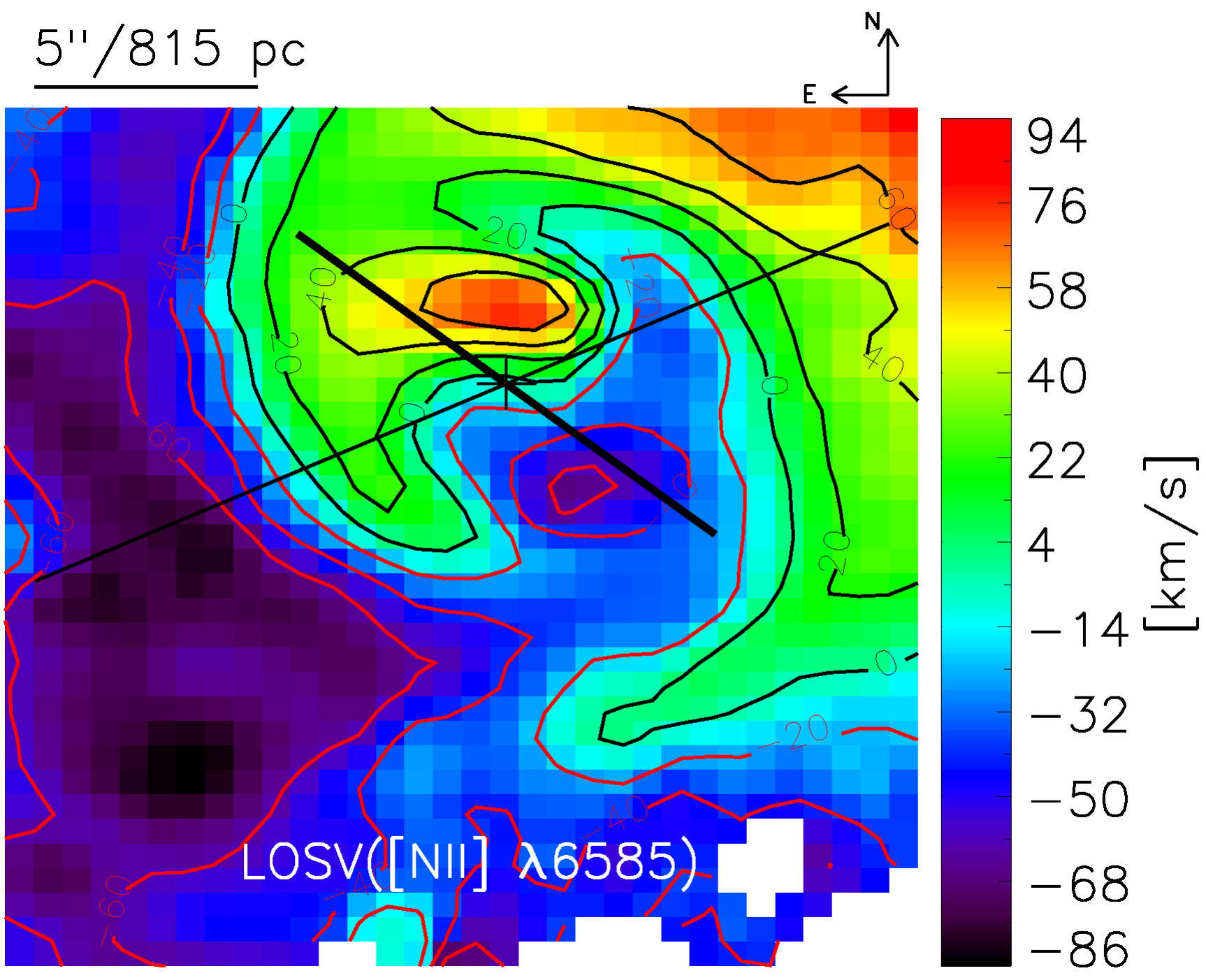}
   \includegraphics[width=0.33\textwidth]{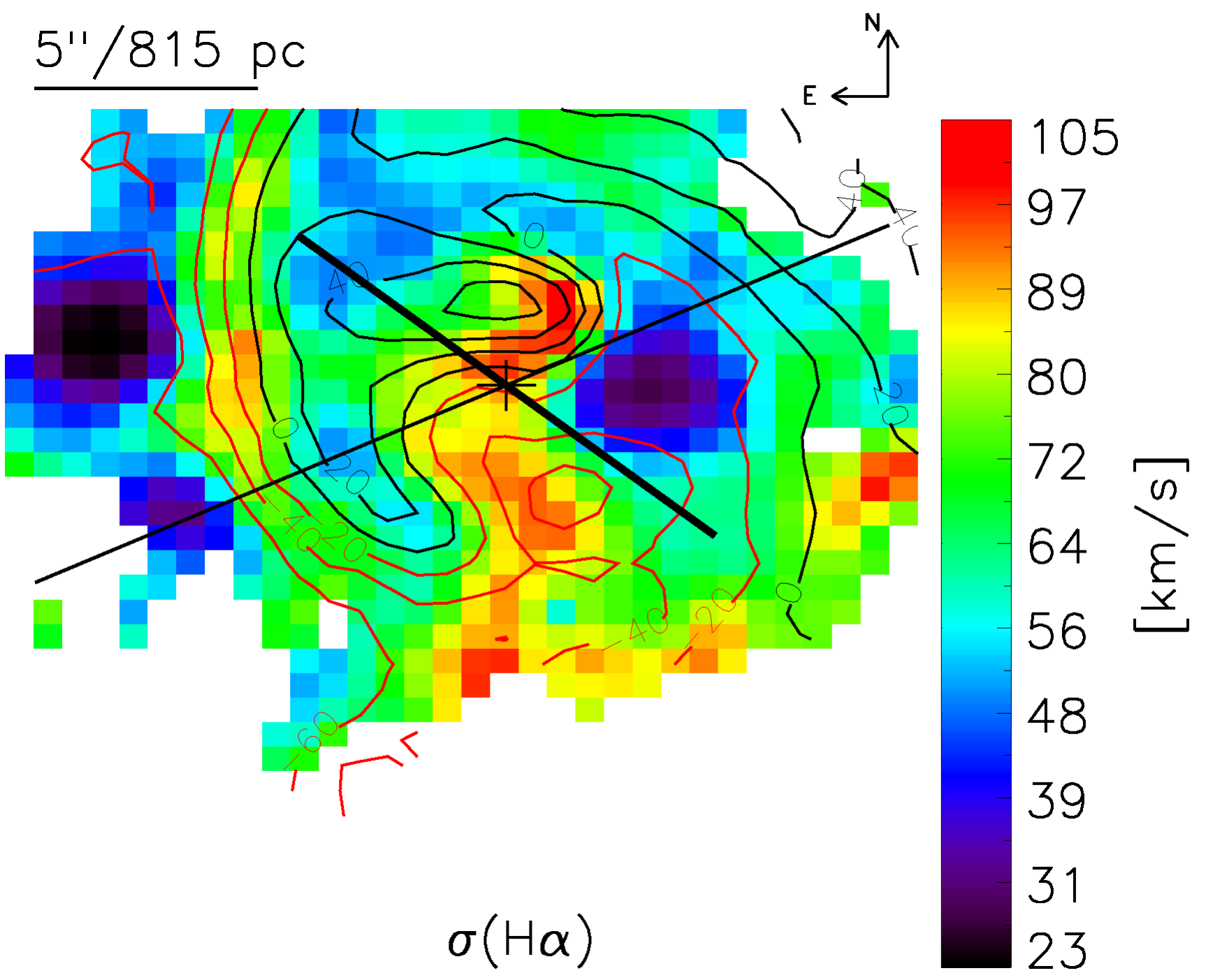}
   \includegraphics[width=0.33\textwidth]{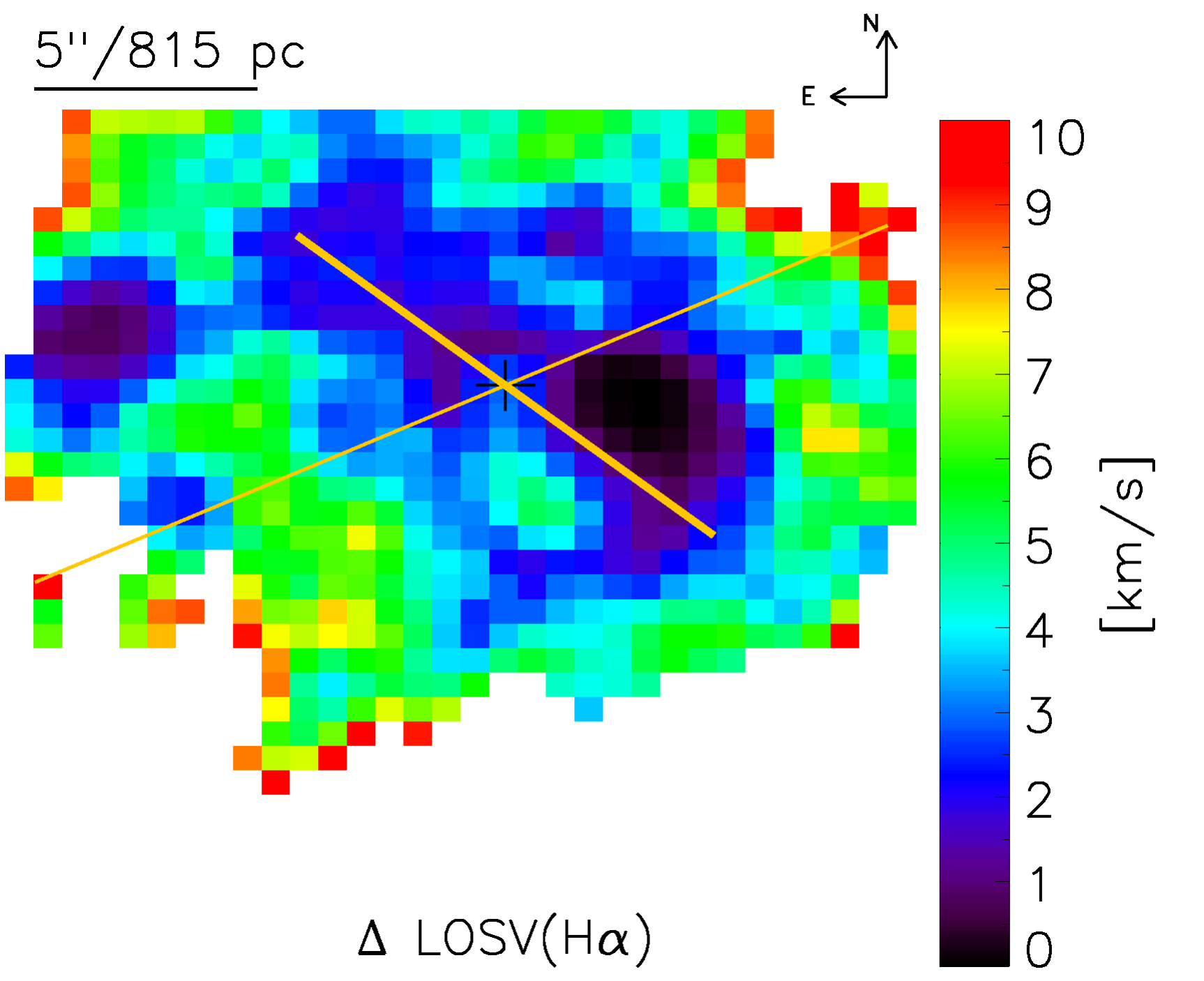} 
   \includegraphics[width=0.33\textwidth]{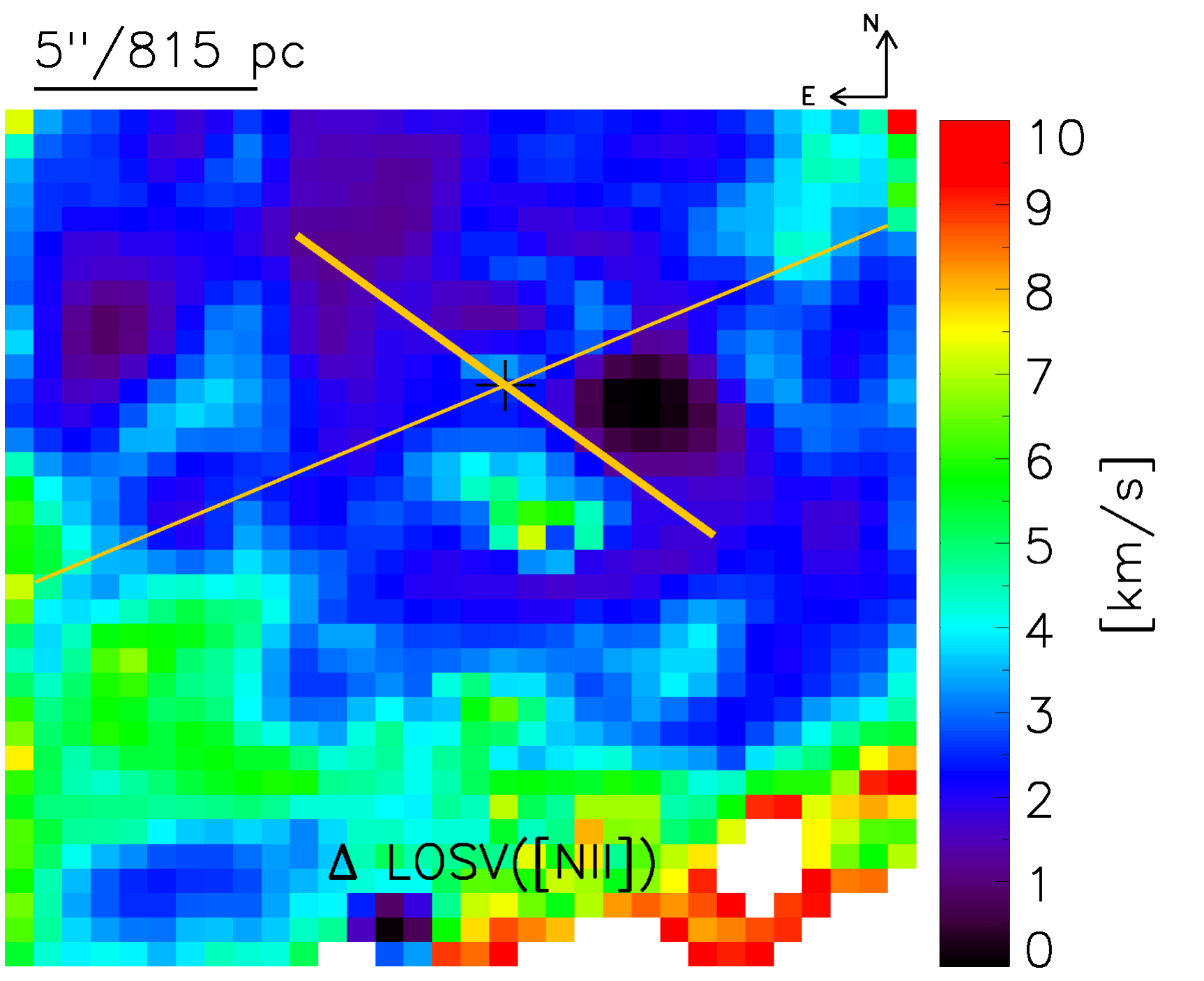}  
   \includegraphics[width=0.33\textwidth]{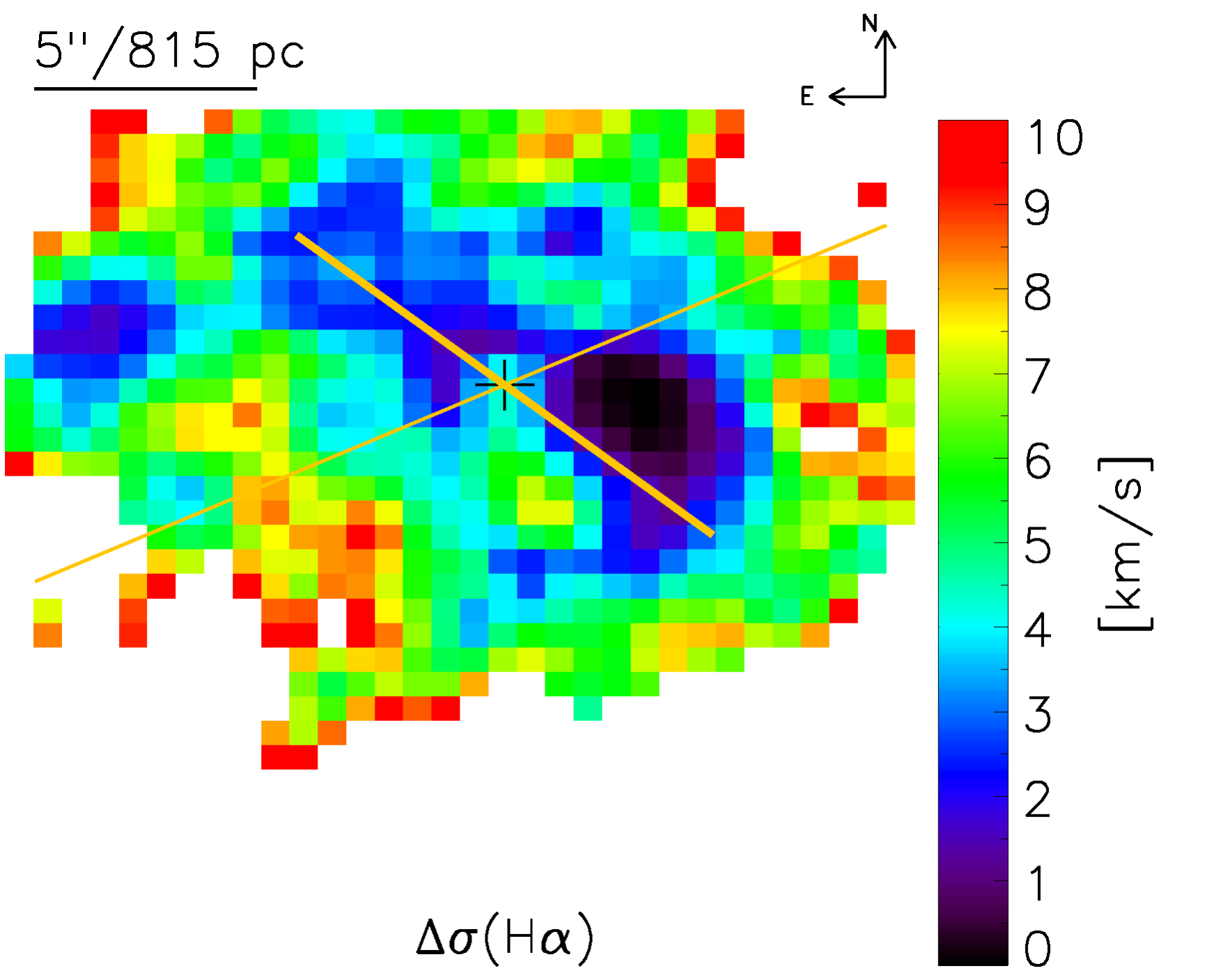}    
   \caption{Upper row (left): H$\alpha$ line of sight velocity maps with overplotted major axis of the primary (thin line) and the secondary bar (thick line). The contours mark the velocities $-60$, $-40$, $-20$, $0$, $20$, and $40~\mathrm{km~s^{-1}}$. Contours of negative velocities are red. (center): [\ion{N}{ii}] $\lambda6585$ line of sight velocity map. It has been derived \emph{before} the subtraction of the stellar continuum and clipping of spaxels according to the criteria outlined in the text. Both velocity maps show the same morphology. (right): H$\alpha$ velocity dispersion map corrected for instrumental resoulution. The isovelocity contours of H$\alpha$ line emission are overplotted. Lower row: The error maps corresponding to the maps in the upper row.} 
   \label{fig:gas_kinematics}
   \end{figure*}

The line of sight velocity (LOSV) maps of H$\alpha$ and [\ion{N}{ii}] $\lambda6585$ are shown in Fig. \ref{fig:gas_kinematics}. All emission lines, including those whose kinematics are not displayed here, show s-shaped isovelocity contours in the center. The nucleus is at relative rest, as expected. Areas of negative velocities are found in a ring-like structure in the east and southeast of the FOV, while positive velocities are predominantly found in the north and northwest.\

To complete the kinematic picture on a larger scale, we additionally display the [\ion{N}{ii}] $\lambda 6585$ LOSV map in Fig.\ref{fig:gas_kinematics}. It was derived \emph{before} we subtracted the stellar continuum. Consequently, this map has not been subject to any clipping for fitting quality reasons. The values are consistent with the H$\alpha$ velocities and do not deviate much from those of [\ion{N}{ii}] $\lambda 6858$ that are derived \emph{after} the subtraction of the stellar continuum, since [\ion{N}{ii}] is only weakly affected by the stellar population. 

In the inner $3\arcsec$ of NGC 5850, the LOSV field is characterized by a positive and negative velocity peak. These peaks are located almost symmetrically with respect to the galaxy center. The negative peak southwest of the center and the positive peak in the northeast of the center have almost the same absolute velocity and lie on a straight line through the center. We estimate the position angle to be $\mathrm{P.A._{kin,ion} \approx 25 \degr}$. Consequently, the kinematic major axis is tilted with respect to the major axis of the secondary bar ($\mathrm{P.A._{secondary}\approx 49.3 \degr}$; \citet{Lourenso_stellar}). Furthermore, the velocity peaks do not have the same distance to the center.\

In most pixels, the [\ion{N}{ii}] $\lambda6585$ emission line is the strongest line in our spectra. Since it is contaminated by but corrected for a weak sky line, we prefer to show the H$\alpha$ velocity dispersion map in Fig. \ref{fig:gas_kinematics}. After the correction for the instrumental resolution, the H$\alpha$ velocity dispersion map is characterized by regions of medium to high dispersion that coincide with strong velocity gradients. The high dispersion areas are aligned approximately in the north-south direction, extending over but not peaked on the center.\

We notice the spatial coincidence of increased LOSV dispersion and strong LOSV gradients. This is a symptom of beam smearing. To evaluate the influence of this effect, we calculated the average dispersion value in the FOV $(\approx 65~\mathrm{km~s^{-1}})$ and compared it with the dispersion peak values ($95-105~\mathrm{km~s^{-1}})$. From this, we estimate the dispersion of the potentially unresolved component of about $70$ to $80~\mathrm{km~s^{-1}}$. This value is similar to the observed LOSV differences in seeing elements on both sides of areas with increased dispersion values. Hence, the increase in line width might be dominated by the change in LOSV within a seeing element, and it appears likely that we are observing gas at different velocities within one spatial element.\

\begin{figure*}
  \centering
  \includegraphics[width=\textwidth]{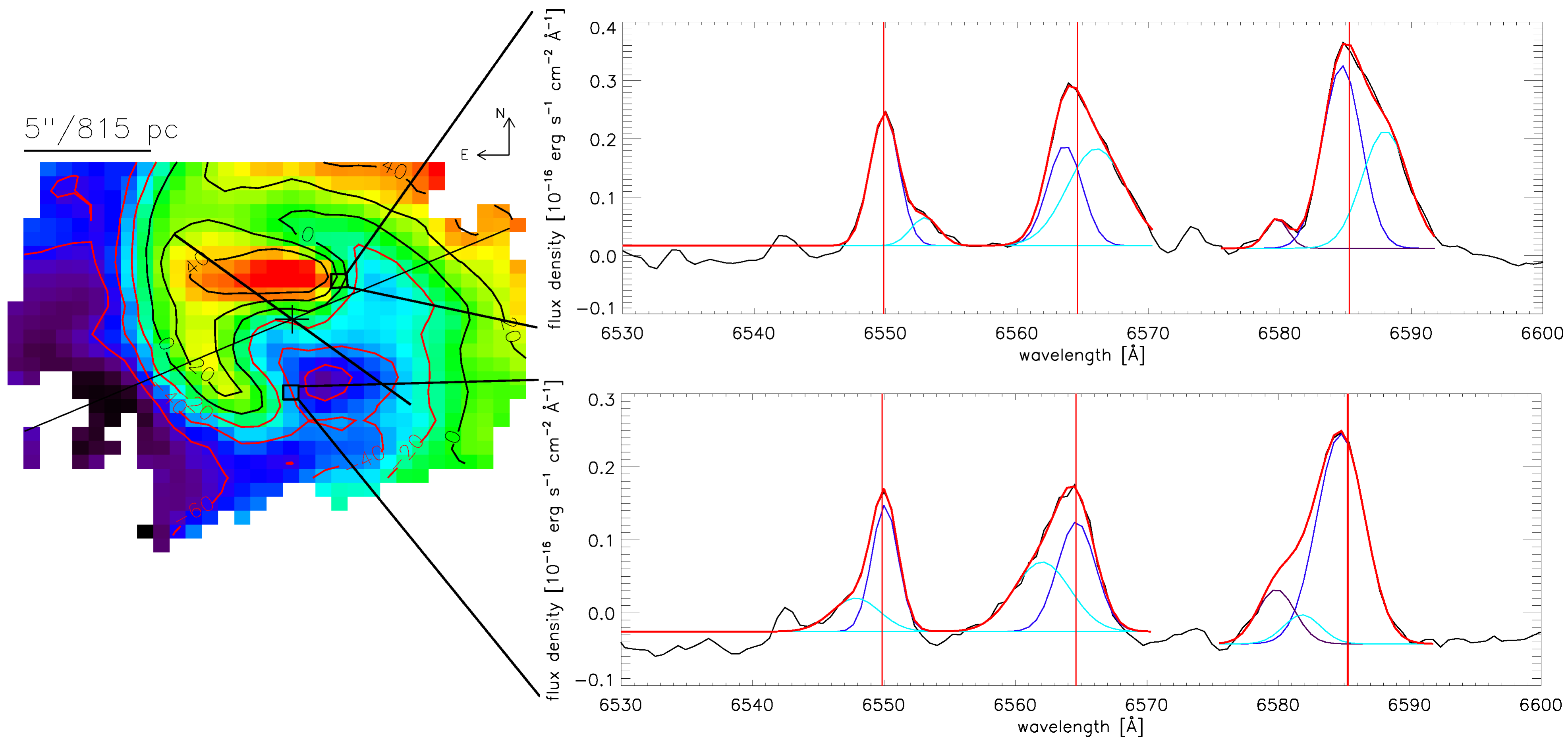}
  \caption{H$\alpha$ line of sight-velocity map (left) as in Fig. \ref{fig:gas_kinematics}. On the right hand side, we display the red grating spectra of the regions highlighted by black boxes in the velocity map. The different components of the attempted line fits are plotted in the spectra as well. In red, the overall fit is displayed. The third (and left-most) fitting component in the [\ion{N}{ii}] $\lambda 6585$ emission line is the fit on the sky line contamination. The fitting results are reported in Table \ref{table:double_fits}. The vertical red lines denote the rest wavelength of the emission line.}
  \label{fig:asymmetries}
\end{figure*}

The \ion{H}{ii} emission regions have the lowest dispersion in the whole map with $\sim 30~\mathrm{km~s^{-1}}$ as is expected for star forming regions \citep[e.g.,][]{Wisnioski_HII_sigma}. After a close inspection of all emission lines, we find asymmetries in their shape. They show blue line wings south of the core. The wing moves to the red flank of the line, while the emission is traced to the north of the center. As a global multi-component fit is too uncertain, we only show example spectra for the regions, where the lines show the strongest asymmetries. These example regions and the fits to their respective emission line shapes are plotted in Fig. \ref{fig:asymmetries}, and the fit results are reported in Table \ref{table:double_fits}.  

\begin{table*}
\caption{Results of the double Gaussian emission line-fit attempts for regions with strong line asymmetry.}
\label{table:double_fits}
\centering
\begin{tabular}{llcccc}
\hline\hline
Location & Emission line					& Fit component 		&	$v~\mathrm{[km~s^{-1}]}$	&	$\mathrm{\sigma}~\mathrm{[km~s^{-1}]}$	& Flux [$10^{-16}\mathrm{erg~s^{-1}~cm^{-2}}$] \\
\hline
		& $\mathrm{H\beta}$				& main	 				&	$11$		&	$55$		&	$0.5$\\
		& 								& $2^{\mathrm{nd}}$ 	&	$200$		&	$33$		&	$0.2$\\
		& [\ion{O}{iii}] $\lambda5008$	& main 					&	$10$		&	$57$		&	$0.3$\\
	 	& 								& $2^{\mathrm{nd}}$ 	&	$76$		&	$134$		&	$0.5$\\
 north	& [\ion{N}{ii}] $\lambda6549$	& main					&	$3$			&	$28$		&	$0.6$\\
	 	& 								& $2^{\mathrm{nd}}$ 	&	$143$		&	$26$		&	$0.1$\\
		& $\mathrm{H\alpha}$			& main					&	$-44$		&	$43$		&	$0.6$\\
		& 								& $2^{\mathrm{nd}}$ 	&	$66$		&	$93$		&	$1.0$\\
		& [\ion{N}{ii}] $\lambda6585$	& main					&	$-25$		&	$52$		&	$1.2$\\
		& 								& $2^{\mathrm{nd}}$ 	&	$124$		&	$68$		&	$0.9$\\
\hline		
		& $\mathrm{H\beta}$				& main	 				&	$-18$		&	$110$		&	$0.5$\\
		& 								& $2^{\mathrm{nd}}$ 	&	$-259$		&	$\mathrm{-}$&	$\mathrm{-}$\\
		& [\ion{O}{iii}] $\lambda5008$	& main 					&	$-11$		&	$132$		&	$0.7$\\
	 	& 								& $2^{\mathrm{nd}}$ 	&	$-268$		&	$76$		&	$0.1$\\
 south	& [\ion{N}{ii}] $\lambda6549$	& main					&	$10$		&	$21$		&	$0.4$\\
	 	& 								& $2^{\mathrm{nd}}$ 	&	$-92$		&	$72$		&	$0.2$\\
		& $\mathrm{H\alpha}$			& main					&	$2$			&	$52$		&	$0.5$\\
		& 								& $2^{\mathrm{nd}}$ 	&	$-118$		&	$90$		&	$0.5$\\
		& [\ion{N}{ii}] $\lambda6585$	& main					&	$-24$		&	$73$		&	$1.3$\\
		& 								& $2^{\mathrm{nd}}$ 	&	$-158$		&	$50$		&	$0.1$\\
\hline
\end{tabular}
\\
\end{table*}

\begin{figure}
  \centering

  \includegraphics[width=0.24\textwidth]{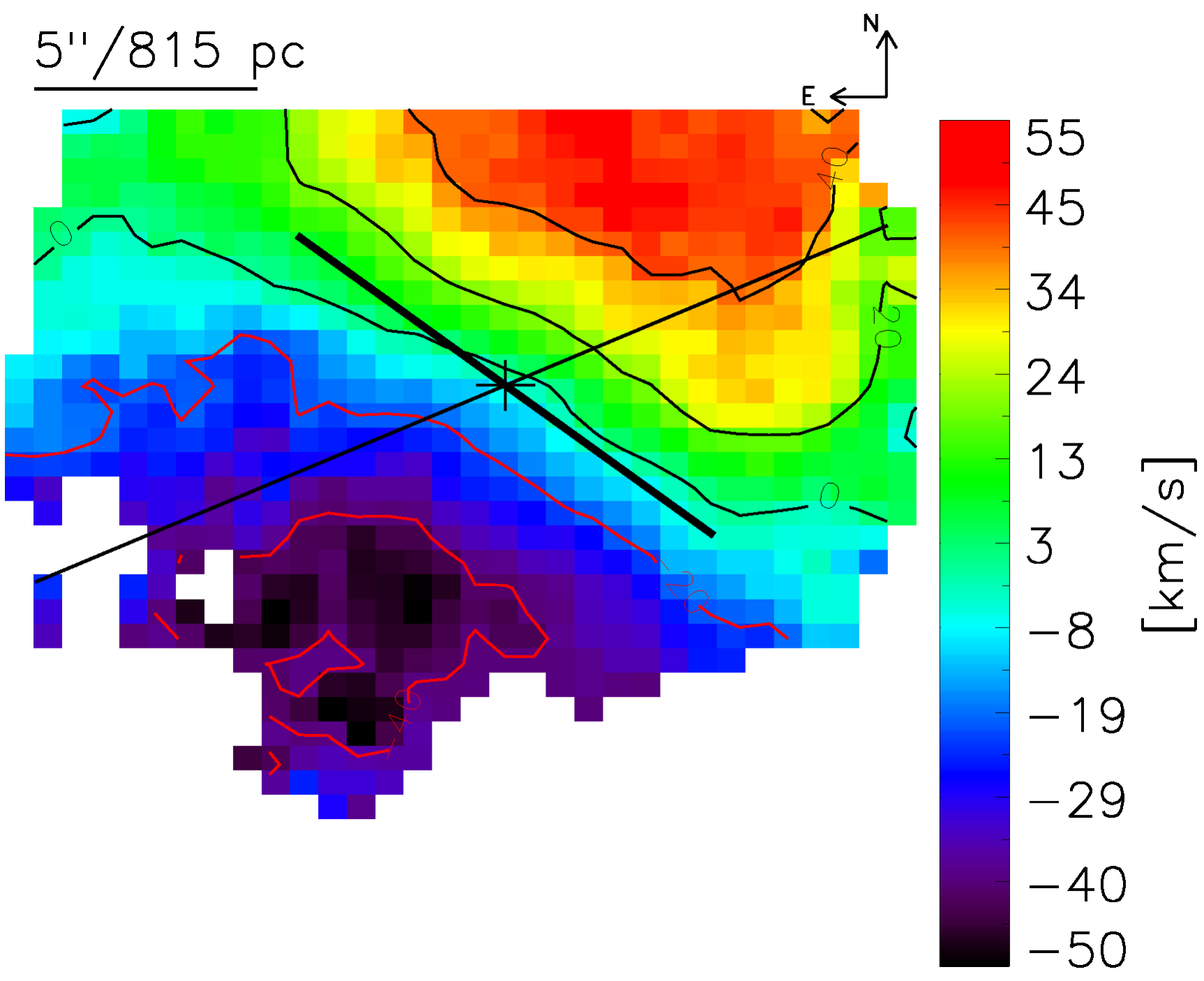}
  \includegraphics[width=0.24\textwidth]{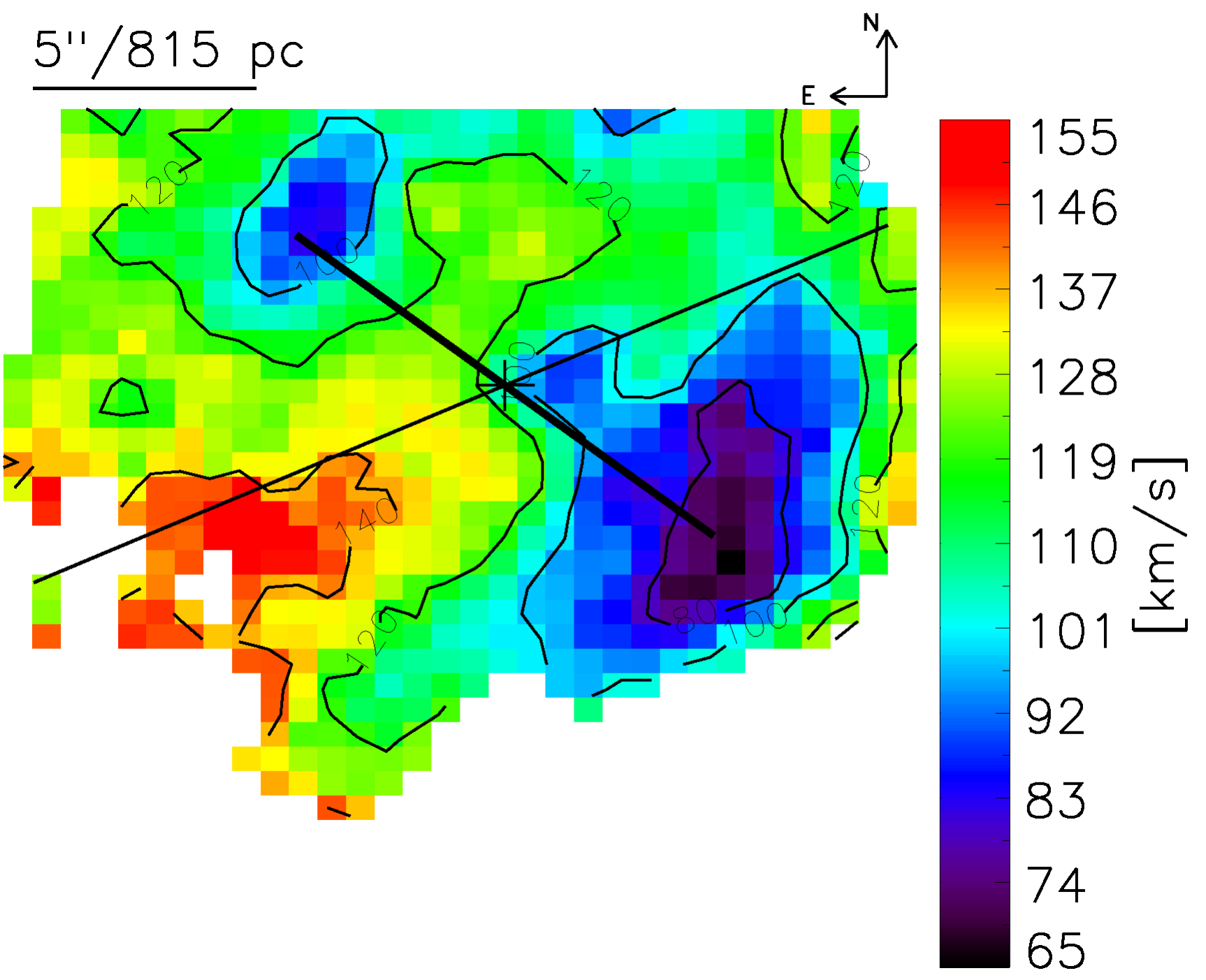}
  \caption{Stellar maps derived by STARLIGHT. The major axis of the primary (thin line) and the secondary bar (thick line) are overplotted. Left: Stellar line of sight velocity map with overplotted isovelocity contours at -40, -20 (red), 0, 20, 40 (black) $\mathrm{km~s^{-1}}$. Right: Stellar velocity dispersion map.}
  \label{fig:stellar_kinematics}
\end{figure}

The stellar LOSV dispersion map in the right panel of Fig. \ref{fig:stellar_kinematics} is a byproduct of the STARLIGHT code used to subtract the stellar contribution to the galaxy spectrum. The spatial distribution and the values of the dispersions agree with \citet{doublebarred_structure_moissev} and \citet{sigma_hollows_lorenzo}. Furthermore, we can confirm the detection of the $\sigma$-hollows at the edges of the secondary bar, which were observed by \citet{sigma_hollows_lorenzo}.\
 
 The stellar LOSV map derived by STARLIGHT in the left panel of Fig. \ref{fig:stellar_kinematics} is very regular. It hardly deviates from the spider diagram typical of circular rotation. The influence of the secondary bar is negligible. This also can be seen in Fig. A7 in \citet{sigma_hollows_lorenzo} with a general agreement in the velocity values.\

\subsection{Excitation}
\label{excitation}

\begin{figure*}
   \centering
   \includegraphics[width=0.33\textwidth]{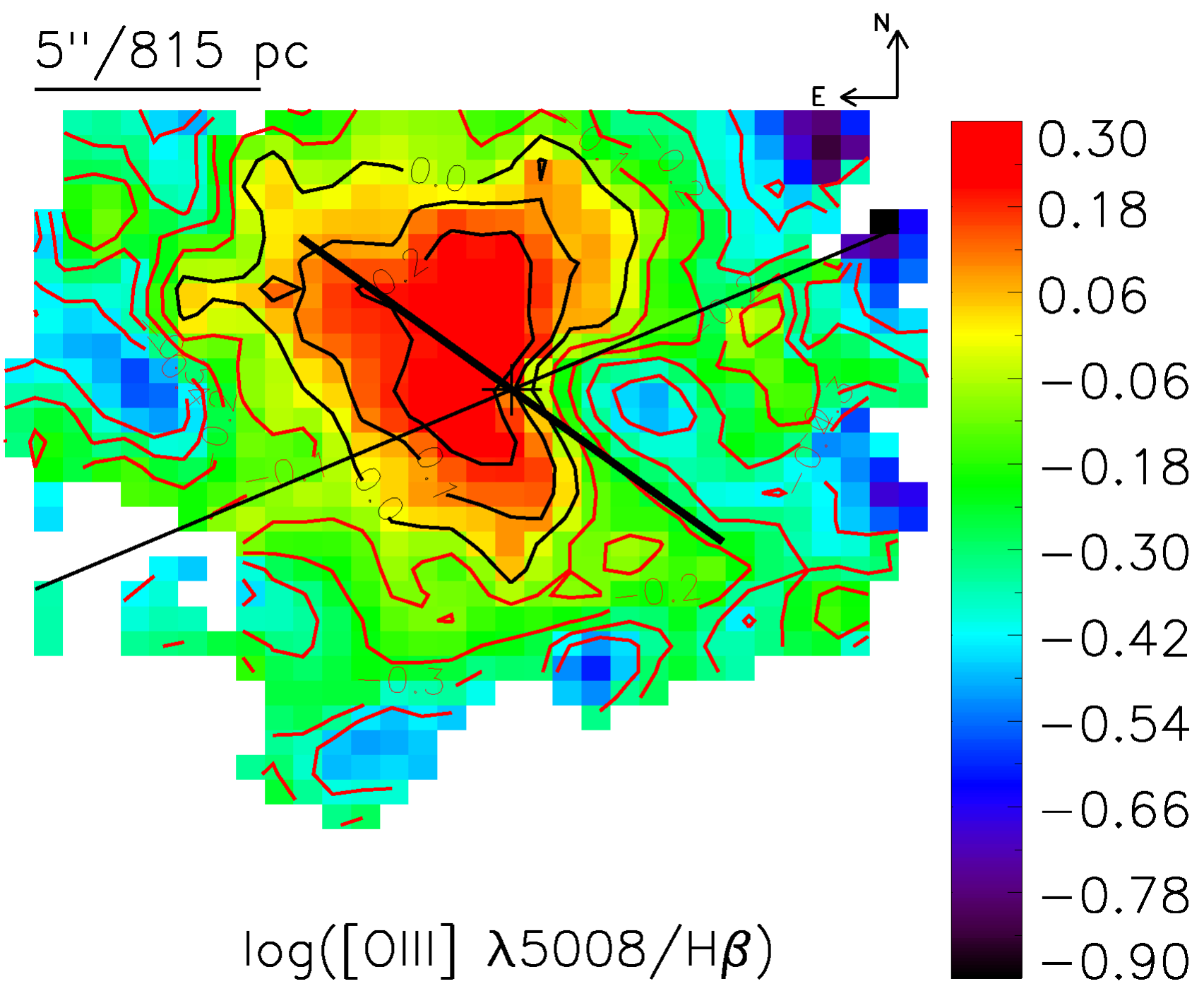}
   \includegraphics[width=0.33\textwidth]{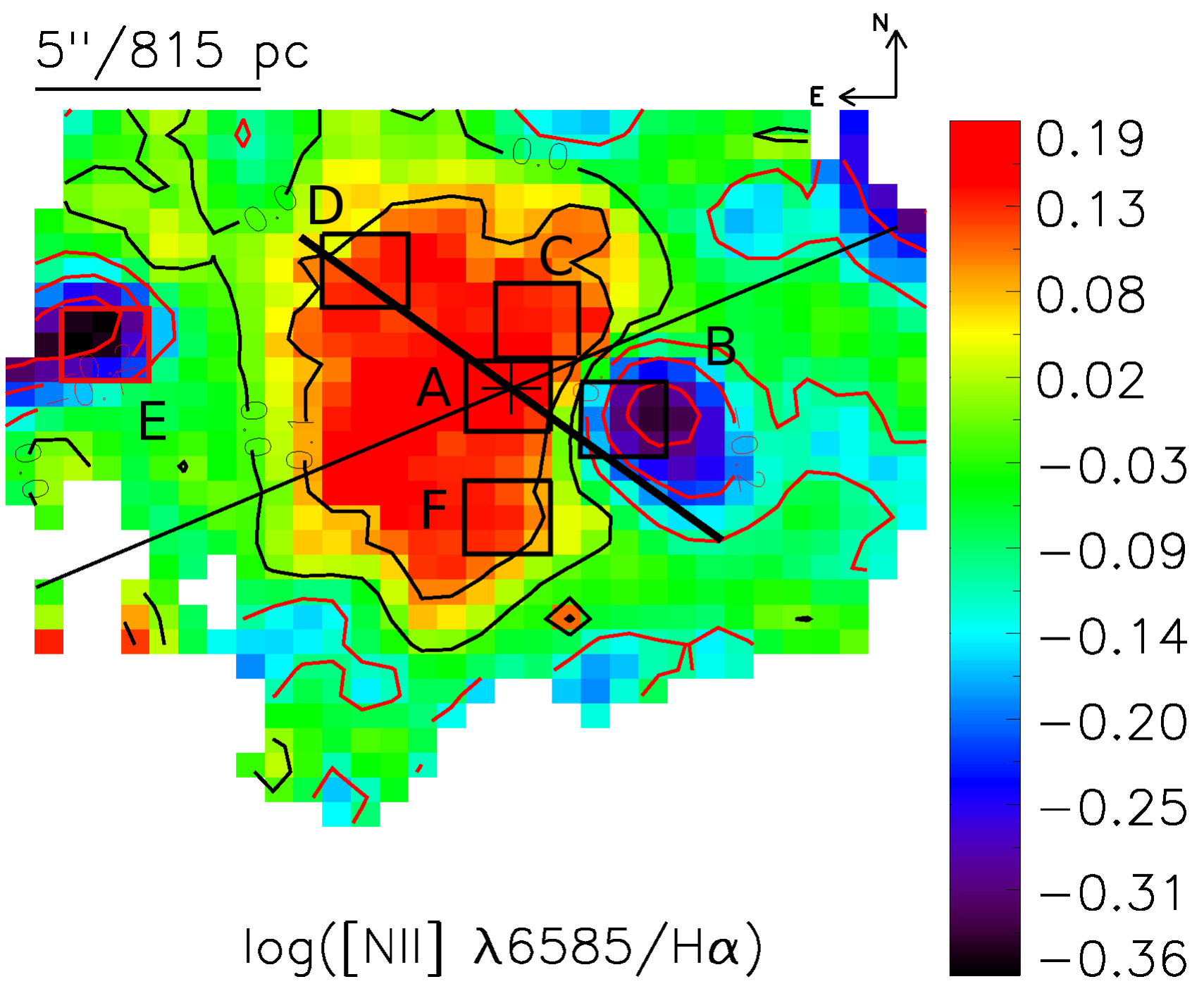}
   \includegraphics[width=0.33\textwidth]{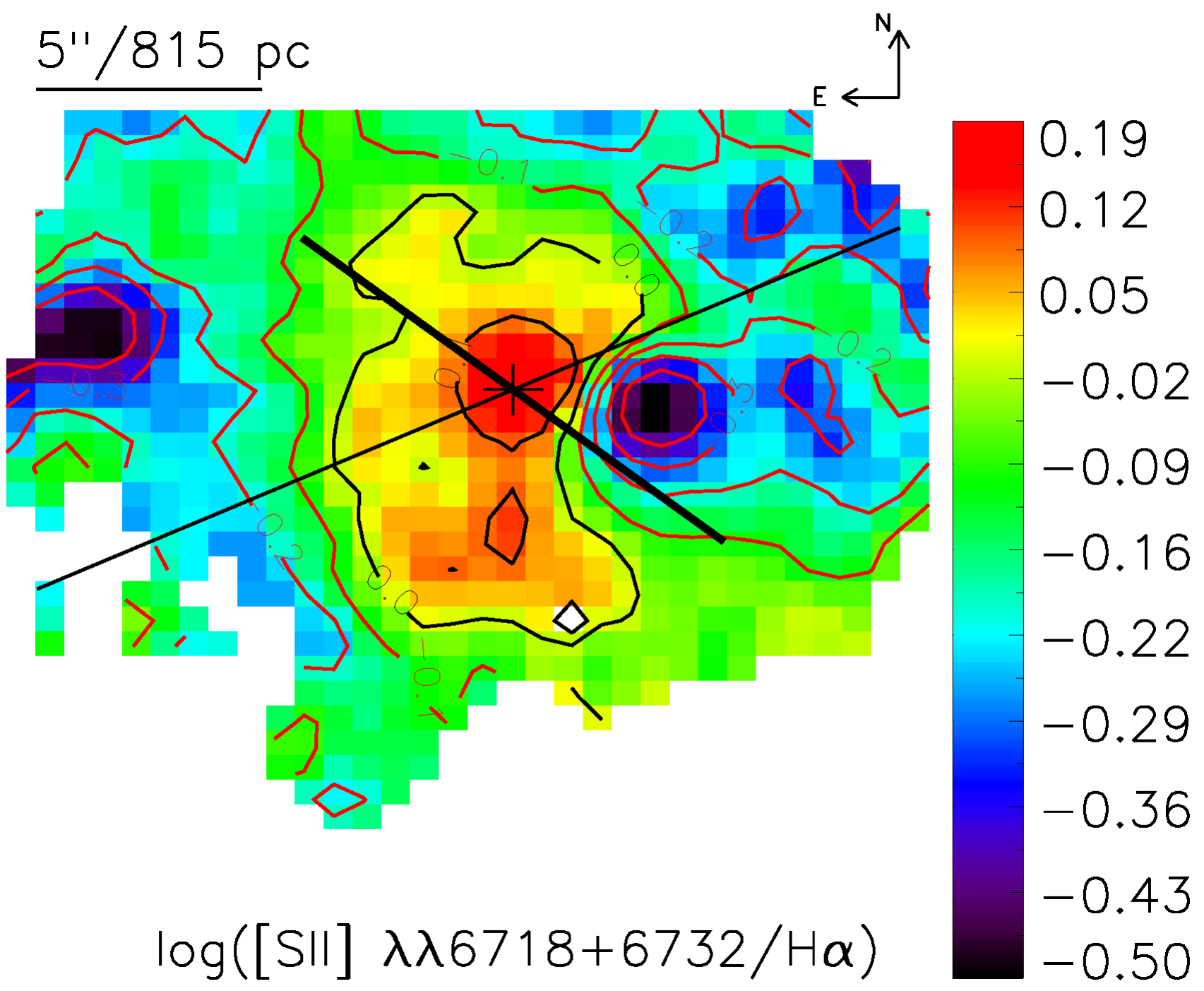}
   \includegraphics[width=0.33\textwidth]{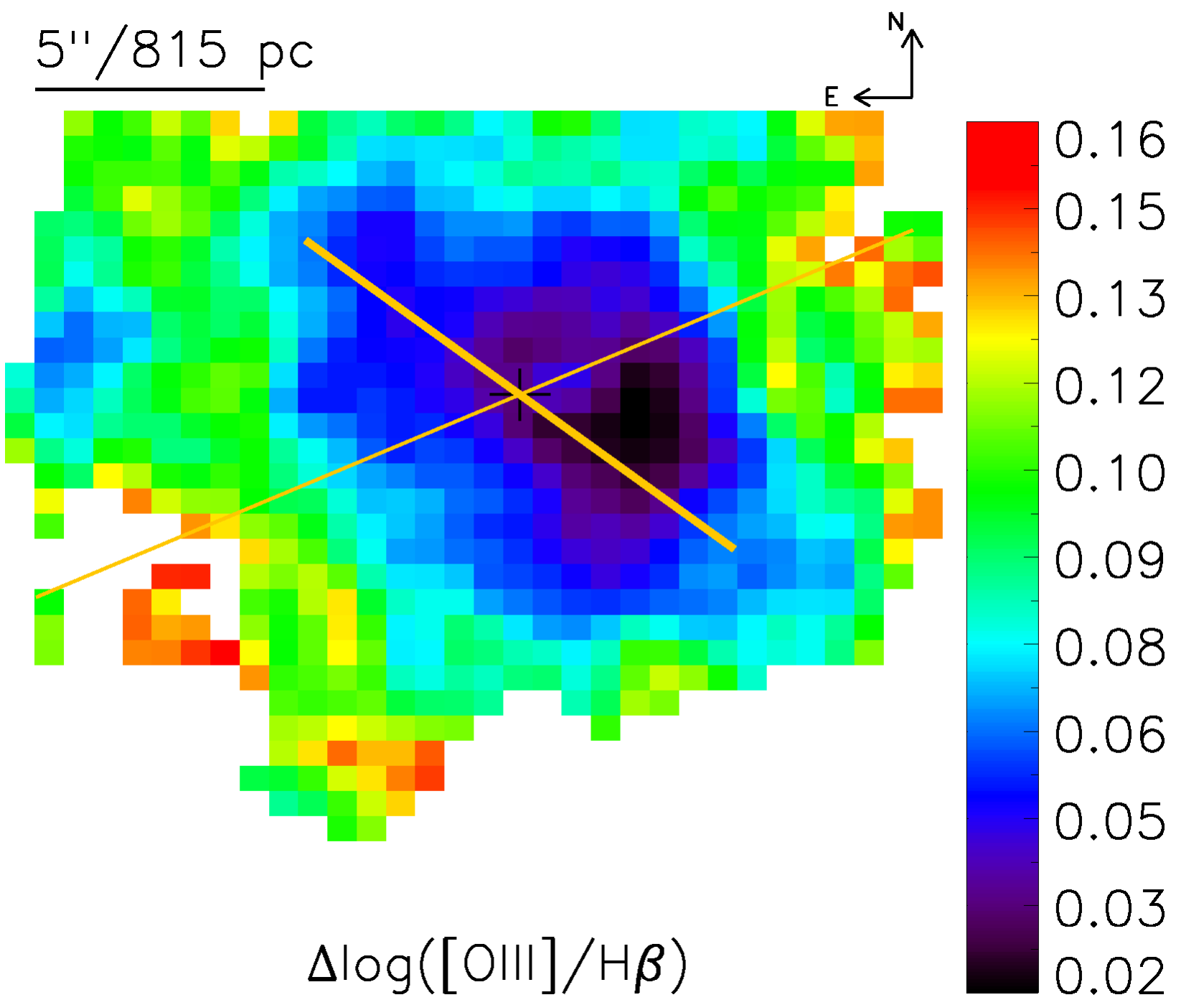}
   \includegraphics[width=0.33\textwidth]{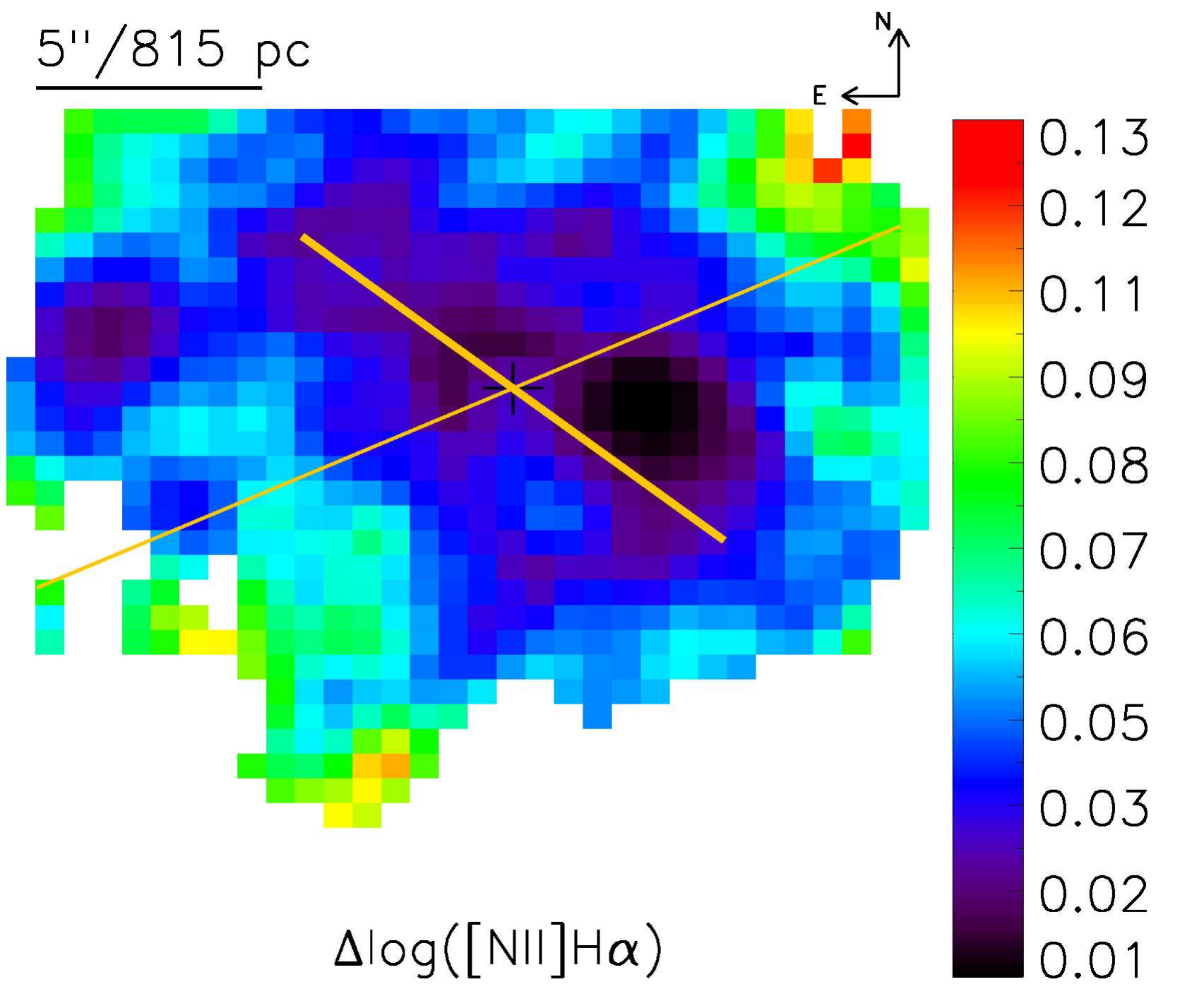}
   \includegraphics[width=0.33\textwidth]{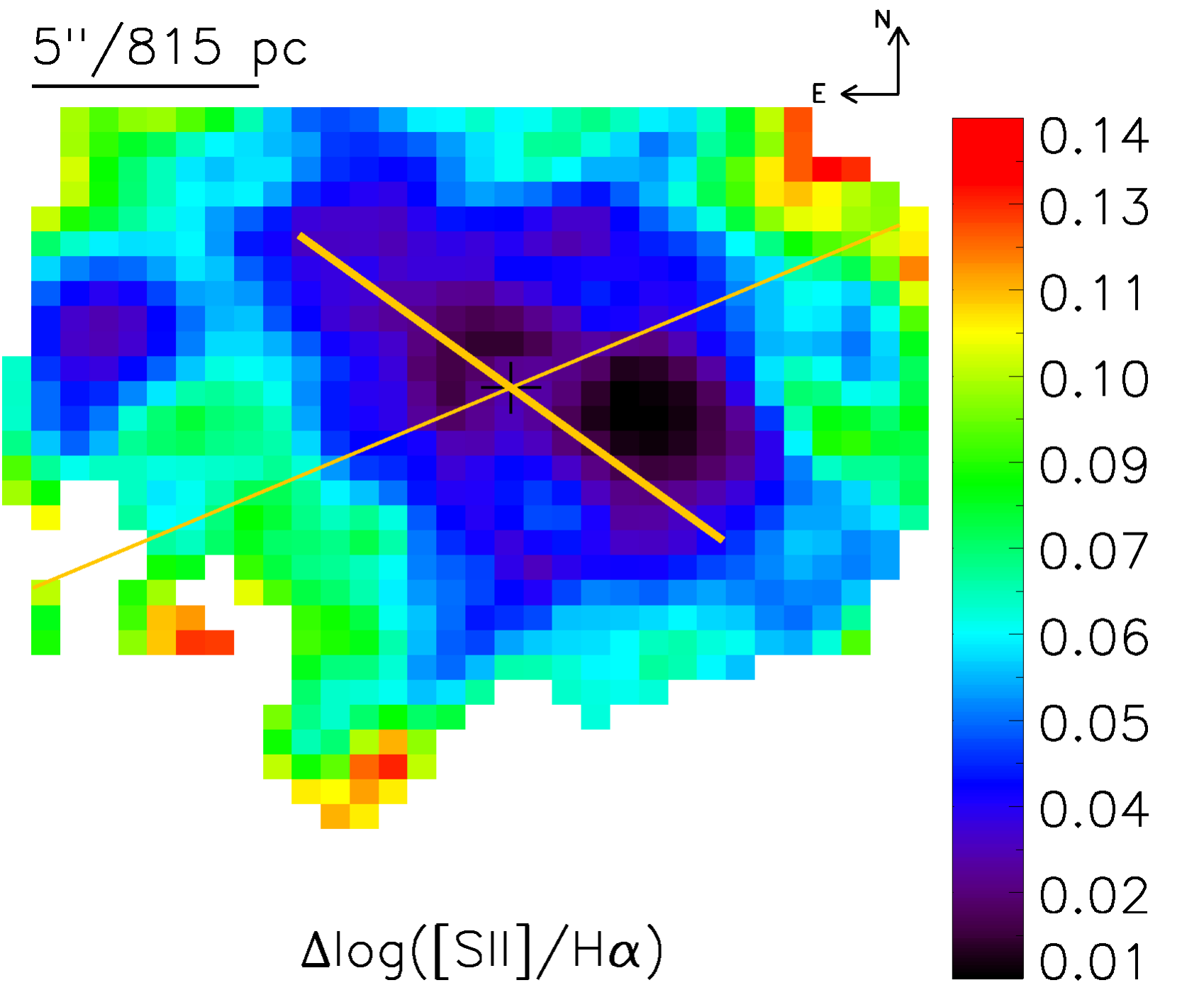}
   \caption{Upper row: Logarithmic line ratio maps of lines used for diagnostic diagrams. Black (red) contours mark regions of positive (negative) logarithmic line ratios. For the explanation of the denoted regions see text in Sect. \ref{excitation}. Bottom row: The error maps corresponding to the maps in the upper row. In all panels, the major axis of the primary (thin line) and the secondary bar (thick line) are overplotted.} 
   \label{fig:diagnostic_maps}
\end{figure*}

Maps of the observed line ratios in NGC 5850 are shown in Fig. \ref{fig:diagnostic_maps}. The highest line ratios are consistent with a LINER classification. The LINER emission is not confined to the location of the continuum peak but extended. In the $\log\left(\left[\mathrm{\ion{N}{ii}}\right]/\mathrm{H\alpha}\right)$ map, the high line ratios extend mainly from the center to the northeast with additional components to the northwest of the core's location and to the south. In this map (Fig. \ref{fig:diagnostic_maps} middle top), we denoted several regions of interest for which we stacked the spectra in a $3\times3$ spaxel aperture for further analysis. The first region is directly at the location of the continuum peak (region A). The second one (B) is the inner \ion{H}{ii}-region. With (C), we denoted an area nearby and northwards of the center, showing a strong velocity gradient (see Fig. \ref{fig:gas_kinematics}). The region in the northwestern extension of the LINER field close to the tip of the secondary bar is marked with (D). The \ion{H}{ii}-region in the far east of the FOV is marked with (E). In $\log\left(\left[\ion{S}{ii}\right]/\mathrm{H\alpha}\right)$, the map (right panel of Fig. \ref{fig:diagnostic_maps}) shows a dominant peak on the galaxy core, but high line ratios are also found south of the center (region F).\
The center of the galaxy does not coincide with the line ratio peak in the $\log\left(\left[\ion{O}{iii}\right]/\mathrm{H\beta}\right)$ map (left panel of Fig. \ref{fig:diagnostic_maps}). This is likely a contrast effect induced by the \ion{H}{ii}-region, because the [\ion{O}{iii}] peak clearly coincides spatially with the continuum peak. Nevertheless, the center has an elevated value consistent with the classification as LINER.\

\begin{figure*}
  \centering

  \includegraphics[width=0.33\textwidth]{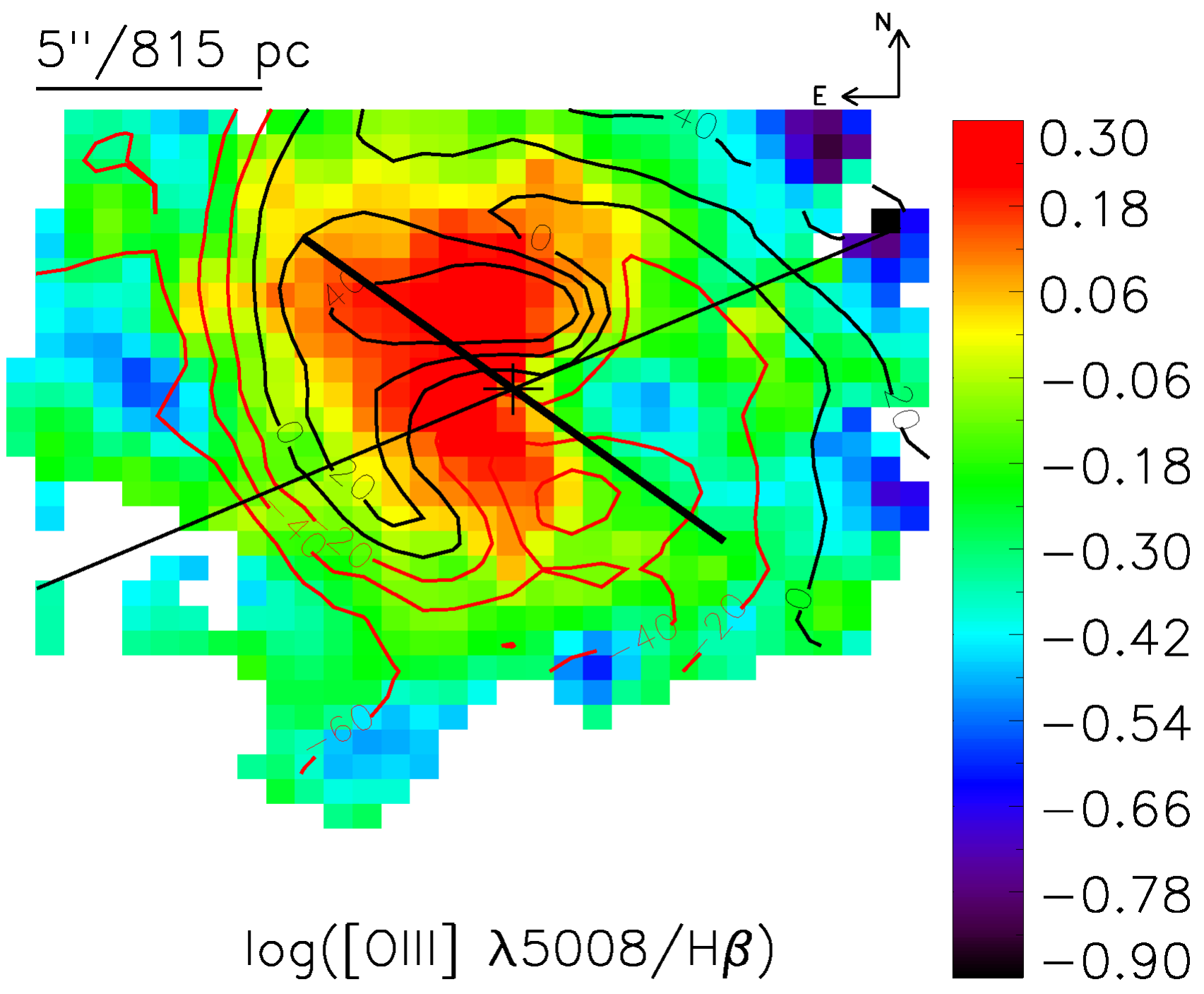}
  \includegraphics[width=0.33\textwidth]{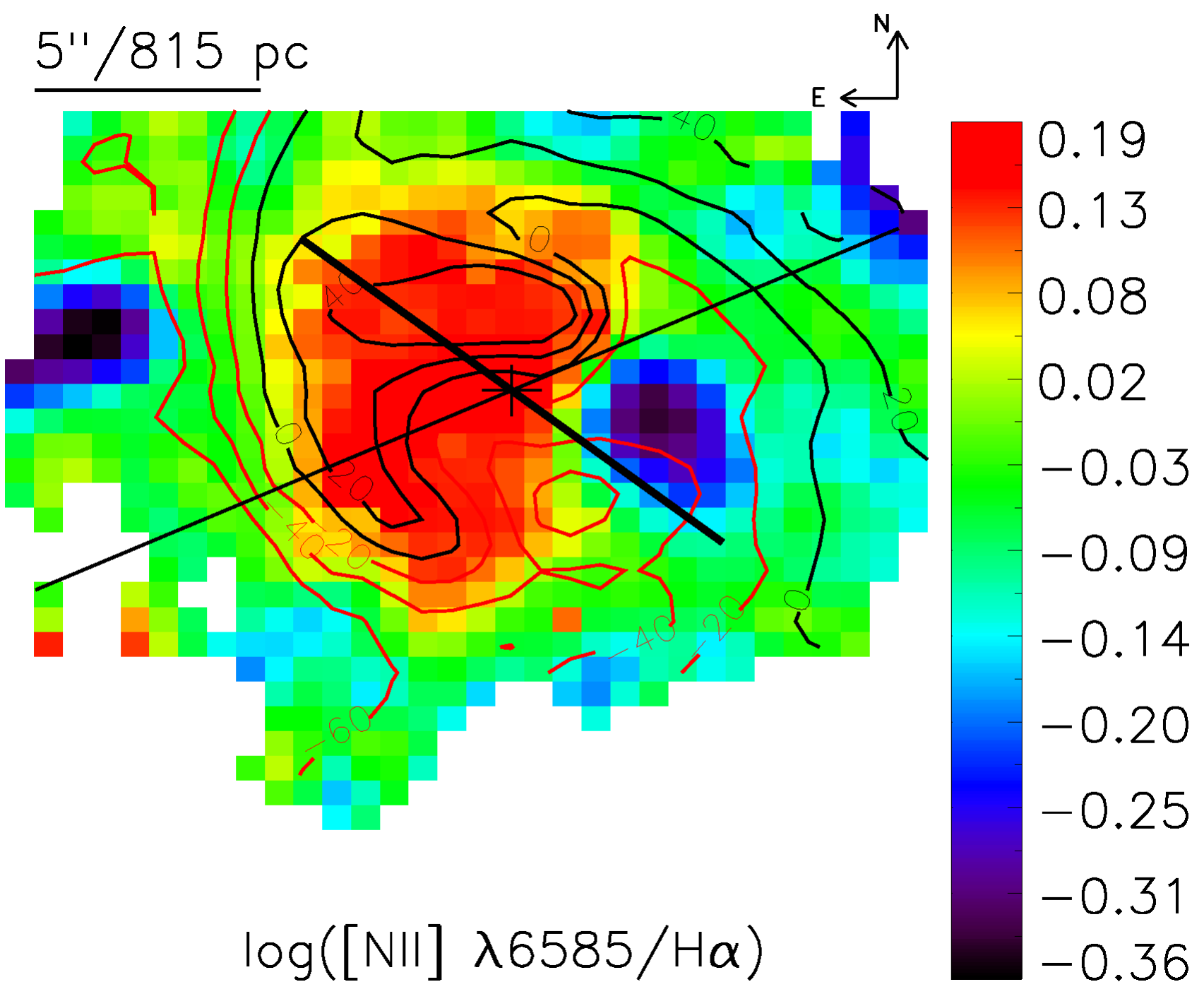}
  \includegraphics[width=0.33\textwidth]{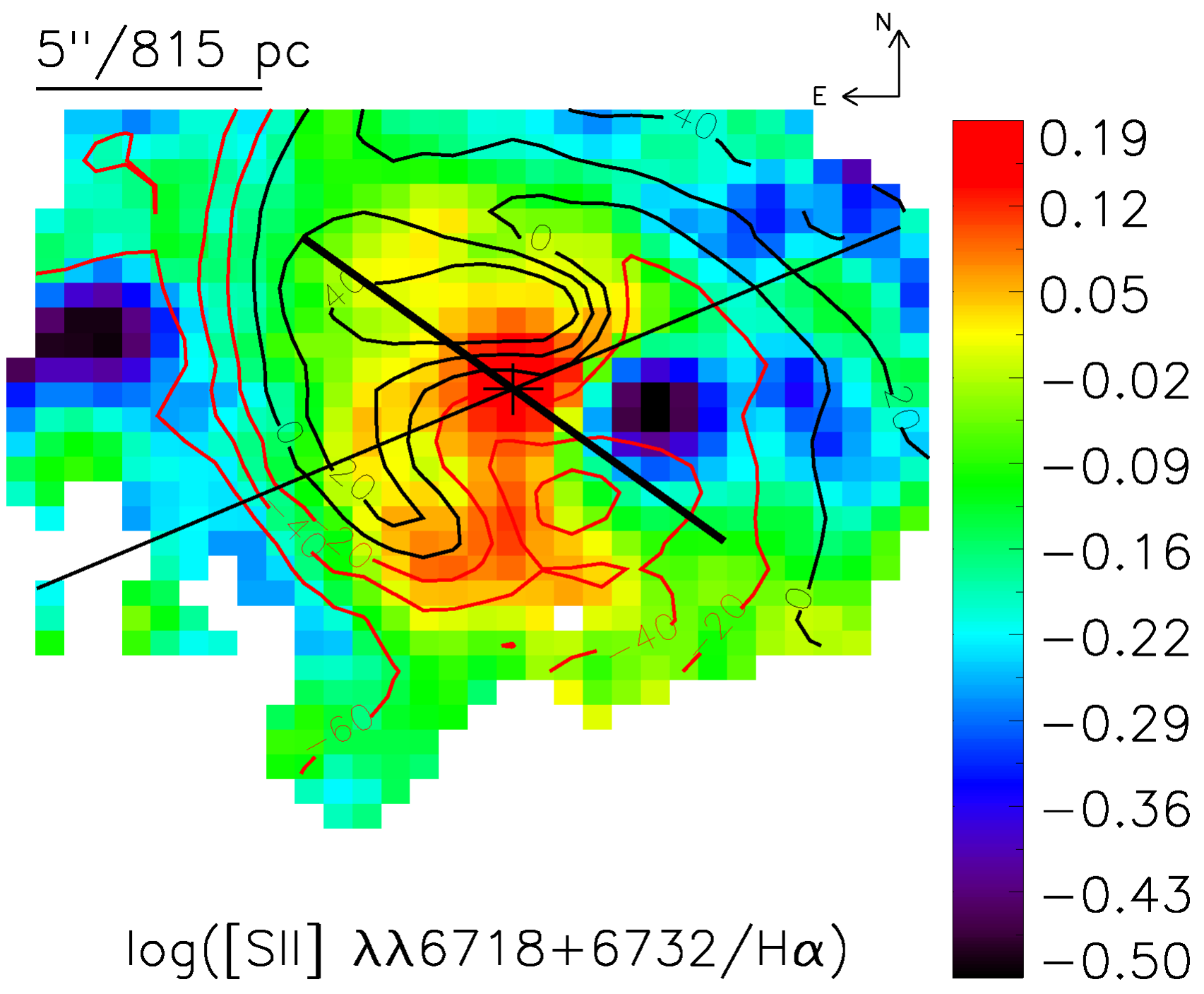}
  \caption{Map of logarithmic line ratios with overplotted H$\alpha$ line of sight velocity contours as in Fig. \ref{fig:gas_kinematics}. The high line ratios are largely confined by steep velocity gradients. The major axis of the primary (thin line) and the secondary bar (thick line) are overplotted.}
  \label{fig:line_ratios_vel_cont}
\end{figure*}

Comparison of the shape of the $\log\left(\left[\mathrm{\ion{N}{ii}}\right]/\mathrm{H\alpha}\right)$ map with the $\mathrm{H\alpha}$ LOSV (left panel of Fig. \ref{fig:line_ratios_vel_cont}) shows that the high emission line ratios largely follow the steep velocity gradients. In $\log\left(\left[\ion{S}{ii}\right]/\mathrm{H\alpha}\right)$, the spatial coincidence of the high values and the strong velocity gradients are only prominent in the center and south of it (see right panel of Fig. \ref{fig:line_ratios_vel_cont}).\

We note further the tail of lower line ratios indicated in all maps by the contours in Fig. \ref{fig:diagnostic_maps}. They start at the inner \ion{H}{ii}-region and are oriented radially outwards to the west. The average line ratios corresponding to the apertures indicated in Fig. \ref{fig:diagnostic_maps} are plotted in the diagnostic diagrams of Fig. \ref{fig:pixel_diagnostics} with the line ratio of each individual pixel.\

\begin{figure*}
   \includegraphics[width=0.33\textwidth]{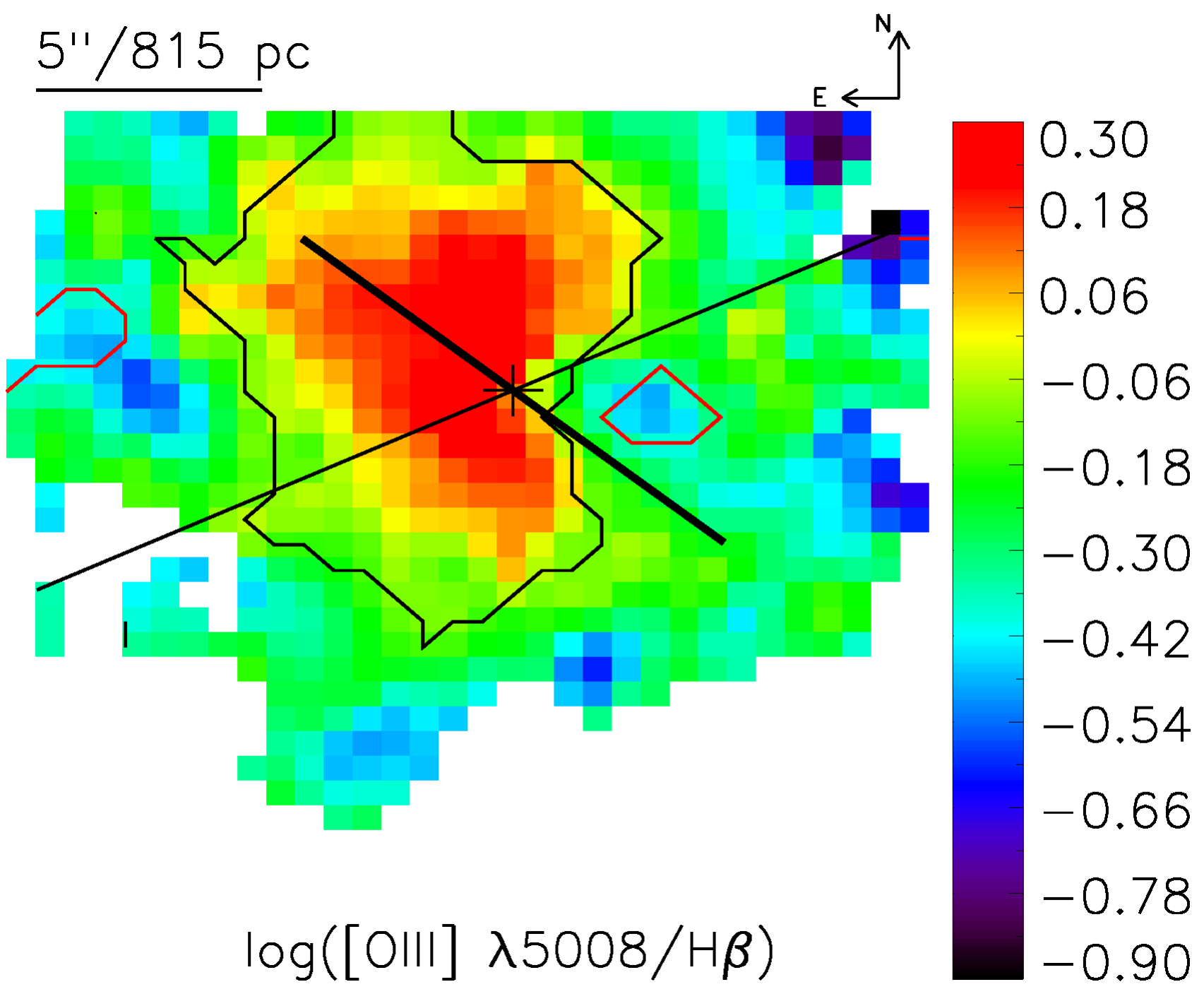}
   \includegraphics[width=0.33\textwidth]{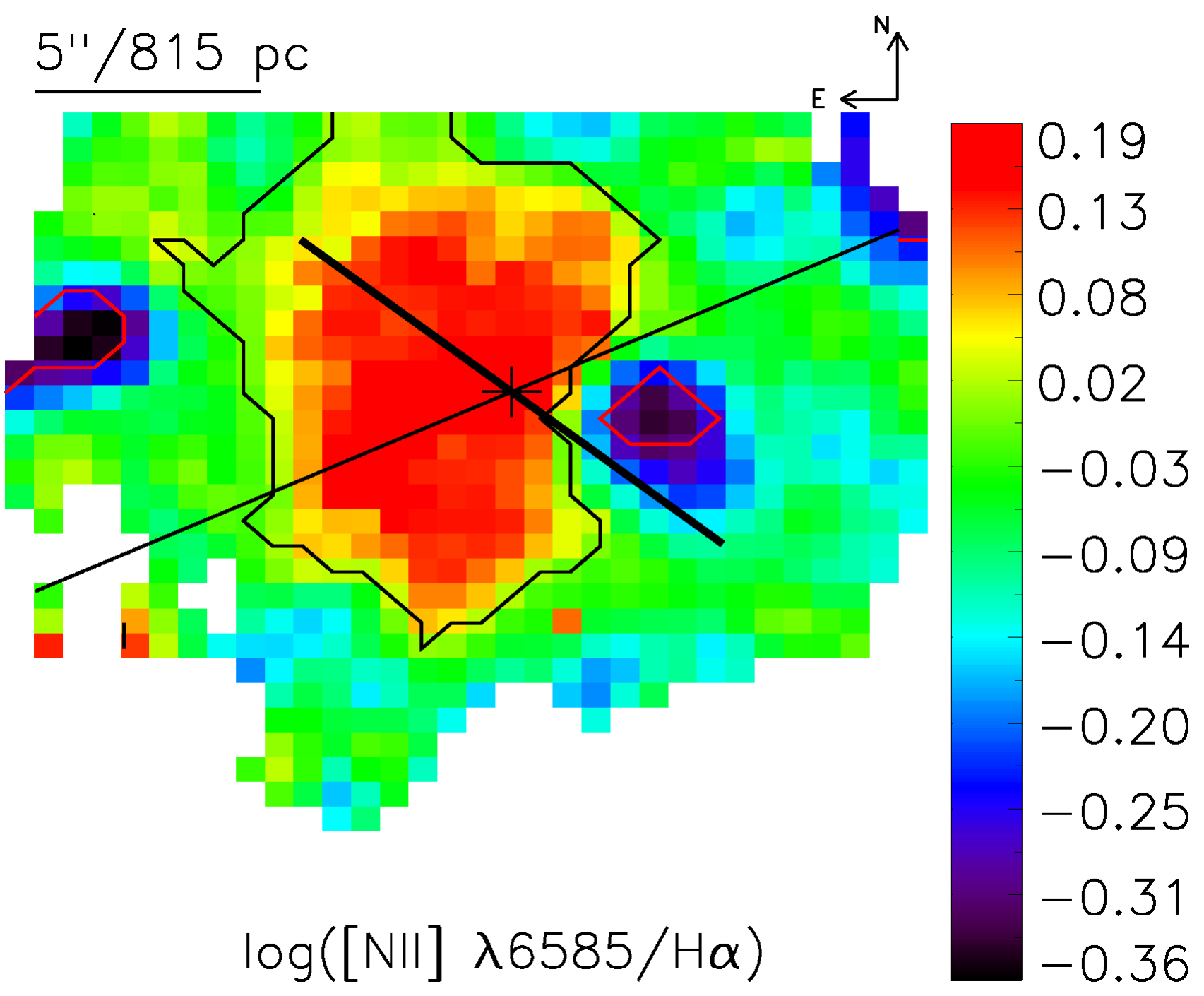}
   \includegraphics[width=0.33\textwidth]{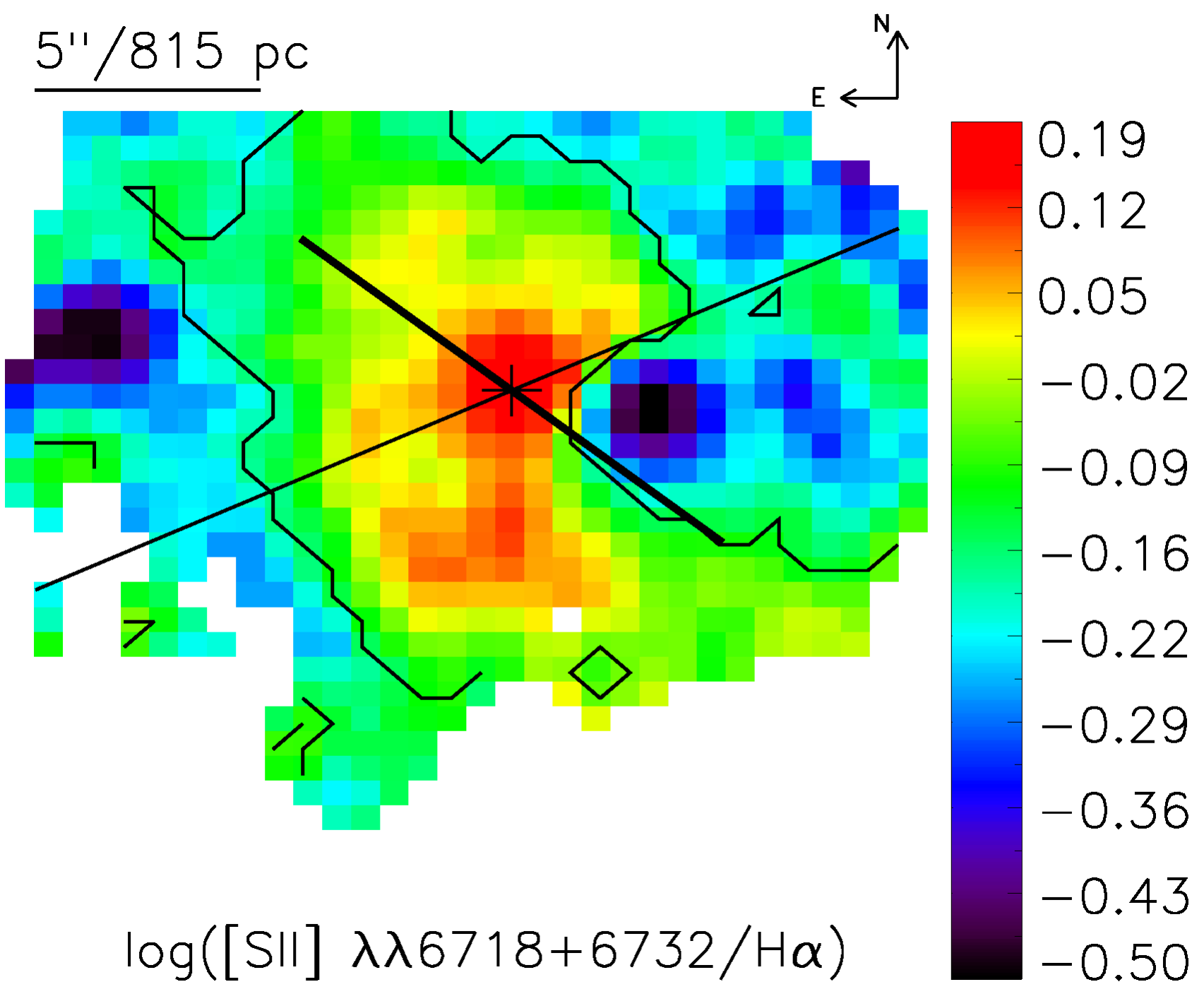}

  \caption{Line ratio maps overplotted with classification contours and the major axis of the primary (thin line) and the secondary bar (thick line). Left and center: Areas surrounded by the black contour line correspond to AGN-like ionisation. Areas enclosed by the red contour line contain regions classified as star forming. Other regions are dominated by composite-type line ratios. The classification was made according to the diagnostic diagram, the extreme starburst demarcation line of \citet{Kewley_maximum_starburst}, and the pure star-formation demarcation line of \citet{Kauffmann_pure_SF}. Right: Areas encircled by the black contour line are classified as LINER-like within the $\log\left(\left[\ion{S}{ii}\right]/\mathrm{H\alpha}\right)$ diagnostic diagram using the LINER-Seyfert demarcation line of \citet{Kewley_Sey_LINER} and the extreme starburst line.}
  \label{fig:fov_diagnostics}
\end{figure*}

\begin{figure*}
   \includegraphics[width=0.49\textwidth]{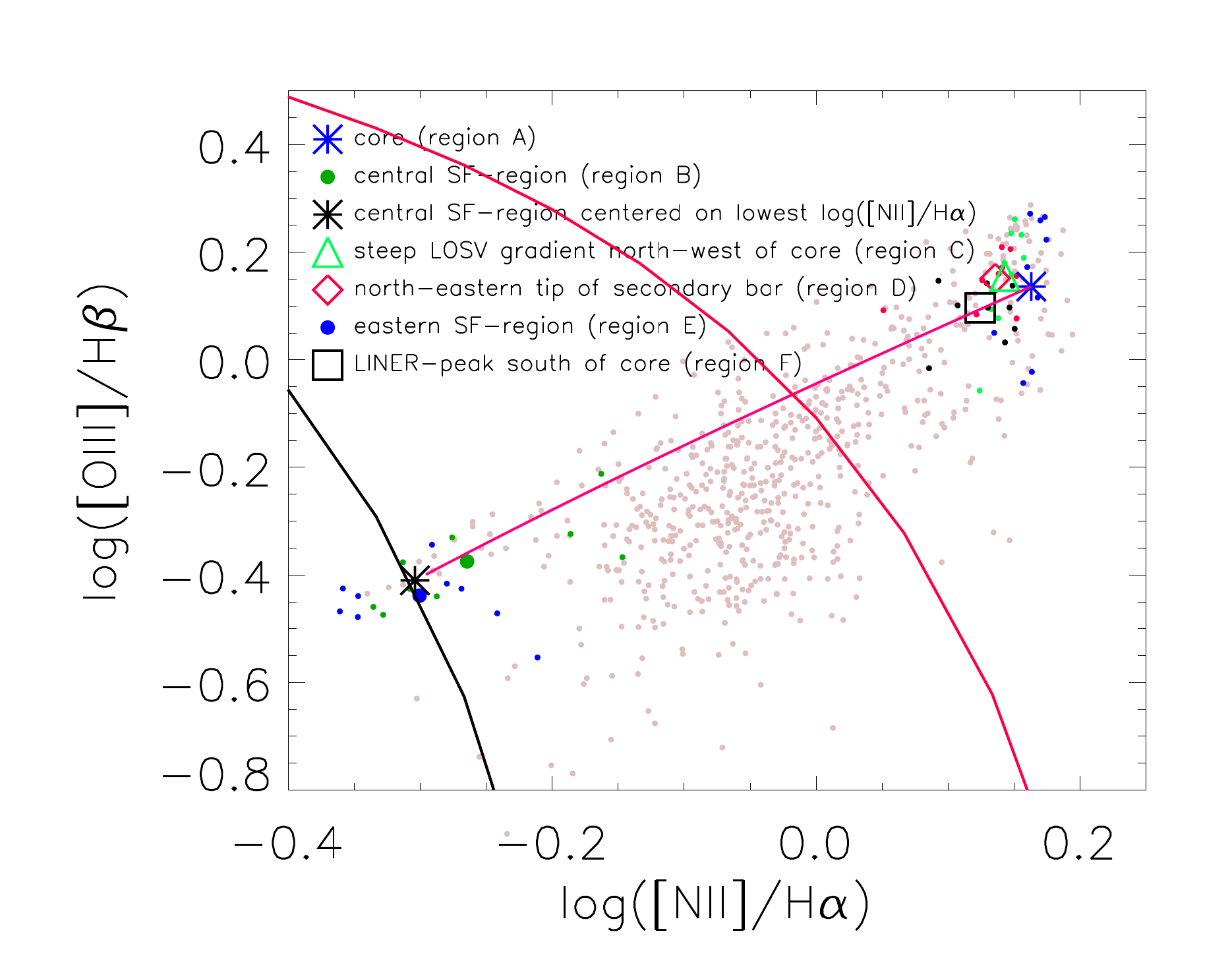}
   \includegraphics[width=0.49\textwidth]{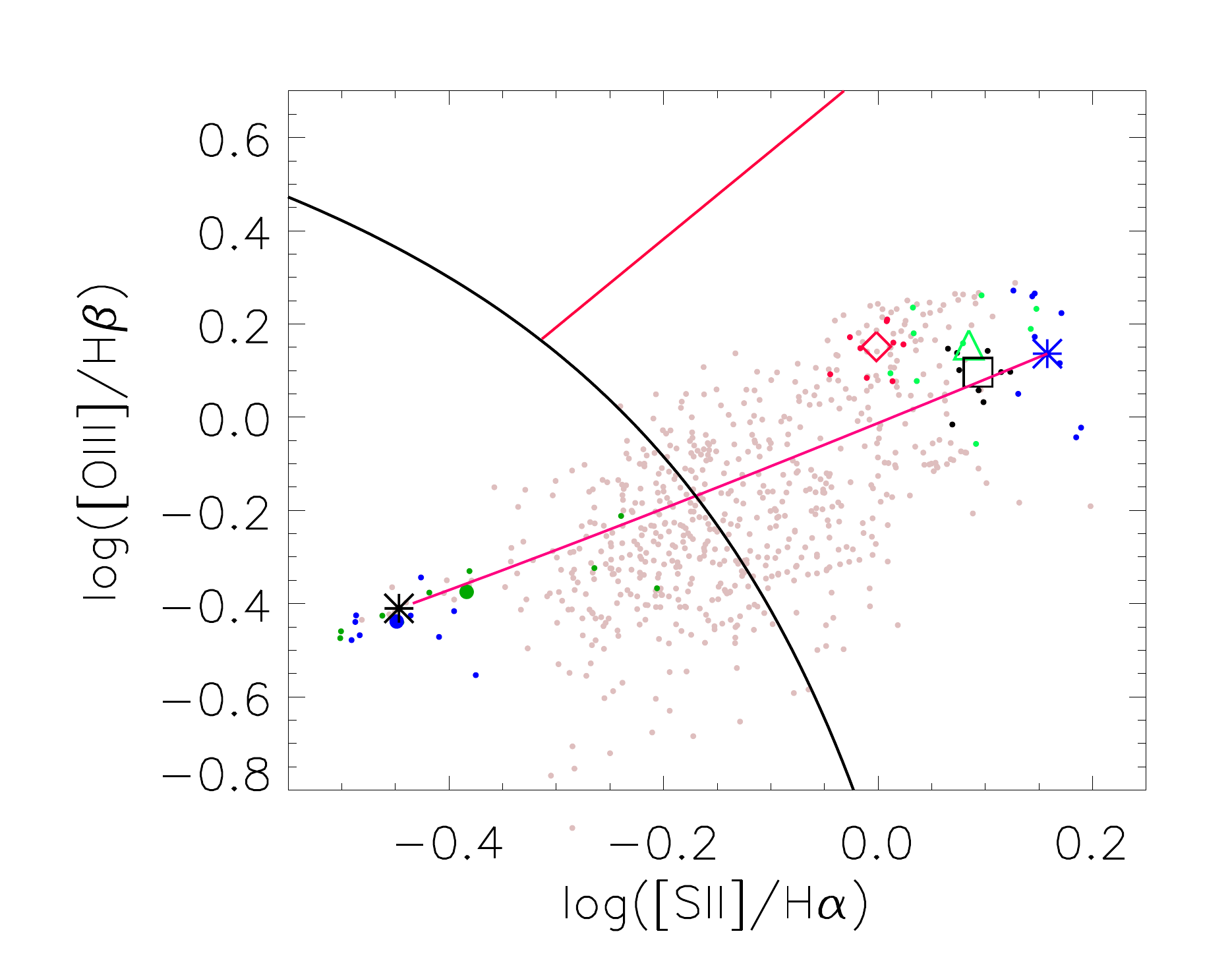}

  \caption{Diagnostic diagrams: The grey dots mark the positions in the diagnostic diagrams of spaxel for which we have all necessary line measurements. The big colored symbols mark areas in the FOV, consisting of stacked spectra of 3 by 3 spaxels. The small colored dots are the corresponding single spectra of which the average spectrum of a region was calculated. Medium blue asterisk: core region (A); dark green filled circle: central \ion{H}{ii}-region centered on spaxel with brightest H$\alpha$ emission (B); black asterisk: central \ion{H}{ii}-region centered on spaxel with lowest $\log\left(\left[\mathrm{\ion{N}{ii}}\right]/\mathrm{H\alpha}\right)$ (B); light green triangle: steep velocity gradient area northwest of core (C); red square: northeastern LINER-region close to tip of secondary bar (D); dark blue filled circle: eastern \ion{H}{ii}-region (E); and black square: LINER peak south of core (F). The solid magenta line is the SF-AGN mixture line, as explained in the text. Left: $\log\left(\left[\mathrm{\ion{O}{iii}}\right]~\lambda5008 / \mathrm{H\beta}\right)$ vs. $\log\left(\left[\mathrm{\ion{N}{ii}}\right]\lambda6585 / \mathrm{H\alpha}\right)$ with the extreme starburst line of \citet{Kewley_maximum_starburst} overplotted in red and the pure star-formation line of \citet{Kauffmann_pure_SF} in black. Right: $\log\left(\left[\mathrm{\ion{O}{iii}}\right]~\lambda5008 / \mathrm{H\beta}\right)$ vs. $\log\left(\left[\mathrm{\ion{N}{ii}}\right]~\lambda6585 / \left[\mathrm{\ion{S}{ii}}\right]~\lambda\lambda6718+6732\right)$ with the LINER-Seyfert demarcation line of \citet{Kewley_Sey_LINER} in red and the extreme starburst line in black.}
  \label{fig:pixel_diagnostics}
\end{figure*}

Figure \ref{fig:pixel_diagnostics} confirms the LINER nature of the core region A and of regions C, D, and F. In addition to the demarcation lines by \citet{Kauffmann_pure_SF} and \citet{Kewley_maximum_starburst, Kewley_Sey_LINER}, we added a mixture line to evaluate the LINER influence on the inner \ion{H}{ii}-region and the FOV (see Sec \ref{HII}). The line was calculated as the linear combination of line ratio fractions of the inner SF-region and the LINER core (each from an aperture of $3\times3$ pixels), i.e., $$\left(\frac{\left[\ion{O}{iii}\right]}{\mathrm{H\beta}}\right)_{\mathrm{mix}} = x \times \left(\frac{\left[\ion{O}{iii}\right]}{\mathrm{H\beta}}\right)_{\mathrm{SF}} + y \times \left(\frac{\left[\ion{O}{iii}\right]}{\mathrm{H\beta}}\right)_{\mathrm{LINER}} \mathrm{,}$$ where $x+y=1$ and $x=0.01,0.02,\ldots,0.99$.\
The average spectra of both \ion{H}{ii}-regions, B and E, are very close to the demarcation line that separates the SF classification from the composite one. However, they are not classified as SF, but the apertures comprise SF classified pixels. It is also interesting to note that the remaining parts of the FOV are dominated by composite type spectra.\

\section{Discussion}
\label{discussion}

\subsection{Composite emission}
\label{EELR}

In NGC 5850, we observe spatially extended regions with spectra containing narrow emission lines, which are not purely originating in the photoionization by an AGN. Large continuous parts of the FOV are classed as composite, as shown in Fig. \ref{fig:fov_diagnostics}. Within the diagnostic diagram in Fig. \ref{fig:pixel_diagnostics}, the composite region is significantly populated and forms a sequence that connects the \ion{H}{ii}-regions with the LINER-regions. With the line ratio maps, this might indicate that the importance of sources capable of producing LINER-like emission increases with decreasing distance to the center. The mixing line in the $\log\left([\ion{O}{iii}]/\mathrm{H\beta}\right)$ vs. $\log\left([\ion{S}{ii}]/\mathrm{H\alpha}\right)$ diagram, which is introduced in Sect. \ref{excitation} and connects the classification of the core (blue asterisk) with the strongest SF-like spectrum of the inner \ion{H}{ii}-region (black asterisk), confirms the hypothesis that the bulk of the gas in the FOV is ionized by a mixture of SF and a mechanism that produces LINER-like spectra. However, the mixing line in the $\log[\ion{O}{iii}]/\mathrm{H\beta}$ vs. $\log[\ion{N}{ii}]/\mathrm{H\alpha}$ diagram is located at the upper edge of the point cloud. The proximity of the central \ion{H}{ii}-region to the nucleus leads to an increase of the ionization parameter of the SF-region and lifts its position within the diagnostic diagram into the composite area. This effect might appear over-pronounced by the extent of the point-cloud toward low [\ion{O}{iii}]/H$\beta$ values. Some of these spaxels have relatively high uncertainties in [\ion{O}{iii}], indicating the line measurement to be close to the $3\sigma$ detection limit. These spaxels are commonly outside the KDC, as can be seen in the $\log([\ion{O}{iii}]/\mathrm{H}\beta)$-error map (Fig. \ref{fig:diagnostic_maps}) .\

This would not be the first discovery of an extended narrow line region (ENLR) that is classified as composite in terms of diagnostic diagrams and that is interpreted as the result of a mixture of different ionizing sources. \citet{Julia_EELR} found composite emission in the circum-nuclear region and between \ion{H}{ii}-regions in the ring of HE 2211-3903. They concluded the ionization to stem from star formation and from the AGN. \citet{Rich_galaxy_shocks} shows how star formation and slow shocks resemble the composite emission of two nearby luminous infrared galaxies. The composite emission in NGC 5850 does not appear to stem from a mixture of SF and slow shocks. Given that the velocity of slow shocks is of the order of $100$-$200~\mathrm{km~s^{-1}}$ \citep{Rich_galaxy_shocks} and that our observed LOSV dispersion values are mostly below this range, the contribution of shocks to the composite emission seems to be of low relevance.

\subsection{HII regions}
\label{HII}

Within our FOV, NGC 5850 contains two SF-regions, as identified by means of diagnostic diagrams. In the following, we report their properties. In Table \ref{table:HII}, we list the star-formation rates (SFR) as derived from the H$\alpha$ luminosity according to \citet{Kennicutt_SFR} and \citet{Calzetti_SFR} \citep[see also][]{Leroy_SFR} for the east and the central SF-region within a $3\times3$ pixels aperture. Note that the extinction correction is a source of uncertainty for the derived star formation rates. For the faint eastern HII region, the data suggest a Balmer decrement that is smaller than the case B value, and therefore, the extinction was assumed to be zero. This could be the result of inaccuracies in the choice of the stellar templates during the stellar continuum subtraction process. Another possibility could be scattering, which is not considered by the assumed dust screen model of the applied extinction law. The $\mathrm{H\alpha}$ emission in the central SF-region is affected by LINER-like photoionization, as indicated by its classification as `composite' in the diagnostic diagram. This extra radiation enhances the forbidden emission line intensity stronger than the Balmer line intensity. Furthermore, it prevents us from estimating the influence of diffuse ionized gas, which can boost the Balmer emission \citep{DIG}, and therefore, we might overestimate the calculated SFR. Scaling the \citet{Calzetti_SFR} star-formation rate of the central SF-region to an area of $\mathrm{1~kpc^{2}}$ yields a SFR-density of $\Sigma_{\mathrm{SFR}} \approx 0.06\mathrm{~M_{\sun}~yr^{-1}~kpc^{-2}}$, which is lower than the median value of $0.2\mathrm{~M_{\sun}~yr^{-1}~kpc^{-2}}$ for circumnuclear ($0.2~\mathrm{kpc} \leq 2~\mathrm{kpc}$) SF in galaxies that are the same Hubble type and that are located at roughly the same distance ($z=0.00344-0.0344$) \citep{Shi_CNSF} \citep[see also][]{Monik}. 

\begin{table*}
\caption{Star-formation rates and oxygen abundances in the \ion{H}{ii}-regions of NGC 5850.}
\label{table:HII}
\centering
\begin{tabular}{lccccc}
\hline\hline
Region	&	SFR \tablefootmark{a} &	SFR \tablefootmark{b}	&	Oxygen abundance \tablefootmark{c} &	Oxygen abundance \tablefootmark{d}	& Oxygen abundance \tablefootmark{e}\\
 		&	$[10^{-3}M_{\sun}/yr]$		   &	$[10^{-3}M_{\sun}/yr]$			&	$12+\log\left(\mathrm{O/H}\right)$ &	$12+\log\left(\mathrm{O/H}\right)$	& $12+\log\left(\mathrm{O/H}\right)$\\
\hline
\ion{H}{ii} east	&	$1.0\pm0.1$\tablefootmark{f}	&	$0.7\pm0.1$\tablefootmark{f}	&	$9.02^{+0.03}_{-0.02}$	&	$\gtrsim9.14$ & $8.77\pm0.04$ \\
\ion{H}{ii} center\tablefootmark{g}	& $9.3\pm3.7$	& 	$6.5\pm2.5$						&	$8.99^{+0.04}_{-0.04}$	&	$\gtrsim9.20$ & $8.77\pm0.06$ \\
\hline
\end{tabular}
\tablefoot{The oxygen abundance is given in values for $12+\log\left(\mathrm{O/H}\right)$. Details are described in the text.\\
\tablefoottext{a}{calculated following \citet{Kennicutt_SFR}.}
\tablefoottext{b}{Calculated following \citet{Calzetti_SFR} assuming a \citet{Kroupa_IMF} initial mass function, stellar masses in the range of $0.1~M_{\sun} < M < 100~M_{\sun}$, continuous star formation lasting for $\tau\geq6 ~\mathrm{Myr}$ with a temperature of $T_e=10^{4}~\mathrm{K}$ and an electron density of $n_e=100~\mathrm{cm^{-3}}$.}
\tablefoottext{c}{Calculated from [\ion{O}{iii}]/H$\beta$.}
\tablefoottext{d}{Calculated from [\ion{N}{ii}]/H$\alpha$.}
\tablefoottext{e}{Calculated from O3N2.}
\tablefoottext{f}{Zero extinction has been assumed.}
\tablefoottext{g}{Aperture has been centered on brightest $\mathrm{H\alpha}$ spaxel.}}\\
\end{table*}

To get an impression of the nebular metallicity\footnote{In the context of nebulae, we use metallicity and oxygen abundance synonymously. However, they are not strictly the same as metallicity commonly describes the iron-to-hydrogen ratio, and the oxygen abundance is the oxygen-to-hydrogen ratio.}, we make use of the connection between the optical emission line ratios and $12+\log\left(\mathrm{O/H}\right)$, as shown in Fig. 6 in \citet{Maiolino_metallicity}. Our work focuses on the gas present in NGC 5850. Thus, we do not use the stellar metallicity \citep{5850_lorenzo} and choose the strong-lines method instead. We use $\log\left(\left[\ion{O}{iii}\right]~\lambda 5008/\mathrm{H\beta}\right)$ and $\log\left(\left[\ion{N}{ii}\right]~\lambda 6585/\mathrm{H\alpha}\right)$ again for the close separation between the emission lines involved. We apply the method on $3\times3$ pixel apertures on both \ion{H}{ii}-regions and report the results in Table \ref{table:HII}. The values derived from different line ratios are not consistent with each other but they indicate the gas in both regions to be characterized by super-solar abundances with values roughly around $9$, which correspond to an oxygen abundance $\left(\mathrm{O/H}\right) \gtrsim 1.8\left(\mathrm{O/H}\right)_{\sun}$. For comparison, the solar value is $12+\log\left(\mathrm{O/H}\right)_{\sun} = 8.69$ \citep{Asplund_solar_O_abundance}. Please note that the metallicity calculated from $\log\left(\left[\ion{O}{iii}\right]~\lambda 5008/\mathrm{H\beta}\right)$ in \citet{Maiolino_metallicity} is double-valued. We constrained the range of possible solutions by considering the value derived from $\log\left(\left[\ion{N}{ii}\right]~\lambda 6585/\mathrm{H\alpha}\right)$. For the abundance calculated from $\log\left(\left[\ion{N}{ii}\right]~\lambda 6585/\mathrm{H\alpha}\right)$, we can only give a lower limit since our line ratios intersect the value space of the fit by \citet{Maiolino_metallicity} only with the negative error bar.This high value is possibly the signature of a nitrogen overabundance, a feature commonly associated to AGN \citep[e.g.][]{Julia_EELR}.\

Additionally, we calculated the oxygen abundance from $\mathrm{O3N2}\equiv\log \lbrace\left(\left[\ion{O}{iii}\right]~\lambda5008/\mathrm{H\beta}\right) / \left(\left[\ion{N}{ii}\right]~\lambda6585/\mathrm{H\beta}\right)\rbrace$ \citep{Pettini_metallicity} that yielded values of about $8.77$ for both \ion{H}{ii}-regions, which are close to the solar value.\

To complete the picture, we refer the reader to the results of \citet{5850_lorenzo} on the stellar metallicity. They found a systematically increased value in NGC 5850 $(\mathrm{Z\approx1.3~Z_{\sun}})$ within the secondary bar compared to the remaining parts of the kinematically decoupled core. The central \ion{H}{ii}-region is located well within the stripe of increased metallicity of the secondary bar. The eastern \ion{H}{ii}-region, in contrast, is not positioned within the secondary bar but nevertheless shows values higher than solar in its approximate region. Due to the projected proximity of the central \ion{H}{ii}-region to the LINER-center, we expect signatures of mutual contamination of their spectra. Indeed, the outer parts of the central \ion{H}{ii}-region are diagnosed as composite and as LINER-like while approaching the galactic center (Fig. \ref{fig:diagnostic_maps}). 

On the other hand, the [\ion{O}{iii}] and [\ion{N}{ii}] emission in the central SF-region do not appear to originate in the gas irradiated by young stars only, but it comprises the contribution of other distributed ionizing sources and/or the AGN. This becomes clear if we look at Fig. \ref{fig:pixel_diagnostics}. Though close to the pure-SF-demarcation line, harder radiation lifted the average line ratios of the inner SF-region (aperture centered on the brightest H$\alpha$ pixel, thick green dot in Fig. \ref{fig:pixel_diagnostics}) to the composite area. Note that the aperture centered on the strongest H$\alpha$ pixel is slightly closer to the nucleus than the aperture centered on the lowest $\log\left(\left[\ion{N}{ii}\right]/\mathrm{H\alpha}\right)$ ratio of the central SF-region (black asterisk), but the latter is more `SF-like' in the diagnostic diagrams.\\

The line ratios of the outer SF-region are almost coinciding with the position of the black asterisk within the diagnostic diagrams, i.e. the least core contaminated average spectrum of the central SF-region. Both show approximately the same gas abundance.\\

\begin{figure}
  \includegraphics[width=0.5\textwidth]{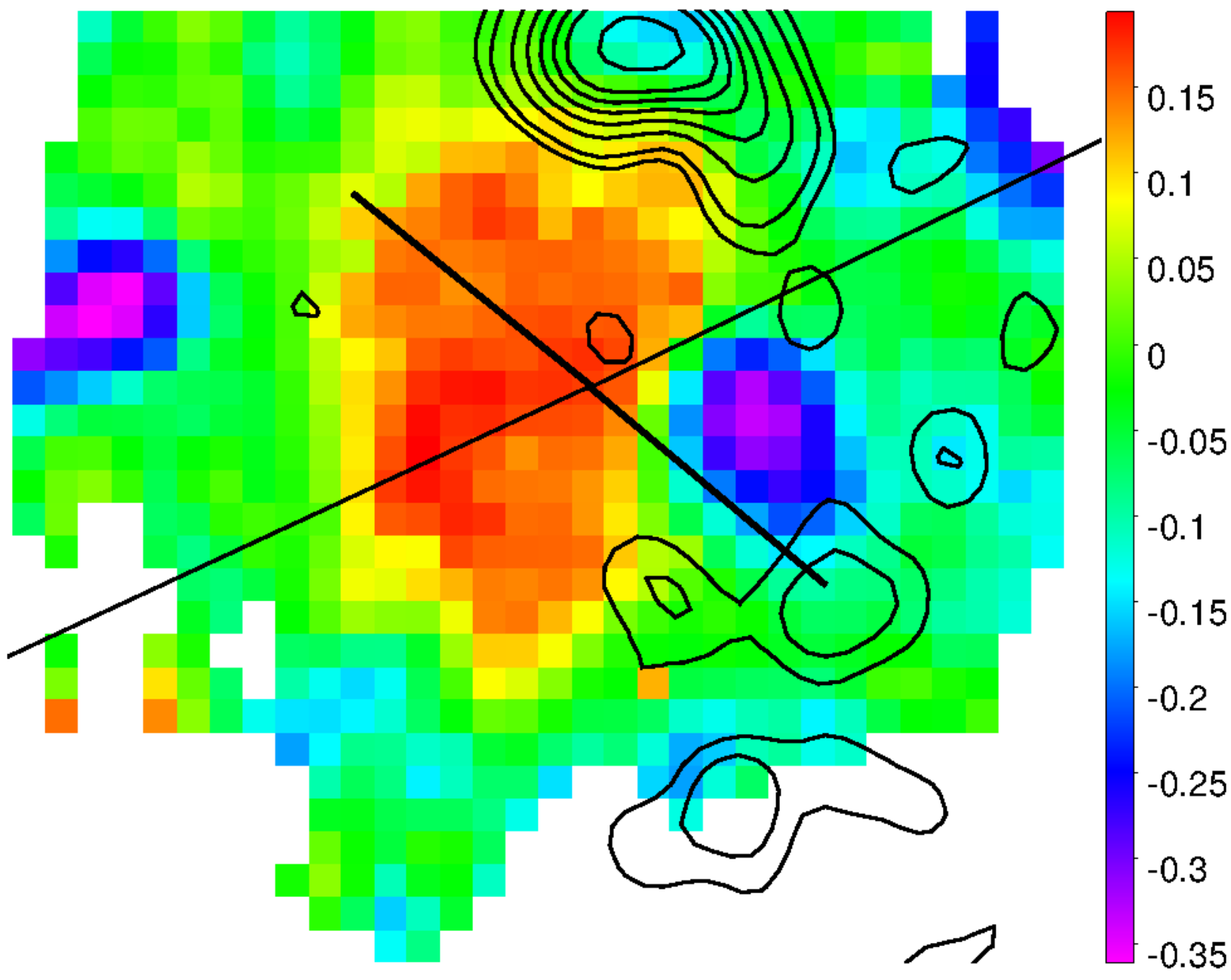}
  \caption{The same as in the $\log(\mathrm{[\ion{N}{ii}]/H\alpha})$ map in Fig. \ref{fig:fov_diagnostics} but the CO intensity contours at 0.006, 0.01, 0.02, 0.03, 0.04, 0.05, 0.06, 0.08 and 0.1 Jy/beam are additionally superimposed in black .}
  \label{fig:line_ratio_CO_contours}
\end{figure}

Possibly the interplay of radiation originating in the core and in the central SF-region is not only confined to the space in between them. The wedge-shaped line ratio contours in all of the three diagnostic maps (Fig. \ref{fig:diagnostic_maps}) indicate reduced line ratios radially outwards from the inner SF-region. This hints at a shielding effect, where the parts of the galaxy located behind the \ion{H}{ii}-region are protected from the LINER-like emission from the more central region. However, the wedge more likely represents an area of  enhanced SF, although weaker than in the central SF-region. This wedge is unlikely caused by an extinction effect. CO is detected weakly in this area as can be seen in Fig. \ref{fig:line_ratio_CO_contours}, where we display $\log\left(\left[\mathrm{\ion{N}{ii}}\right]/\mathrm{H\alpha}\right)$ overplotted with CO emission contours from the NUGA observation by \citet{LeonCombes2000}. CO emission is related to the hydrogen column density which again is linked proportionally to extinction \citep{Heiles_hydro_extinct, Predehl_hydro_extinct}. The extinction map derived from the $\mathrm{H\alpha/H\beta}$ Balmer decrement shows no extinction there. Furthermore, the emission lines used to calculate the line ratio are spectrally close to each other so that reddening effects can be neglected. Unfortunately the differences between the 'shielded' area and its immediate surroundings are small, so this effect disappears in the uncertainties. In the $\log\left(\left[\mathrm{\ion{O}{iii}}\right]/\mathrm{\ion{S}{ii}}\right)$ map (Fig. \ref{fig:SII_maps}), this feature is much better visible, though this line ratio map might be more prone to extinction effects than the others despite our effort to correct for them. However, it is remarkable that the feature can be seen in all diagnostic maps.

\subsection{Photoionization by AGN-contribution}
\label{AGN}

Generally, we confirm the classification of the nucleus of NGC 5850 to be a LINER, following the classical diagnostic diagrams. However, considering the extent of the LINER-like emission (Fig. \ref{fig:pixel_diagnostics}) over the few central arcseconds leads to the question of whether the emission really originates in an AGN, which is a point source that is supposed to be unresolved. In this section, we examine the question of whether the gas in the central few arcseconds is photoionized by an AGN.\\

Extended line emission has been observed in all kinds of galaxies (see Sect. \ref{Intro}). For example \citet{Yan_nature_of_LINERS} agreed on AGNs to be a possible origin. Further, they pointed out that the line surface brightness depends on the photoionizing flux profile, the gas volume filling factor, the spatial distribution of the gas clouds and the gas density as function of distance to the point source.\

In NGC 5850, one could imagine the elongated extent of the forbidden line emission could be attributed to an increased gas reservoir within the secondary bar. Especially in the $\log\left(\left[\ion{S}{ii}\right]~\lambda6718+6732/\mathrm{H\alpha}\right)$ map, if there were not the central \ion{H}{ii}-region, the impression of the LINER-like emission elongated along the complete bar would be striking. Nevertheless, this scenario is not supported by the electron density distribution, which is roughly constant throughout the complete FOV.\

In the $\log\left(\left[\ion{S}{ii}\right]~\lambda6718+6732/\mathrm{H\alpha}\right)$ map, the central region of increased line ratios is not so largely extended but almost unresolved. The second peak in the south of the core coincides spatially with the `horn' of the velocity field, and therefore, we do not connect it to the photoionization by the putative AGN. \

Surprisingly, the $\log\left(\left[\ion{O}{iii}\right]~\lambda5008/\mathrm{H\beta}\right)$ map does not peak on the center though the continuum peak position is still classified as LINER-like. Outflows are usually made responsible for such behaviour \citep[e.g.][]{outflow_geometry}. The possible existence of an outflow, as discussed later in this section, would argue in favour of the single AGN model. However, the proximity of the central SF-region to the core makes a contamination of the core's emission probable.\\

In general, the detection of a compact X-ray and/or radio source at the position of the nucleus is a strong evidence of the presence of an AGN \citep{Ho_review, Nagar_LLAGN_radio}. The faint X-ray luminosity of NGC 5850 ($L_{\mathrm{X}}(0.5-3~\mathrm{keV})\approx10^{40.36}~\mathrm{erg~s^{-1}}$; \citet{5850_Lx}) suggests the nucleus to be in a state of low activity. Unfortunately, the value is integrated over a radius of $80\arcsec$, and therefore, it does not allow a statement on the compactness of the source. Further, it could comprise other X-ray emitting sources, such as supernova remnants, accreting binaries, or a hot interstellar medium \citep{5850_Lx}. Hence, the X-ray luminosity can only be used as an upper limit for the activity state of the nucleus, and as such, it would be consistent with luminosities expected for LINERs. Consequently, \citet{5850_Lx} considered NGC 5850 as unlikely to be AGN dominated. \citet{5850_radio} classified the luminosity of NGC 5850 in their radio continuum survey at $1.4~\mathrm{GHz}$ as star formation driven, though one should keep the beam size of $FWHM=45\arcsec$ here in mind as well. The beam easily comprises the central and the eastern SF-region. The beam used by \citet{Hummel_radio} at $1.49 \mathrm{~GHz}$ was narrower (half power beam width$=1\farcs3$), but the flux measurement yielded a value below the detection limit ($S_{\nu}\lesssim 0.6\mathrm{~mJy}$).\\

The bolometric luminosity of the core region can also give a hint of the presence of an AGN. We measured the extinction corrected luminosity of [\ion{O}{iii}] $\lambda5008$ in a circular aperture of $3 \times FWHM\mathrm{(PSF)}$, corresponding to $\sim 6\farcs04$ in diameter. We derive from this aperture $\log(L_{\left[\ion{O}{iii}\right]}/\mathrm{erg~s^{-1}}) = 39.1$. Making use of the estimation of \citet{Shen_L_bol}, the bolometric luminosity is then $\log (L_{\mathrm{bol}}/\mathrm{erg~s^{-1}}) = \log (3200 \times L_{\left[\ion{O}{iii}\right]}/\mathrm{erg~s^{-1}})=42.6$ which corresponds to $L_{\mathrm{bol}}\approx 3.7\times10^9 ~\mathrm{L}_{\sun}$. We emphasize that this large aperture comprises parts of the central \ion{H}{ii}-region. This conversion has been derived from a sample of quasars and might not be applicable to LINER galaxies. Therefore, we include the calculation by \citet{Netzer_L_bol}, who gave a conversion for Seyfert 2 and LINER 2 objects relying on both the [\ion{O}{iii}] and the $\mathrm{H\beta}$ emission lines: 
$$\log (L_{\mathrm{bol}})=\log (L(\mathrm{H}\beta)) + x + \mathrm{max}\left[0.0,~0.3(\log(\left[\ion{O}{iii}\right]/\mathrm{H}\beta)-0.6)\right],$$ 
where $x=3.75$ for galactic reddening. We calculate $\log(L_{\mathrm{bol}}) = 42.8\pm0.6$. The uncertainty of the scaling relation is mentioned to be $0.3-0.4~\mathrm{dex}$. Here, the central \ion{H}{ii}-region also influences the calculation. Therefore, the bolometric luminosity can only be an upper limit. However, the derived value agrees with the one derived by the scaling relation of \citet{Shen_L_bol}, and both are within the range expected for LINER AGN \citep{Ho_review}.\

In summary, there is no clear evidence of the presence of an AGN in NGC 5850 from X-ray and radio observations, but the calculated upper limit of bolometric luminosity cannot be used to reject the presence of an AGN either.\\

\subsection{Outflows}
\label{outflows}

Outflows can become apparent in the shape of ionization cones, which are elongated structures that present an increased degree of ionization. Generally, outflows can have two different origins. Either they are driven by AGN or starbursts. \citet{Sharp_cones} found that in AGN-galaxies the wind filaments are predominantly photoionized by the AGN. In starburst galaxies on the other hand, the mechanical interaction with the ISM (i.e., shocks) is made responsible for their ionization. As mentioned in Sect. \ref{Intro}, shocks can be associated with spectra showing LINER-like line emission ratios \citep[e.g.][]{Dopita_shock_producers}. Furthermore, outflows leave a kinematic signature in the form of an increased velocity dispersion and/or additional line components with different velocities compared to the main component. \citet{Walsh_Outflow_LINER} showed that outflows can have a significant impact on the kinematics of NLRs.\\

In NGC 5850, the spatial distribution of LINER-like emission is strongly extended, resembling a structure that could be interpreted as an ionization cone mainly oriented along the secondary bar (Fig. \ref{fig:fov_diagnostics}). Only some regions in our velocity dispersion map show increased values, possibly due to beam smearing (Sect. \ref{kinematics}). A wing component in the emission lines was detected north and south of the core (regions C and F) but also spatially coincided with both steep gradients and strong twists in the velocity field. We could not properly fit these wings with multicomponent fits of Gaussian profiles in the few spectra concerned. The procedure either tended to fit the continuum or encountered the parameter limits. Hence, no uncertainties could be derived. Stacking the spectra of $3\times3$ pixels apertures could not alleviate the problem. We display sample spectra and attempted fits in Fig. \ref{fig:asymmetries} with the derived kinematic values in Table \ref{table:double_fits} in Sect. \ref{kinematics}. They show velocities up to $260 \mathrm{~km~s^{-1}}$, which is a common value for outflows. The LOSV and the LOSV dispersion values of different lines within one spectrum do not agree with each other. Since the spatial extent of the emission lines that show wing components is not oriented along the major axis of the LINER-like emission it is likely that they are not directly linked to each other. As discussed before, the asymmetries coincide with steep velocity gradients that point to beam smearing effects. \

In the case of NGC 5850 the differentiation of a potential outflow by its ionization mechanism is problematic. Assuming the presence of a LINER-AGN, the outflow would manifest itself with LINER-like line ratios, making it indistinguishable from a starburst-driven and therefore shock-ionized wind \citep[e.g][]{outflow_geometry}. The SF-region close to the center also makes the presence of a starburst-driven wind a viable option.\\

In Sect. \ref{AGN}, we inferred the bolometric luminosity. In combination with the Eddington luminosity, this can give us a hint on the likelihood of the presence of an outflow. To calculate the Eddington luminosity, we first derived the mass of the central black hole ($M_{\mathrm{BH}}$).\ 

We used the M-$\mathrm{\sigma}$ relation \citep{Gebhardt_M_sigma, Ferrarese_M_sigma} for the different morphologic subsamples of galaxy bulges of \citet{Gueltekin_M_sigma}. These subsamples do not consider the highly peculiar environment of NGC 5850 and therefore, they pose a coarse approximation at best for the black hole mass. We set a circular aperture centered on the continuum peak with a radius of $3\times FWHM$ of the PSF ($\approx 6\arcsec$). The covered area excludes the mentioned $\mathrm{\sigma}$-hollows (Sect. \ref{kinematics}) but comprises the lower $\sigma$-values of the central SF-region. From the mean stellar velocity dispersion in that aperture, we infer $\mathrm{\sigma_{*}} = (113.7 \pm 17.9)~\mathrm{km~s^{-1}}$. Hence, the mass calculates to  $\log\left(\mathrm{M_{BH}/M_{\sun}}\right) = 7.4\pm 0.3$ , using  $\log(M_{\mathrm{BH}} / M_{\sun}) = \alpha + \beta \times \log(\sigma/ 200\mathrm{~km~s^{−1}})$ with $\alpha = 7.67 \pm 0.115$ and $\beta = 1.08 \pm 0.751$ from the barred subsample of \citet{Gueltekin_M_sigma}, which provided the highest value.\

It is not clear yet, whether NGC 5850 has an active nucleus or not (Sect. \ref{AGN}). Therefore, we compare our $M_{\mathrm{BH}}$ value with those of 32 inactive galaxies from \citet{Graham_M_sigma} and the M-$\mathrm{\sigma}$ relation from \citet{Park_M_sigma} for local active galaxies. In the M-$\mathrm{\sigma}$-plane of inactive galaxies, the $M_{\mathrm{BH}}$ of NGC 5850 is close to the regression line. The M-$\mathrm{\sigma}$ relation for local active galaxies results in $\log(M_{\mathrm{BH}}/M_{\sun}) = (7.1 \pm 0.6)$. Hence, our values agree with both samples. We adopt the value of $\log(M_{\mathrm{BH}}/M_{\sun}) = 7.1$ and infer consequently an Eddington luminosity of $(\mathrm{L_{Edd}/\mathrm{~erg~s^{-1}}}) = 9.6\times10^{44}$ and an Eddington ratio of $L_{\mathrm{bol}}/L_{\mathrm{Edd}}\leq 0.007^{+0.009}_{-0.006}$.\

In \citet{Ho_review}, our Eddington ratio is above the upper end of the distribution amongst LINER-galaxies. Furthermore, there is a non-zero probability for the existence of an outflow \citep{Ho_review} with a luminosity just below $1~\%$ of $L_{\mathrm{Edd}}$. Besides their polar disk scenario, \citet{doublebarred_structure_moissev} also speculated on an outflow. \citet{Masegosa_Halpha_LINER} classed $42~\%$ of their sample of $36$ LINERs as outflow candidates. Interestingly, they found a slight trend. Their classification of the galaxy center morphologies from core-halo (i.e., unresolved point source) via outflow to disky systems appears to be accompanied by a decrease in Eddington ratio. Ratios on the order of $10^{-3}$ are all core-halo classified. For NGC 5850, we derive a similar value. \\

\begin{figure*}
  \includegraphics[width=1.0\textwidth]{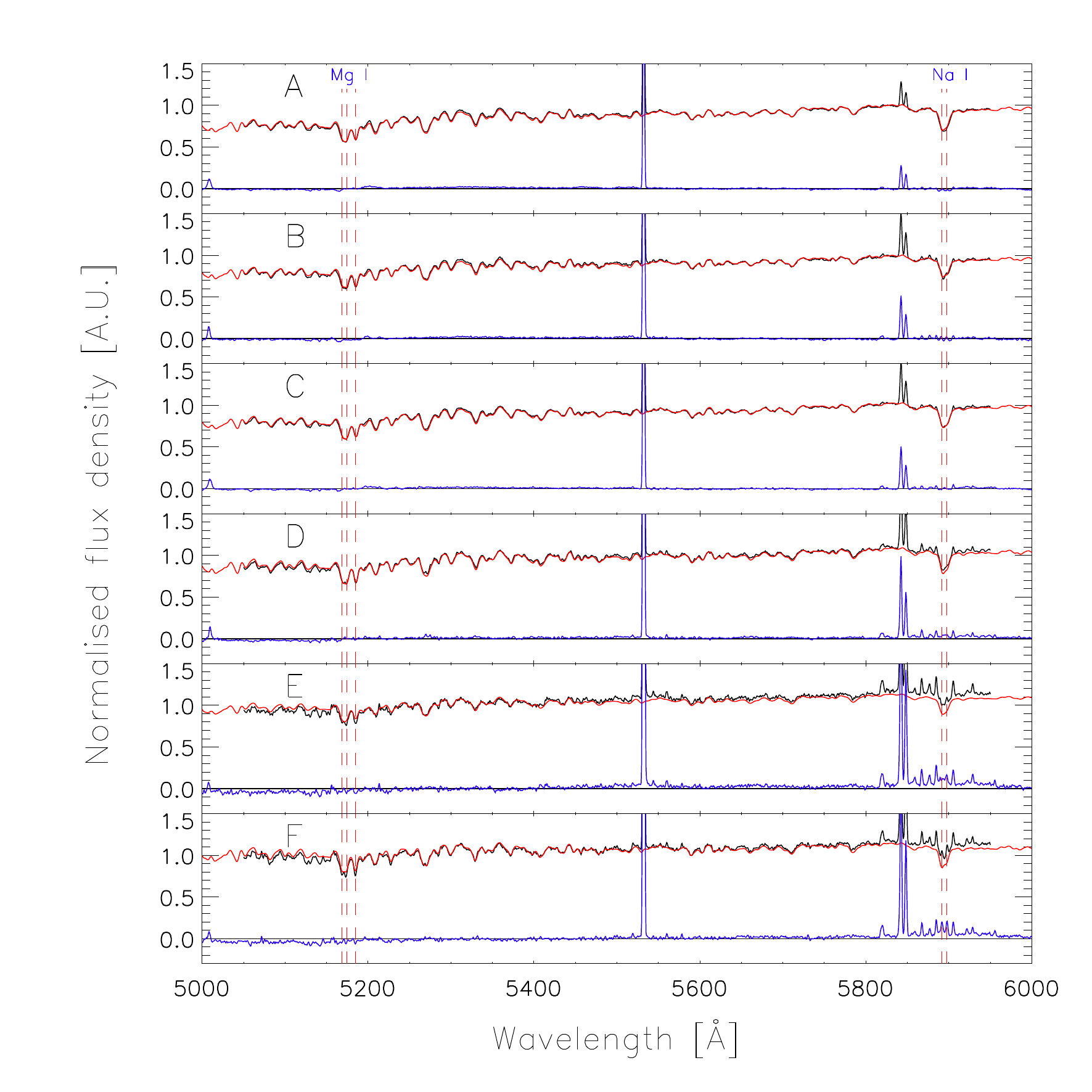}
  \caption{The spectra taken from  $3\times3$ spaxels apertures centered on the regions described in Fig. \ref{fig:diagnostic_maps}. The spectra are denoted accordingly. The vertical lines mark the rest wavelengths of the Mg b and the Na D absorption features.}
  \label{fig:NaD}
\end{figure*}

Another indication of the presence of an outflow is the blueshift of the absorbed sodium doublet (Na D) \citep[e.g.][]{Rupke_NaD, Martin_Outflows_NaD}. The blueshift is the signature of interstellar gas entrained in the outflow. Figure \ref{fig:NaD} shows the $5000$ to $6000~\AA$ spectra taken from $3\times3$ spaxels apertures centered on the different regions described in Fig. \ref{fig:diagnostic_maps}. The proper subtraction of the stellar fraction in Na D is of high importance to examine the absorption due to the interstellar gas. Of course, the subtraction works best on the regions, which have a higher S/N ratio. We compare the residuals of the stellar continuum subtraction in Na D with those in the purely stellar Mg b absorption and find them to be of the same order. In areas of lower S/N, the residuals are dominated by weak sky emission lines that contaminate the spectra. From this, we conclude that either a possible outflow component is not sufficiently significant to be detected in our data or NGC 5850 lacks the interstellar gas necessary to show the Na D absorption feature. \\

Considering the asymmetric line shapes and the Eddington ratio, the presence of an outflow appears possible but is not further supported by the presence of blueshifted Na D absorption features. Anyhow, a possible outflow is likely not linked to the extended LINER-like emission for orientation reasons. We come back to this topic again in Sect. \ref{kinematic_scenarios}.

\subsection{Photoionization by p-AGB stars}
\label{pAGB}

The photoionization of gas to LINER-like line ratios is not exclusive to AGNs, and indeed \citet{Liner_vs_retired} pointed out that a nonnegligible number of LINER classified galaxies in the SDSS are retired or passive galaxies dominated by the old stellar population in the post-asymptotic giant branch (p-AGB).

\begin{figure}
   \includegraphics[width=0.5\textwidth]{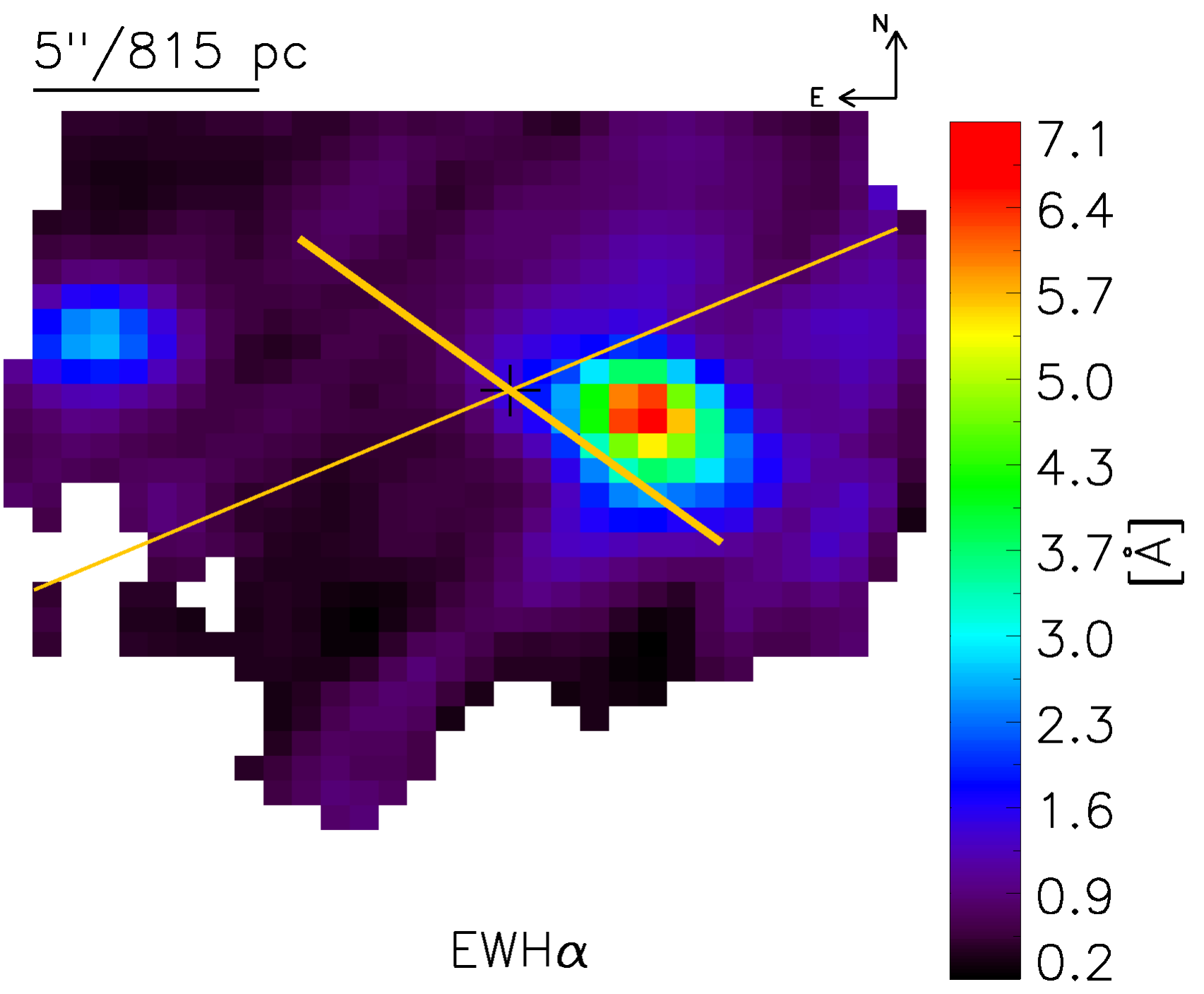}

  \caption{Map of the equivalent width of the H$\alpha$ emission line with the major axis of the primary (thin line) and the secondary bar (thick line) overplotted. The EWH$\alpha$ was calculated by using the emission after the subtraction of the stellar continuum. The continuum value was measured in the spectrum before the stellar population was subtracted.}
  \label{fig:EW_H_alpha}
\end{figure}

\citet{Liner_vs_retired} developed an alternative diagnostic diagram, which relies on the equivalent width of H$\alpha$ (from here on denoted as EWH$\alpha$) instead. With this new tool, it turned out that a number of formerly LINER classified galactic nuclei seem to be retired or passive galaxies. Empirically, \citet{Cid_EL_taxonomy} found a demarcation threshold at EWH$\alpha = 3~\AA$, where values below represent retired galaxies. We created an EWH$\alpha$ map of NGC 5850 (Fig. \ref{fig:EW_H_alpha}) to evaluate the importance of p-AGB stars as a possible ionization mechanism.\

On average, EWH$\alpha = (0.9\pm0.8)~\AA$ across the FOV. We note that only the inner \ion{H}{ii}-region shows values well above the threshold, indicating p-AGB stars are not the dominant ionization source. This agrees with our previous classification in the diagnostic diagrams, in which it was found to be dominated by SF. The eastern \ion{H}{ii}-region falls a little below the threshold value. In contrast to the central \ion{H}{ii}-region, the core region has EWH$\alpha$s well below $3~\AA$.\\	

We further explore the possibility of the presence of spatially distributed ionization sources (e.g. shocks, p-AGB stars, etc.). We follow the arguments of \citet{Yan_nature_of_LINERS}, where they tested whether increasing line ratios can be a result from a central ionizing source or not. They took advantage of two relations: the sensitivity of $\log\left(\left[\ion{O}{iii}\right]~\lambda5008/\left[\ion{S}{ii}\right]~\lambda\lambda6718+6732\right)$ to the ionization parameter $U$ and the electron density in giant ellipticals as a power law of the stellar density profile, $n_{\mathrm{e}} \propto \rho_\ast^{1/2}$. With the gas density profile $\sim r^{-1}$, the presence of a central ionizing source, and the flux scaling with $r^{-2}$, they conclude the ionization parameter and the line ratio have to decrease outwards, which contrasts their observations. These findings convinced them of distributed ionization sources being dominant rather than AGNs. A similar suggestion was made by \citet{Shields_nearby_nuclei}, also based on ionization gradients.\

\begin{figure*}
   \includegraphics[width=0.49\textwidth]{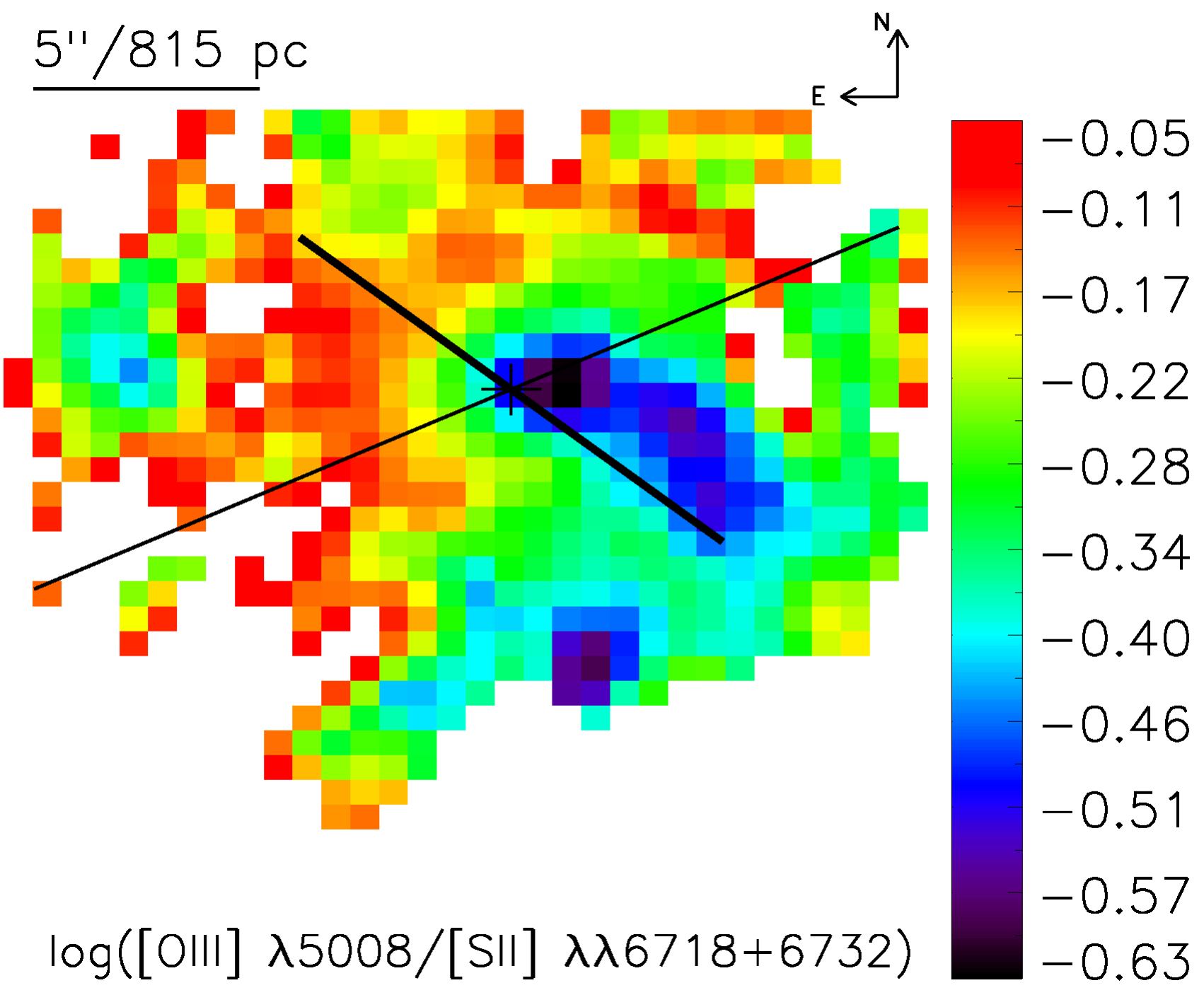}
   \includegraphics[width=0.49\textwidth]{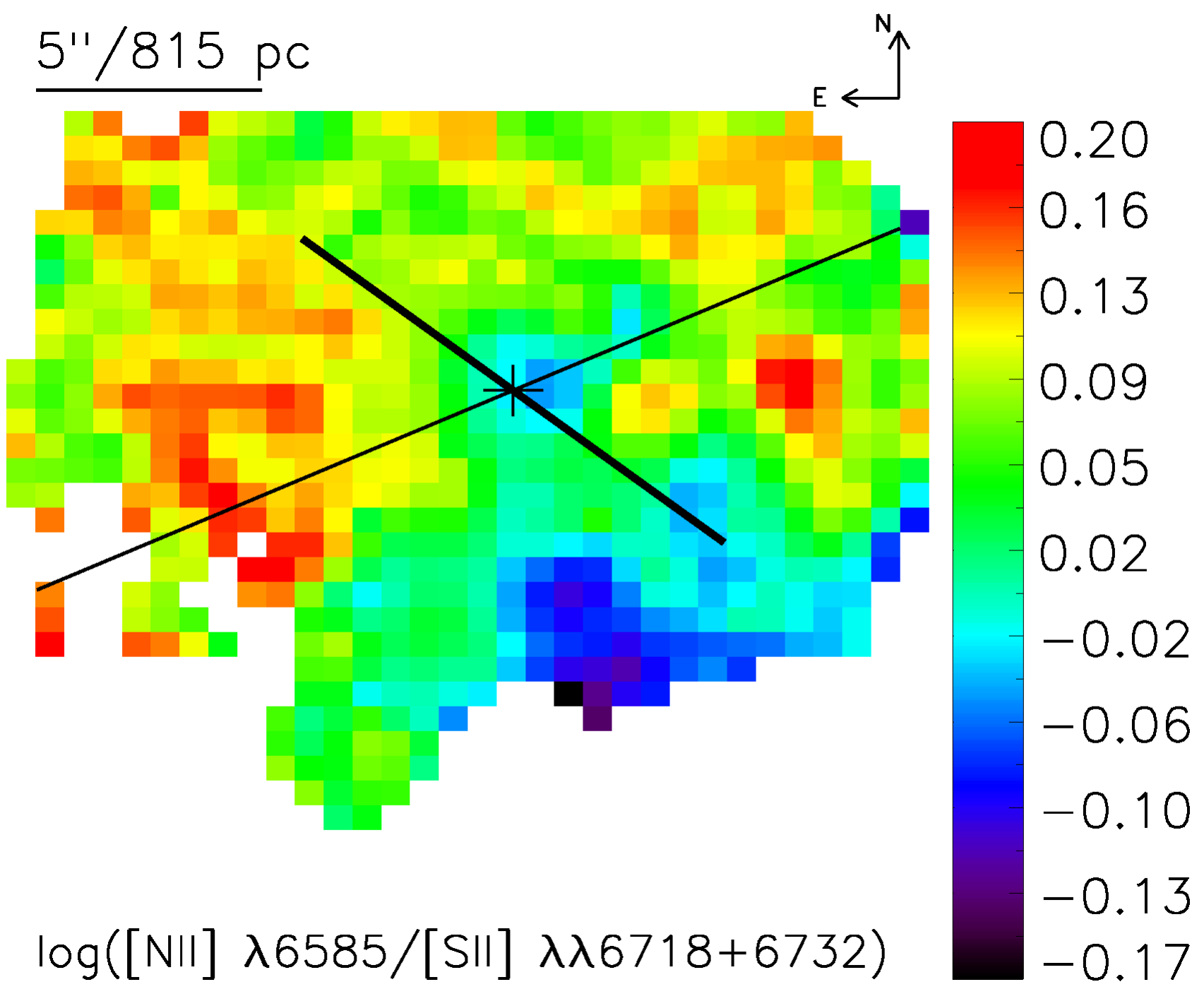}

  \caption{Logarithmic line ratio maps. The major axis of the primary (thin line) and the secondary bar (thick line) are overplotted.}
  \label{fig:SII_maps}
\end{figure*}

The line ratios in our data as presented in Fig. \ref{fig:fov_diagnostics} generally slightly decrease with distance to the core. In Fig. \ref{fig:SII_maps}, we show the line ratio maps of $\log\left(\left[\ion{O}{iii}\right]~\lambda5008/\left[\ion{S}{ii}\right]~\lambda\lambda6718+6732\right)$ and $\log\left(\left[\ion{N}{ii}\right]~\lambda6585/\left[\ion{S}{ii}\right]~\lambda\lambda6718+6732\right)$. We note that they indeed increase with distance to the center. We calculate the ionization parameter for the $3\times3$ pixels aperture on the nucleus with $U=Q(\mathrm{H})/(4 \pi r^{2} n_e c)$, where $Q(\mathrm{H})$ is the number of photons per second with an energy above the ionization energy of hydrogen, $r$ is the distance to the nucleus, $n_{\mathrm{e}}$ corresponds to the electron density, and the light speed $c$ is a factor used to make the ionization parameter dimensionless. If we consider that the electron density is radially almost constant within the KDC $(n_{\mathrm{e}}\sim 150~\mathrm{cm^{3}})$, then $U$ has to decrease with $r^{-2}$. This would result in a negative gradient of [\ion{O}{iii}]/[\ion{S}{ii}] and [\ion{N}{ii}]/[\ion{S}{ii}] as \citet{Yan_nature_of_LINERS} argue. This is in contrast to our observations. Consequently, $U$ must decrease shallower or even increase with distance. This cannot be achieved by a point source but by distributed ionization sources, possibly in combination with a point source. For an overview and model calculation on the radial behaviour of the ionization parameter in the presence of distributed ionization sources, we refer the reader to \citet{Yan_nature_of_LINERS}.\

The combination of the information on the radial behaviour of $U$ with the EWH$\alpha$ diagnostics make a strong case in favor of the stellar origin of ionization. A possible contribution of X-ray emitting gas to the ionization budget cannot be estimated, because there is no information on the presence of an X-ray halo around NGC 5850.\

\subsection{Kinematic scenarios}
\label{kinematic_scenarios}

Large scale bars are believed to remove angular momentum from the interstellar medium inside co-rotation, enabling it to fall inwards until it is stopped on the inner Lindblad resonance at a distance of the order of $100~\mathrm{pc}$. The matter there can form a ring \citep{Shlosman_bar_in_bar, Friedli_bar_in_bar}. The gas can proceed further inwards by an inner, secondary bar formed by gravitational instabilities or galaxy encounters \citep[e.g]{Hunt_NUGA}. Eventually, viscous processes could take over the transport at scales on the order of $10~\mathrm{pc}$. Viscous torques could also feed the central engine alternatively to the bar by forming a (temporary) gaseous disk that spans the complete distance from the nuclear ring to the SMBH \citep{Garcia-Burillo_torque}. Simulations by \citet{Hopkins_feeding} showed that lopsidedness ($m = 1$ mode) in the galaxy's gas distribution in the inner few $\mathrm{pc}$ could also support the flow to the center. The existence of the secondary bar in NGC 5850 has been shown by isophotal analysis in several papers (see Sect. \ref{Intro}). However, is the kinematic impact of the secondary bar responsible for the extended LINER-like emission?\

Shocks are almost always present in galaxies. The LINER-like areas northwest (region C) and south (region F) of the center could be interpreted as shock regions resulting from the kinematics of the secondary bar. This hypothesis would be supported by the steep H$\alpha$ velocity gradients. However, the LOSV dispersion of the gas appears flat and shows only a few values that would be consistent with shocks (i.e., $100-200\mathrm{~km~s^{-1}}$ for slow shocks). The number distribution of the dispersion values is not bimodal and cannot give a clear hint of the presence of shocks \citep{Rich_galaxy_shocks}.\\

At the northeastern tip of the inner bar (region D), H$\alpha$ shows signs of beam smearing with LOSV gradients not as steep as in the center. In this region, the sharp transition from the kinematically decoupled core (KDC) to the large scale kinematics of the galaxy takes place. Similar kinematic structures can be found at the opposite end of the secondary bar, though it is not that clear. Both features should be parts of a ring delineating the spatial limit of the core region. Observations with higher S/N ratio should show this feature more clearly.\ 

The LOSV field of the KDC appears to be rotational with a superimposed spiral perturbation. The position angle of the LOSV field of the ionized gas is $\mathrm{P.A._{kin,center} \approx 25 \degr}$ and thus very similar to the kinematic P.A. of \ion{H}{i} in the core region \citep[see Fig. 15a][]{Higdon_encounter}. In the outer parts of NGC 5850, the kinematic P.A. of \ion{H}{i} is about $\mathrm{P.A._{kin,outer} \approx -25\degr}$, comparable to the P.A. of our stellar LOSV field (Fig. \ref{fig:stellar_kinematics}). Since the minor axis of the stellar LOSV field is almost aligned with the nuclear bar, we do not expect any perturbations due to elongated orbits. Indeed, the stellar kinematics are very regular. If the ionized gas disk inside the ring would share the same plane with the outer gas disk, then we would not expect any perturbations. What then is the origin of the perturbations in the ionized gas? \

\begin{figure}
   \includegraphics[width=0.49\textwidth]{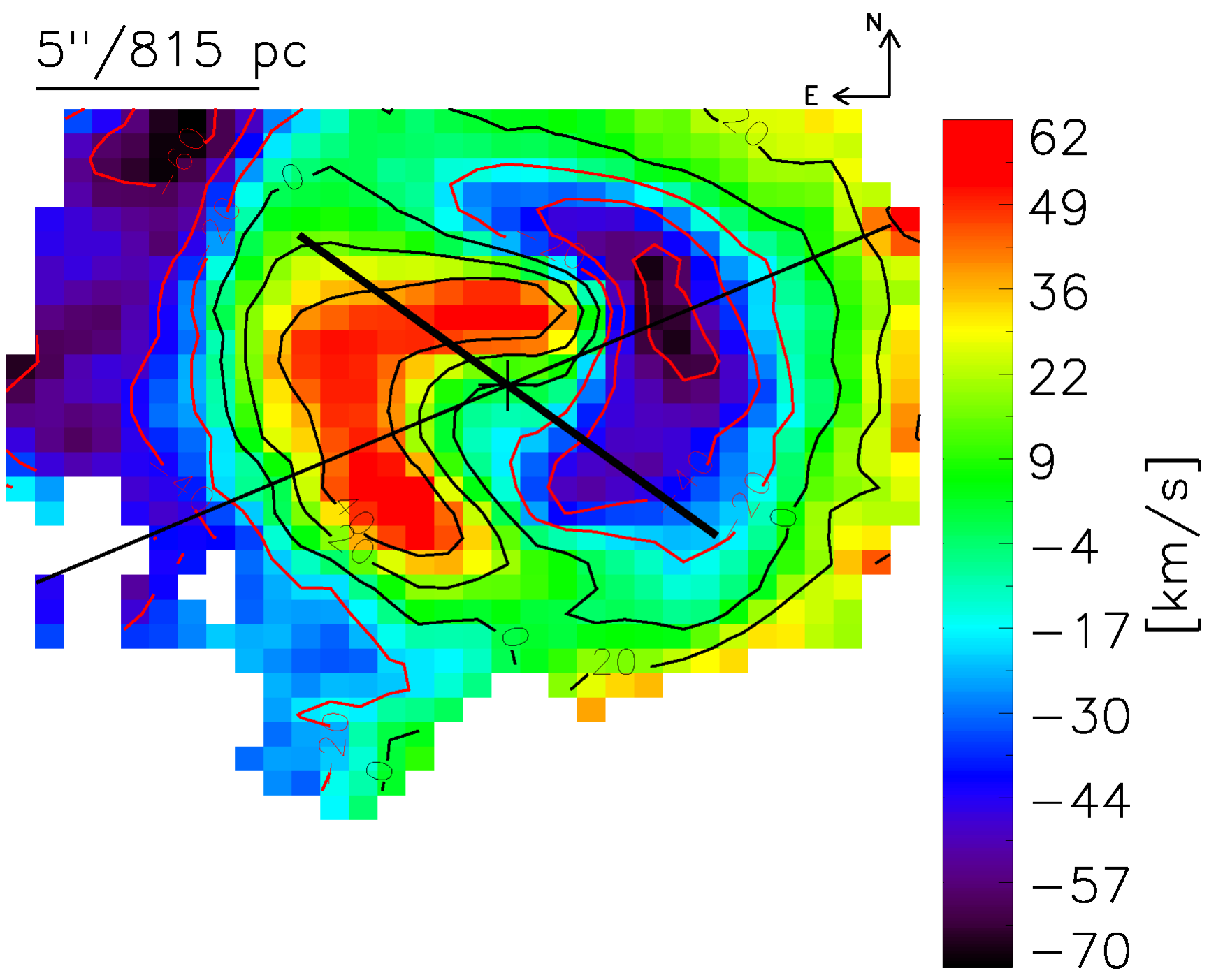}
  \caption{Differential LOSV map $\mathrm{\left(LOSV_{GAS}-LOSV_{STELLAR}\right)}$. The major axis of the primary (thin line) and the secondary bar (thick line) are overplotted. The isovelocity contours are superimposed for $0$, $20$, and $40\mathrm{~km~s^{-1}}$ in black and for $-20$, $-40$, and $-60\mathrm{~km~s^{-1}}$ in red. The spatial extend of the KDC is given by its circular structure. Within the KDC the underlying gas disk is counter-rotating with respect to the large scale gas distribution outside of the KDC.}
  \label{fig:wind}
\end{figure}

Though we have a weak SF region in the center (Sec. \ref{HII}) it is unlikely to be strong enough to perturb the gas sufficiently. The presence of outflows is discussed already in Sec. \ref{outflows} and could be a viable mechanism. If the outflow would be oriented along the plane of the inner disk, we would however expect a signature in the LOSV dispersion map with values of $\sim200~\mathrm{km~s^{-1}}$, which we do not observe. An outflow oriented along the nuclear bar would disturb the ring-like structure of the KDC that can be seen in the LOSV maps in Fig. \ref{fig:gas_kinematics} and \ref{fig:wind}. Therefore, the outflow has an orientation that does not allow interaction with the central disk.\ 

The perturbation observed in the LOSV maps could be due to the ionized gas disk inside the ring being tilted with respect to the outer disk. In \ion{H}{i}, the inclination for both gas disks is $i=30\degr$. With $\mathrm{P.A._{kin,center}}$ and $\mathrm{P.A._{kin,outer}}$, we calculate a separation of about $24\degr$. Since both planes are close in inclination the KDC precesses very quickly towards alignment and would eventually form a counter-rotating disk with respect to the outer one. This scenario contradicts the proposed presence of a gaseous polar disk \citep{doublebarred_structure_moissev}, since it requires the planes to be perpendicular to each other.\ 

The counter-rotation might have triggered the lopsidedness $(m=1)$ observed in the \ion{H}{i} and \element[][12]{CO} distribution \citep{Higdon_encounter,LeonCombes2000}, as simulated by \citet{NUGA_counter_rot}. The counter-rotation itself has its origin not only in the kinematic impact of the primary and the secondary bar but also requires a companion galaxy which drives the gas violently toward the center by tidal forces. As mentioned before, NGC 5850 experienced a high-velocity encounter with NGC 5846. The encounter might have triggered the accretion of the gas in the nuclear region, which then could have become counter-rotating. In Fig. \ref{fig:wind}, we present a differential LOSV image $\mathrm{\left(LOSV_{GAS}-LOSV_{STELLAR}\right)}$ to emphasize the peculiar gas motion and the kinematic distinctiveness of the KDC.\

It remains open whether the bar kinematics can sufficiently excite the central few arcseconds without a strong central ionizing source. To answer this question detailed calculations of the energy balance and modeling of the kinematics is required. This is beyond the scope of this paper.\

\section{Summary}
\label{summary}

We analyzed optical IFU data of the central $\sim 4~\mathrm{kpc}$ of NGC 5850 to examine the ionization mechanisms that cause LINER-like spectra and to assess the kinematic impact of the bar on the circumnuclear region. We found no clear indication of the presence of one single dominant ionization mechanism. Evidence is present of photoionization by distributed ionization sources, which is probably p-AGB stars. Nevertheless, the presence of a weak AGN and an outflow cannot be ruled out. We summarize our results as follows:

\begin{enumerate}
\item{Although extended, the LINER-like emission is peaked on the center. Neither radio nor X-ray observations can securely confirm the presence of an AGN. Using the $M$-
$\sigma$ relation, we find $\log(M_{\mathrm{BH}}/M_{\sun}) \approx 7.1$. Application of $L_{[\ion{O}{iii}]}$ as a proxy for $L_{\mathrm{bol}}$ yields the Eddington ratio to be $L_{\mathrm{bol}}/L_{\mathrm{Edd}}\leq10^{-2.2}$.}
\item{The EWH$\alpha$ is smaller than $3~\AA$ throughout the areas of extended LINER-like emission, except in the central SF-region. Furthermore, $\log([\ion{O}{iii}]/[\ion{S}{ii}])$ and $\log([\ion{N}{ii}]/[\ion{S}{ii}])$ radially increase. From this conclude on the presence of an old (p-AGB) stellar population. However, the discrimination to X-ray emitting gas is not possible due to the lack of suitable X-ray data.}
\item{Almost all regions not classified as LINER-like show composite ionization properties, which are likely due to a mixture of photoionization by SF with those by an AGN and LINER-like ionization.}
\item{We confirm the presence of a kinematically decoupled core (KDC) first observed by \citet{LeonCombes2000}.}
\item{The ionized-gas LOSV field possesses a strong spiral feature that hints at gas inflow toward the center. The perturbation is likely due to the ionized gas disk being tilted with respect to the outer disk. We calculated a tilting angle of about $24\degr$. Both planes precess quickly toward alignment. The counter-rotation might be triggered by the lopsided gas distribution, which could be a result from the high-speed encounter with NGC 5846.}
\item{We found indications for the presence of an outflow extending from south to the north of the center. However, the wing components are weak and coincide with strong velocity gradients. More observations are needed to confirm this. We do not find a connection between a possible outflow and the extended LINER-like emission.}
\item{A starforming region close to the center ($\sim 346~\mathrm{pc}$) was detected showing super-solar metallicities.}
\item{Possible evidence of shielding of the outer ISM by the central \ion{H}{ii}-region were found. However, more likely these decreased line ratios that form a tail oriented radially outward from the central \ion{H}{ii}-region are explained by enhanced SF.}
\item{The detection of the $\sigma$-hollows by \citet{sigma_hollows_lorenzo} was confirmed by us.}
\end{enumerate}

The lack of evidence to distinguish between the ionization by an old stellar population and an X-ray emitting gas make X-ray observations with higher spatial resolution necessary. Higher resolution observations in the optical regime are necessary to disentangle the beam-smearing problem from possible intrinsically increased velocity dispersion in regions of steep velocity gradients. This could answer the question of how much shocks are involved in the matter transport within the central few hundred parsec of NGC 5850. In the future, MUSE at the VLT will provide the capabilities necessary.\

Since NGC 5850 has been only weakly detected in \element[][12]{CO}, this work shows how well optical integral field spectroscopy can complement the NUGA survey. The impact of the secondary bar on the kinematics and, consequently, on the feeding mechanisms (the matter near to NUGA's heart) are of particularly high interest. NGC 5850 certainly deserves further investigation by theoretical models, which are going to be the subject of a future paper.\

\begin{acknowledgements}
M. B. likes to thank the anonymous referee for constructive comments that made the paper more complete. Furthermore, M.B. thanks L. Christensen for the support in the initial part of the data reduction process and gladly acknowledges the support of C. Iserlohe, M. Vitale and G. Busch. Also, M.B. wishes to thank the participants (especially L. Kewley and R. Yan) of the conference 'Dissecting Galaxies with 2D Wide-field Spectroscopy` for fruitful discussions and its organizers for a very interesting event. This work is carried out within the Collaborative Research Council SFB 956, sub-project [A2], funded by the Deutsche Forschungsgemeinschaft (DFG). M.B., J.S, J.Z. and S.F. are grateful for the travel support granted by the Group of Eight (Go8) of Australia and the German Academic Exchange Service (DAAD) (Go8 Australia-Germany Joint Research Co-operation Scheme). M.V.-S. thanks the support of the European Union Seventh Framework Programme Seven under grant agreement n° 312789. This research has made use of the NASA/IPAC Extragalactic Database (NED) which is operated by the Jet Propulsion Laboratory, California Institute of Technology, under contract with the National Aeronautics and Space Administration. Funding for the SDSS and SDSS-II has been provided by the Alfred P. Sloan Foundation, the Participating Institutions, the National Science Foundation, the U.S. Department of Energy, the National Aeronautics and Space Administration, the Japanese Monbukagakusho, the Max Planck Society, and the Higher Education Funding Council for England. The SDSS Web Site is http://www.sdss.org/. The SDSS is managed by the Astrophysical Research Consortium for the Participating Institutions. The Participating Institutions are the American Museum of Natural History, Astrophysical Institute Potsdam, University of Basel, University of Cambridge, Case Western Reserve University, University of Chicago, Drexel University, Fermilab, the Institute for Advanced Study, the Japan Participation Group, Johns Hopkins University, the Joint Institute for Nuclear Astrophysics, the Kavli Institute for Particle Astrophysics and Cosmology, the Korean Scientist Group, the Chinese Academy of Sciences (LAMOST), Los Alamos National Laboratory, the Max-Planck-Institute for Astronomy (MPIA), the Max-Planck-Institute for Astrophysics (MPA), New Mexico State University, Ohio State University, University of Pittsburgh, University of Portsmouth, Princeton University, the United States Naval Observatory, and the University of Washington.

\end{acknowledgements}

\bibliographystyle{aa} 
\bibliography{32_NGC5850_Rev_1_LE_1}

\begin{thebibliography}{93}
\expandafter\ifx\csname natexlab\endcsname\relax\def\natexlab#1{#1}\fi

\bibitem[{{Arribas} {et~al.}(2001){Arribas}, {Colina}, \&
  {Clements}}]{outflow_geometry}
{Arribas}, S., {Colina}, L., \& {Clements}, D. 2001, \apj, 560, 160

\bibitem[{{Asplund} {et~al.}(2009){Asplund}, {Grevesse}, {Sauval}, \&
  {Scott}}]{Asplund_solar_O_abundance}
{Asplund}, M., {Grevesse}, N., {Sauval}, A.~J., \& {Scott}, P. 2009, \araa, 47,
  481

\bibitem[{{Binette} {et~al.}(1994){Binette}, {Magris}, {Stasi{\'n}ska}, \&
  {Bruzual}}]{Binette_old_stars}
{Binette}, L., {Magris}, C.~G., {Stasi{\'n}ska}, G., \& {Bruzual}, A.~G. 1994,
  \aap, 292, 13

\bibitem[{{Blanc} {et~al.}(2009){Blanc}, {Heiderman}, {Gebhardt}, {Evans}, \&
  {Adams}}]{DIG}
{Blanc}, G.~A., {Heiderman}, A., {Gebhardt}, K., {Evans}, II, N.~J., \&
  {Adams}, J. 2009, \apj, 704, 842

\bibitem[{{Bruzual} \& {Charlot}(2003)}]{BC03}
{Bruzual}, G. \& {Charlot}, S. 2003, \mnras, 344, 1000

\bibitem[{{Buta} \& {Crocker}(1993)}]{Buta_5850}
{Buta}, R. \& {Crocker}, D.~A. 1993, \aj, 105, 1344

\bibitem[{{Calzetti}(1997)}]{Calzetti_extinction}
{Calzetti}, D. 1997, \aj, 113, 162

\bibitem[{{Calzetti}(2012)}]{Calzetti_SFR}
{Calzetti}, D. 2012, ArXiv e-prints

\bibitem[{{Cardelli} {et~al.}(1989){Cardelli}, {Clayton}, \&
  {Mathis}}]{Cardelli_extinction}
{Cardelli}, J.~A., {Clayton}, G.~C., \& {Mathis}, J.~S. 1989, \apj, 345, 245

\bibitem[{{Cid Fernandes} {et~al.}(2005){Cid Fernandes}, {Mateus}, {Sodr{\'e}},
  {Stasi{\'n}ska}, \& {Gomes}}]{starlight}
{Cid Fernandes}, R., {Mateus}, A., {Sodr{\'e}}, L., {Stasi{\'n}ska}, G., \&
  {Gomes}, J.~M. 2005, \mnras, 358, 363

\bibitem[{{Cid Fernandes} {et~al.}(2009){Cid Fernandes}, {Schlickmann},
  {Stasinska}, {Asari}, {Gomes}, {Schoenell}, {Mateus}, \&
  {Sodr{\'e}}}]{Cid_Fernandes_disconnection}
{Cid Fernandes}, R., {Schlickmann}, M., {Stasinska}, G., {et~al.} 2009, in
  Astronomical Society of the Pacific Conference Series, Vol. 408, The
  Starburst-AGN Connection, ed. W.~{Wang}, Z.~{Yang}, Z.~{Luo}, \& Z.~{Chen},
  122

\bibitem[{{Cid Fernandes} {et~al.}(2011){Cid Fernandes}, {Stasi{\'n}ska},
  {Mateus}, \& {Vale Asari}}]{Liner_vs_retired}
{Cid Fernandes}, R., {Stasi{\'n}ska}, G., {Mateus}, A., \& {Vale Asari}, N.
  2011, \mnras, 413, 1687

\bibitem[{{Cid Fernandes} {et~al.}(2010){Cid Fernandes}, {Stasi{\'n}ska}, {Vale
  Asari}, {Mateus}, {Schlickmann}, {Schoenell}, \&
  {Schoenell}}]{Cid_EL_taxonomy}
{Cid Fernandes}, R., {Stasi{\'n}ska}, G., {Vale Asari}, N., {et~al.} 2010, in
  IAU Symposium, Vol. 267, IAU Symposium, 65--72

\bibitem[{{Condon} {et~al.}(2002){Condon}, {Cotton}, \&
  {Broderick}}]{5850_radio}
{Condon}, J.~J., {Cotton}, W.~D., \& {Broderick}, J.~J. 2002, \aj, 124, 675

\bibitem[{{de Lorenzo-C{\'a}ceres} {et~al.}(2013){de Lorenzo-C{\'a}ceres},
  {Falc{\'o}n-Barroso}, \& {Vazdekis}}]{5850_lorenzo}
{de Lorenzo-C{\'a}ceres}, A., {Falc{\'o}n-Barroso}, J., \& {Vazdekis}, A. 2013,
  \mnras, 431, 2397

\bibitem[{{de Lorenzo-C{\'a}ceres} {et~al.}(2008){de Lorenzo-C{\'a}ceres},
  {Falc{\'o}n-Barroso}, {Vazdekis}, \&
  {Mart{\'{\i}}nez-Valpuesta}}]{sigma_hollows_lorenzo}
{de Lorenzo-C{\'a}ceres}, A., {Falc{\'o}n-Barroso}, J., {Vazdekis}, A., \&
  {Mart{\'{\i}}nez-Valpuesta}, I. 2008, \apjl, 684, L83

\bibitem[{{Dopita} \& {Sutherland}(1995)}]{Dopita_shock_producers}
{Dopita}, M.~A. \& {Sutherland}, R.~S. 1995, \apj, 455, 468

\bibitem[{{Dopita} \& {Sutherland}(1996)}]{Dopita_1996}
{Dopita}, M.~A. \& {Sutherland}, R.~S. 1996, \apjs, 102, 161

\bibitem[{{Dudik} {et~al.}(2009){Dudik}, {Satyapal}, \&
  {Marcu}}]{Dudik_LINER_AGN}
{Dudik}, R.~P., {Satyapal}, S., \& {Marcu}, D. 2009, \apj, 691, 1501

\bibitem[{Edlén(1966)}]{Edlen}
Edlén, B. 1966, Metrologia, 2, 71

\bibitem[{{Fabbiano} {et~al.}(1992){Fabbiano}, {Kim}, \&
  {Trinchieri}}]{5850_Lx}
{Fabbiano}, G., {Kim}, D.-W., \& {Trinchieri}, G. 1992, \apjs, 80, 531

\bibitem[{{Fabian}(1994)}]{Fabian_Cooling_Flows}
{Fabian}, A.~C. 1994, \araa, 32, 277

\bibitem[{{Farage} {et~al.}(2010){Farage}, {McGregor}, {Dopita}, \&
  {Bicknell}}]{Farage_spiral_in}
{Farage}, C.~L., {McGregor}, P.~J., {Dopita}, M.~A., \& {Bicknell}, G.~V. 2010,
  \apj, 724, 267

\bibitem[{{Ferrarese} \& {Merritt}(2000)}]{Ferrarese_M_sigma}
{Ferrarese}, L. \& {Merritt}, D. 2000, \apjl, 539, L9

\bibitem[{{Filippenko}(1982)}]{DAR}
{Filippenko}, A.~V. 1982, \pasp, 94, 715

\bibitem[{{Friedli} \& {Martinet}(1993)}]{Friedli_bar_in_bar}
{Friedli}, D. \& {Martinet}, L. 1993, \aap, 277, 27

\bibitem[{{Friedli} {et~al.}(1996){Friedli}, {Wozniak}, {Rieke}, {Martinet}, \&
  {Bratschi}}]{Friedli_5850}
{Friedli}, D., {Wozniak}, H., {Rieke}, M., {Martinet}, L., \& {Bratschi}, P.
  1996, \aaps, 118, 461

\bibitem[{{Garc{\'{\i}}a-Burillo} {et~al.}(2005){Garc{\'{\i}}a-Burillo},
  {Combes}, {Schinnerer}, {Boone}, \& {Hunt}}]{Garcia-Burillo_torque}
{Garc{\'{\i}}a-Burillo}, S., {Combes}, F., {Schinnerer}, E., {Boone}, F., \&
  {Hunt}, L.~K. 2005, \aap, 441, 1011

\bibitem[{{Garc{\'{\i}}a-Burillo} {et~al.}(2000){Garc{\'{\i}}a-Burillo},
  {Sempere}, {Combes}, {Hunt}, \& {Neri}}]{NUGA_counter_rot}
{Garc{\'{\i}}a-Burillo}, S., {Sempere}, M.~J., {Combes}, F., {Hunt}, L.~K., \&
  {Neri}, R. 2000, \aap, 363, 869

\bibitem[{{Gebhardt} {et~al.}(2000){Gebhardt}, {Bender}, {Bower}, {Dressler},
  {Faber}, {Filippenko}, {Green}, {Grillmair}, {Ho}, {Kormendy}, {Lauer},
  {Magorrian}, {Pinkney}, {Richstone}, \& {Tremaine}}]{Gebhardt_M_sigma}
{Gebhardt}, K., {Bender}, R., {Bower}, G., {et~al.} 2000, \apjl, 539, L13

\bibitem[{{Gonz{\'a}lez-Mart{\'{\i}}n}
  {et~al.}(2009){Gonz{\'a}lez-Mart{\'{\i}}n}, {Masegosa}, {M{\'a}rquez}, \&
  {Guainazzi}}]{Gonzalez_LINER_AGN}
{Gonz{\'a}lez-Mart{\'{\i}}n}, O., {Masegosa}, J., {M{\'a}rquez}, I., \&
  {Guainazzi}, M. 2009, \apj, 704, 1570

\bibitem[{{Gonz{\'a}lez-Mart{\'{\i}}n}
  {et~al.}(2006){Gonz{\'a}lez-Mart{\'{\i}}n}, {Masegosa}, {M{\'a}rquez},
  {Guerrero}, \& {Dultzin-Hacyan}}]{Gonzalez_LINER_Xray}
{Gonz{\'a}lez-Mart{\'{\i}}n}, O., {Masegosa}, J., {M{\'a}rquez}, I.,
  {Guerrero}, M.~A., \& {Dultzin-Hacyan}, D. 2006, \aap, 460, 45

\bibitem[{{Graham} {et~al.}(2011){Graham}, {Onken}, {Athanassoula}, \&
  {Combes}}]{Graham_M_sigma}
{Graham}, A.~W., {Onken}, C.~A., {Athanassoula}, E., \& {Combes}, F. 2011,
  \mnras, 412, 2211

\bibitem[{{G{\"u}ltekin} {et~al.}(2009){G{\"u}ltekin}, {Richstone}, {Gebhardt},
  {Lauer}, {Tremaine}, {Aller}, {Bender}, {Dressler}, {Faber}, {Filippenko},
  {Green}, {Ho}, {Kormendy}, {Magorrian}, {Pinkney}, \&
  {Siopis}}]{Gueltekin_M_sigma}
{G{\"u}ltekin}, K., {Richstone}, D.~O., {Gebhardt}, K., {et~al.} 2009, \apj,
  698, 198

\bibitem[{{Hanuschik}(2003)}]{sky_catalog}
{Hanuschik}, R.~W. 2003, \aap, 407, 1157

\bibitem[{{Heckman}(1980)}]{Heckman_Liner}
{Heckman}, T.~M. 1980, \aap, 87, 152

\bibitem[{{Heiles} {et~al.}(1981){Heiles}, {Kulkarni}, \&
  {Stark}}]{Heiles_hydro_extinct}
{Heiles}, C., {Kulkarni}, S., \& {Stark}, A.~A. 1981, \apjl, 247, L73

\bibitem[{{Higdon} {et~al.}(1998){Higdon}, {Buta}, \&
  {Purcell}}]{Higdon_encounter}
{Higdon}, J.~L., {Buta}, R.~J., \& {Purcell}, G.~B. 1998, \aj, 115, 80

\bibitem[{{Ho}(2008)}]{Ho_review}
{Ho}, L.~C. 2008, \araa, 46, 475

\bibitem[{{Ho} {et~al.}(1997){Ho}, {Filippenko}, \& {Sargent}}]{Ho_dwarf}
{Ho}, L.~C., {Filippenko}, A.~V., \& {Sargent}, W.~L.~W. 1997, \apjs, 112, 315

\bibitem[{{Hopkins} \& {Quataert}(2010)}]{Hopkins_feeding}
{Hopkins}, P.~F. \& {Quataert}, E. 2010, \mnras, 407, 1529

\bibitem[{{Hummel} {et~al.}(1987){Hummel}, {van der Hulst}, {Keel}, \&
  {Kennicutt}}]{Hummel_radio}
{Hummel}, E., {van der Hulst}, J.~M., {Keel}, W.~C., \& {Kennicutt}, Jr., R.~C.
  1987, \aaps, 70, 517

\bibitem[{{Hunt} {et~al.}(2008){Hunt}, {Combes}, {Garc{\'{\i}}a-Burillo},
  {Schinnerer}, {Krips}, {Baker}, {Boone}, {Eckart}, {L{\'e}on}, {Neri}, \&
  {Tacconi}}]{Hunt_NUGA}
{Hunt}, L.~K., {Combes}, F., {Garc{\'{\i}}a-Burillo}, S., {et~al.} 2008, \aap,
  482, 133

\bibitem[{{Jullo} {et~al.}(2008){Jullo}, {Christensen}, {Smette}, {Bagnulo},
  {Izzo}, \& {Marconi}}]{Jullo_fringing}
{Jullo}, E., {Christensen}, L., {Smette}, A., {et~al.} 2008, in 2007 ESO
  Instrument Calibration Workshop, ed. A.~{Kaufer} \& F.~{Kerber}, 343

\bibitem[{{Kauffmann} {et~al.}(2003){Kauffmann}, {Heckman}, {Tremonti},
  {Brinchmann}, {Charlot}, {White}, {Ridgway}, {Brinkmann}, {Fukugita}, {Hall},
  {Ivezi{\'c}}, {Richards}, \& {Schneider}}]{Kauffmann_pure_SF}
{Kauffmann}, G., {Heckman}, T.~M., {Tremonti}, C., {et~al.} 2003, Monthly Notes
  of the Royal Astron. Soc., 346, 1055

\bibitem[{{Keel}(1983)}]{Liner_reference}
{Keel}, W.~C. 1983, \apjs, 52, 229

\bibitem[{{Kehrig} {et~al.}(2012){Kehrig}, {Monreal-Ibero}, {Papaderos},
  {V{\'{\i}}lchez}, {Gomes}, {Masegosa}, {S{\'a}nchez}, {Lehnert}, {Cid
  Fernandes}, {Bland-Hawthorn}, {Bomans}, {Marquez}, {Mast}, {Aguerri},
  {L{\'o}pez-S{\'a}nchez}, {Marino}, {Pasquali}, {Perez}, {Roth},
  {S{\'a}nchez-Bl{\'a}zquez}, \& {Ziegler}}]{Kehrig_califa}
{Kehrig}, C., {Monreal-Ibero}, A., {Papaderos}, P., {et~al.} 2012, \aap, 540,
  A11

\bibitem[{{Kennicutt}(1998)}]{Kennicutt_SFR}
{Kennicutt}, Jr., R.~C. 1998, \araa, 36, 189

\bibitem[{{Kewley} {et~al.}(2001){Kewley}, {Dopita}, {Sutherland}, {Heisler},
  \& {Trevena}}]{Kewley_maximum_starburst}
{Kewley}, L.~J., {Dopita}, M.~A., {Sutherland}, R.~S., {Heisler}, C.~A., \&
  {Trevena}, J. 2001, Astroph. J., 556, 121

\bibitem[{{Kewley} {et~al.}(2006){Kewley}, {Groves}, {Kauffmann}, \&
  {Heckman}}]{Kewley_Sey_LINER}
{Kewley}, L.~J., {Groves}, B., {Kauffmann}, G., \& {Heckman}, T. 2006, Monthly
  Notes of the Royal Astron. Soc., 372, 961

\bibitem[{{Kroupa}(2001)}]{Kroupa_IMF}
{Kroupa}, P. 2001, \mnras, 322, 231

\bibitem[{{Lagerholm} {et~al.}(2012){Lagerholm}, {Kuntschner}, {Cappellari},
  {Krajnovi{\'c}}, {McDermid}, \& {Rejkuba}}]{Lagerholm_fringing}
{Lagerholm}, C., {Kuntschner}, H., {Cappellari}, M., {et~al.} 2012, \aap, 541,
  A82

\bibitem[{{Laine} {et~al.}(2002){Laine}, {Shlosman}, {Knapen}, \&
  {Peletier}}]{Bar_fraction}
{Laine}, S., {Shlosman}, I., {Knapen}, J.~H., \& {Peletier}, R.~F. 2002, \apj,
  567, 97

\bibitem[{{Le F{\`e}vre} {et~al.}(2003){Le F{\`e}vre}, {Saisse}, {Mancini},
  {Brau-Nogue}, {Caputi}, {Castinel}, {D'Odorico}, {Garilli}, {Kissler-Patig},
  {Lucuix}, {Mancini}, {Pauget}, {Sciarretta}, {Scodeggio}, {Tresse}, \&
  {Vettolani}}]{VIMOS}
{Le F{\`e}vre}, O., {Saisse}, M., {Mancini}, D., {et~al.} 2003, in Society of
  Photo-Optical Instrumentation Engineers (SPIE) Conference Series, Vol. 4841,
  Society of Photo-Optical Instrumentation Engineers (SPIE) Conference Series,
  ed. {M.~Iye \& A.~F.~M.~Moorwood}, 1670--1681

\bibitem[{{Leon} {et~al.}(2000){Leon}, {Combes}, \& {Friedli}}]{LeonCombes2000}
{Leon}, S., {Combes}, F., \& {Friedli}, D. 2000, in Astronomical Society of the
  Pacific Conference Series, Vol. 197, Dynamics of Galaxies: from the Early
  Universe to the Present, ed. {F.~Combes, G.~A.~Mamon, \& V.~Charmandaris}, 61

\bibitem[{{Leroy} {et~al.}(2012){Leroy}, {Bigiel}, {de Blok}, {Boissier},
  {Bolatto}, {Brinks}, {Madore}, {Munoz-Mateos}, {Murphy}, {Sandstrom},
  {Schruba}, \& {Walter}}]{Leroy_SFR}
{Leroy}, A.~K., {Bigiel}, F., {de Blok}, W.~J.~G., {et~al.} 2012, \aj, 144, 3

\bibitem[{{L{\'{\i}}pari} {et~al.}(2004){L{\'{\i}}pari}, {Mediavilla},
  {Garcia-Lorenzo}, {D{\'{\i}}az}, {Acosta-Pulido}, {Ag{\"u}ero}, {Taniguchi},
  {Dottori}, \& {Terlevich}}]{Lipari_merger_winds}
{L{\'{\i}}pari}, S., {Mediavilla}, E., {Garcia-Lorenzo}, B., {et~al.} 2004,
  \mnras, 355, 641

\bibitem[{{Lourenso} {et~al.}(2003){Lourenso}, {Aguerri}, {Vazdekis},
  {Beckman}, \& {Peletier}}]{Lourenso_stellar}
{Lourenso}, S., {Aguerri}, J.~A.~L., {Vazdekis}, A., {Beckman}, J.~E., \&
  {Peletier}, R.~F. 2003, \apss, 284, 925

\bibitem[{{Maiolino} {et~al.}(2008){Maiolino}, {Nagao}, {Grazian}, {Cocchia},
  {Marconi}, {Mannucci}, {Cimatti}, {Pipino}, {Ballero}, {Calura}, {Chiappini},
  {Fontana}, {Granato}, {Matteucci}, {Pastorini}, {Pentericci}, {Risaliti},
  {Salvati}, \& {Silva}}]{Maiolino_metallicity}
{Maiolino}, R., {Nagao}, T., {Grazian}, A., {et~al.} 2008, \aap, 488, 463

\bibitem[{{Maoz} {et~al.}(1998){Maoz}, {Koratkar}, {Shields}, {Ho},
  {Filippenko}, \& {Sternberg}}]{Maoz_UV_LINER}
{Maoz}, D., {Koratkar}, A., {Shields}, J.~C., {et~al.} 1998, \aj, 116, 55

\bibitem[{{Martin}(2005)}]{Martin_Outflows_NaD}
{Martin}, C.~L. 2005, \apj, 621, 227

\bibitem[{{Masegosa} {et~al.}(2011){Masegosa}, {M{\'a}rquez}, {Ramirez}, \&
  {Gonz{\'a}lez-Mart{\'{\i}}n}}]{Masegosa_Halpha_LINER}
{Masegosa}, J., {M{\'a}rquez}, I., {Ramirez}, A., \&
  {Gonz{\'a}lez-Mart{\'{\i}}n}, O. 2011, \aap, 527, A23

\bibitem[{{Moiseev} {et~al.}(2004){Moiseev}, {Vald{\'e}s}, \&
  {Chavushyan}}]{doublebarred_structure_moissev}
{Moiseev}, A.~V., {Vald{\'e}s}, J.~R., \& {Chavushyan}, V.~H. 2004, \aap, 421,
  433

\bibitem[{{Monreal-Ibero} {et~al.}(2010){Monreal-Ibero}, {Arribas}, {Colina},
  {Rodr{\'{\i}}guez-Zaur{\'{\i}}n}, {Alonso-Herrero}, \&
  {Garc{\'{\i}}a-Mar{\'{\i}}n}}]{Monreal_Ibero_LIRG}
{Monreal-Ibero}, A., {Arribas}, S., {Colina}, L., {et~al.} 2010, \aap, 517, A28

\bibitem[{{Morton}(1991)}]{Morton}
{Morton}, D.~C. 1991, \apjs, 77, 119

\bibitem[{{Nagar} {et~al.}(2005){Nagar}, {Falcke}, \&
  {Wilson}}]{Nagar_LLAGN_radio}
{Nagar}, N.~M., {Falcke}, H., \& {Wilson}, A.~S. 2005, \aap, 435, 521

\bibitem[{{Netzer}(2009)}]{Netzer_L_bol}
{Netzer}, H. 2009, \mnras, 399, 1907

\bibitem[{{Osterbrock} \& {Ferland}(2006)}]{Osterbrock_book}
{Osterbrock}, D.~E. \& {Ferland}, G.~J. 2006, {Astrophysics of gaseous nebulae
  and active galactic nuclei}

\bibitem[{{Park} {et~al.}(2012){Park}, {Kelly}, {Woo}, \&
  {Treu}}]{Park_M_sigma}
{Park}, D., {Kelly}, B.~C., {Woo}, J.-H., \& {Treu}, T. 2012, \apjs, 203, 6

\bibitem[{{Pettini} \& {Pagel}(2004)}]{Pettini_metallicity}
{Pettini}, M. \& {Pagel}, B.~E.~J. 2004, \mnras, 348, L59

\bibitem[{{Pogge} {et~al.}(2000){Pogge}, {Maoz}, {Ho}, \&
  {Eracleous}}]{Pogge_2000}
{Pogge}, R.~W., {Maoz}, D., {Ho}, L.~C., \& {Eracleous}, M. 2000, \apj, 532,
  323

\bibitem[{{Predehl} \& {Schmitt}(1995)}]{Predehl_hydro_extinct}
{Predehl}, P. \& {Schmitt}, J.~H.~M.~M. 1995, \aap, 293, 889

\bibitem[{{Prieto} {et~al.}(1997){Prieto}, {Gottesman}, {Aguerri}, \&
  {Varela}}]{Prieto}
{Prieto}, M., {Gottesman}, S.~T., {Aguerri}, J.-A.~L., \& {Varela}, A.-M. 1997,
  \aj, 114, 1413

\bibitem[{{Rich} {et~al.}(2010){Rich}, {Dopita}, {Kewley}, \&
  {Rupke}}]{Rich_shock_superwind}
{Rich}, J.~A., {Dopita}, M.~A., {Kewley}, L.~J., \& {Rupke}, D.~S.~N. 2010,
  \apj, 721, 505

\bibitem[{{Rich} {et~al.}(2011){Rich}, {Kewley}, \&
  {Dopita}}]{Rich_galaxy_shocks}
{Rich}, J.~A., {Kewley}, L.~J., \& {Dopita}, M.~A. 2011, \apj, 734, 87

\bibitem[{{Rupke} {et~al.}(2005){Rupke}, {Veilleux}, \& {Sanders}}]{Rupke_NaD}
{Rupke}, D.~S., {Veilleux}, S., \& {Sanders}, D.~B. 2005, \apjs, 160, 87

\bibitem[{{Sarzi} {et~al.}(2010){Sarzi}, {Shields}, {Schawinski}, {Jeong},
  {Shapiro}, {Bacon}, {Bureau}, {Cappellari}, {Davies}, {de Zeeuw}, {Emsellem},
  {Falc{\'o}n-Barroso}, {Krajnovi{\'c}}, {Kuntschner}, {McDermid}, {Peletier},
  {van den Bosch}, {van de Ven}, \& {Yi}}]{Sarzi_pAGB}
{Sarzi}, M., {Shields}, J.~C., {Schawinski}, K., {et~al.} 2010, \mnras, 402,
  2187

\bibitem[{{Scharw{\"a}chter} {et~al.}(2011){Scharw{\"a}chter}, {Dopita},
  {Zuther}, {Fischer}, {Komossa}, \& {Eckart}}]{Julia_EELR}
{Scharw{\"a}chter}, J., {Dopita}, M.~A., {Zuther}, J., {et~al.} 2011, \aj, 142,
  43

\bibitem[{{Sharp} \& {Bland-Hawthorn}(2010)}]{Sharp_cones}
{Sharp}, R.~G. \& {Bland-Hawthorn}, J. 2010, \apj, 711, 818

\bibitem[{{Shen} {et~al.}(2011){Shen}, {Richards}, {Strauss}, {Hall},
  {Schneider}, {Snedden}, {Bizyaev}, {Brewington}, {Malanushenko},
  {Malanushenko}, {Oravetz}, {Pan}, \& {Simmons}}]{Shen_L_bol}
{Shen}, Y., {Richards}, G.~T., {Strauss}, M.~A., {et~al.} 2011, \apjs, 194, 45

\bibitem[{{Shi} {et~al.}(2006){Shi}, {Gu}, \& {Peng}}]{Shi_CNSF}
{Shi}, L., {Gu}, Q.~S., \& {Peng}, Z.~X. 2006, \aap, 450, 15

\bibitem[{{Shields} {et~al.}(2007){Shields}, {Rix}, {Sarzi}, {Barth},
  {Filippenko}, {Ho}, {McIntosh}, {Rudnick}, \&
  {Sargent}}]{Shields_nearby_nuclei}
{Shields}, J.~C., {Rix}, H.-W., {Sarzi}, M., {et~al.} 2007, \apj, 654, 125

\bibitem[{{Shlosman} {et~al.}(1989){Shlosman}, {Frank}, \&
  {Begelman}}]{Shlosman_bar_in_bar}
{Shlosman}, I., {Frank}, J., \& {Begelman}, M.~C. 1989, \nat, 338, 45

\bibitem[{{Stasi{\'n}ska} {et~al.}(2008){Stasi{\'n}ska}, {Vale Asari}, {Cid
  Fernandes}, {Gomes}, {Schlickmann}, {Mateus}, {Schoenell}, {Sodr{\'e}}, \&
  {Seagal Collaboration}}]{Stasinska_pAGB}
{Stasi{\'n}ska}, G., {Vale Asari}, N., {Cid Fernandes}, R., {et~al.} 2008,
  \mnras, 391, L29

\bibitem[{{Terlevich} \& {Melnick}(1985)}]{Terlevich_warmers}
{Terlevich}, R. \& {Melnick}, J. 1985, \mnras, 213, 841

\bibitem[{{Trinchieri} \& {di Serego Alighieri}(1991)}]{Trinchieri_pAGB}
{Trinchieri}, G. \& {di Serego Alighieri}, S. 1991, \aj, 101, 1647

\bibitem[{{Valencia-S.} {et~al.}(2012){Valencia-S.}, {Zuther}, {Eckart},
  {Garc{\'{\i}}a-Mar{\'{\i}}n}, {Iserlohe}, \& {Wright}}]{Monik}
{Valencia-S.}, M., {Zuther}, J., {Eckart}, A., {et~al.} 2012, \aap, 544, A129

\bibitem[{{Veilleux} \& {Rupke}(2002)}]{Veilleux_wind_ratio}
{Veilleux}, S. \& {Rupke}, D.~S. 2002, \apjl, 565, L63

\bibitem[{{Voit} \& {Donahue}(1997)}]{Voit_cooling_flow}
{Voit}, G.~M. \& {Donahue}, M. 1997, \apj, 486, 242

\bibitem[{{Walsh} {et~al.}(2008){Walsh}, {Barth}, {Ho}, {Filippenko}, {Rix},
  {Shields}, {Sarzi}, \& {Sargent}}]{Walsh_Outflow_LINER}
{Walsh}, J.~L., {Barth}, A.~J., {Ho}, L.~C., {et~al.} 2008, \aj, 136, 1677

\bibitem[{{Wisnioski} {et~al.}(2012){Wisnioski}, {Glazebrook}, {Blake},
  {Poole}, {Green}, {Wyder}, \& {Martin}}]{Wisnioski_HII_sigma}
{Wisnioski}, E., {Glazebrook}, K., {Blake}, C., {et~al.} 2012, \mnras, 422,
  3339

\bibitem[{{Wozniak} {et~al.}(1995){Wozniak}, {Friedli}, {Martinet}, {Martin},
  \& {Bratschi}}]{Wozniak95}
{Wozniak}, H., {Friedli}, D., {Martinet}, L., {Martin}, P., \& {Bratschi}, P.
  1995, Astron. And Astroph., 111, 115

\bibitem[{{Yan} \& {Blanton}(2012)}]{Yan_nature_of_LINERS}
{Yan}, R. \& {Blanton}, M.~R. 2012, \apj, 747, 61

\end{thebibliography}


%
%
%

\end{document}